\documentclass[twocolumn]{aa}
\usepackage{graphicx}
\usepackage{txfonts}
\usepackage{rotating}
\def\XMM{{\sl XMM-Newton}}
\def\ns{log$N$-log$S$}
\begin{document}
   \title{The XMM-Newton serendipitous survey\footnote{Based on observations
obtained with XMM-newton, an ESA science mission with instruments and
contributions funded by ESA Member States and the USA (NASA).}}

   \subtitle{IV. The AXIS X-ray source counts and angular clustering}

   \author{F.J. Carrera\inst{1}
          \and
          J. Ebrero\inst{1}
          \and
          S. Mateos\inst{1,3}
          \and
          M.T. Ceballos\inst{1}
          \and
          A. Corral\inst{1}
          \and
          X. Barcons\inst{1}
          \and
          M.J. Page\inst{2}
          \and
          S.R. Rosen\inst{2,3}
          \and
          M.G. Watson\inst{3}
          \and
          J. Tedds\inst{3}
          \and
          R. Della Ceca\inst{4}
          \and
          T. Maccacaro\inst{4}
          \and
          H. Brunner\inst{5}
          \and
          M. Freyberg\inst{5}
          \and
          G. Lamer\inst{6}
          \and
          F.E. Bauer\inst{7}
          \and
          Y. Ueda\inst{8}
          }

   \offprints{F.J. Carrera, \email{carreraf@ifca.unican.es}}

   \institute{Instituto de F\'\i sica de Cantabria (CSIC-UC), Avenida de los
Castros, 39005 Santander, Spain
         \and
             Mullard Space Science Laboratory, University College London,
Holmbury St. Mary, Dorking, Surrey, RH5 6NT, UK
         \and
             Department of Physics and Astronomy, University of Leicester, LE1 7RH, UK
        \and
             Osservatorio Astronomico di Brera, via Brera 28, 20121 Milano, Italy
        \and
             Max-Planck-Institut f\"ur extraterrestrische Physik,
Postfach 1312, 85741 Garching, Germany
        \and
             Astrophysikalisches Institut Potsdam, An der Sternwarte 16, 14482
Potsdam, Germany
	\and
             Columbia Astrophysics Laboratory, Columbia University, Pupin
Laboratories, 550 West 120th Street, Room 1418, New York, NY 10027, US
	\and
	     Department of Astronomy, Kyoto University, Kyoto 606-8502, Japan\\
             }

   \date{Received <date> / Accepted <date>}

\abstract{Recent results have revised upwards the total X-ray background (XRB)
intensity below $\sim$10~keV, therefore an accurate determination of the
source counts is needed. There are also contradicting results on the clustering
of X-ray selected sources.}{We have studied the X-ray source counts in four
energy bands soft (0.5-2~keV), hard (2-10~keV), XID (0.5-4.5~keV) and
ultra-hard (4.5-7.5~keV), to evaluate the contribution of sources at different
fluxes to the X-ray background. We have also studied the angular clustering of
X-ray sources in those bands.}{AXIS (An \XMM{} International Survey) is a
survey of 36 high Galactic latitude \XMM{} observations covering 4.8~deg$^2$
and containing 1433 serendipitous X-ray sources detected with 5-$\sigma$
significance.  This survey has similar depth to the \XMM{} catalogues and can
serve as a pathfinder to explore their possibilities.  We have combined this
survey with shallower and deeper surveys, and fitted the source counts with a
Maximum Likelihood technique. Using only AXIS sources, we have studied the
angular correlation using a novel robust technique.}{Our source counts results
are compatible with most previous samples in the soft, XID, ultra-hard and
hard bands. We have improved on previous results in the latter band. The
fractions of the XRB resolved in the surveys used in this work are 87\%, 85\%,
60\% and 25\% in the soft, hard, XID and ultra-hard bands,
respectively. Extrapolation of our source counts to zero flux are not enough to
saturate the XRB intensity. Only galaxies and/or absorbed AGN may be able
contribute the remaining unresolved XRB intensity. Our results are compatible,
within the errors, with recent revisions of the XRB intensity in the soft and
hard bands.  The maximum fractional contribution to the XRB comes from fluxes
within about a decade of the break in the source counts ($\sim 10^{-14}$~cgs),
reaching $\sim$50\% of the total in the soft and hard bands. Angular clustering
(widely distributed over the sky and not confined to a few deep fields) is
detected at 99-99.9\% significance in the soft and XID bands, with no detection
in the hard and ultra-hard band (probably due to the smaller number of
sources). We cannot confirm the detection of significantly stronger clustering
in the hard-spectrum hard sources.}{Medium depth surveys such as AXIS are
essential to determine the evolution of the X-ray emission in the Universe
below 10~keV.}


   \maketitle
%

\section{Introduction}

The X-ray background (Giacconi et al. \cite{Giacconi62}) is a testimony of the
history of accretion power in the Universe (Fabian \& Iwasawa
\cite{FabianIwasawa99}, So\l{}tan \cite{Soltan82}), and as such its sources and
their evolution have attracted considerable attention (Fabian \& Barcons
\cite{FabianBarcons92}).  Deep pencil-beam surveys (Loaring et
al. \cite{Loaring05}, Bauer et al. \cite{Bauer04}, Cowie et al. \cite{Cowie02},
Giacconi et al. \cite{Giacconi02}, Hasinger et al. \cite{Hasinger01}) have
resolved most
($\gse 90$\%) of the soft (0.5-2~keV) XRB into individual
sources. Most of these turn out to be unobscured and obscured Active
Galactic Nuclei (AGN), while at the fainter fluxes a population of
"normal" galaxies starts to contribute significantly to the source
counts (Hornschemeier et al. \cite{Hornschemeier03}). At higher
energies the resolved fraction is smaller (Worsley et
al. \cite{Worsley04}): 50\% between 4.5 and 7.5 keV, and less than
50\% between 7.5~keV and 10~keV. Models for the XRB based on the
Unified Model for AGN reproduce successfully its spectrum with a
combination of unobscured and obscured AGN, the later being the
dominant population (Setti \& Woltjer \cite{Setti89}, Gilli et
al. \cite{Gilli01}). Recent hard X-ray AGN luminosity function results
(Ueda et al. \cite{Ueda03}, La Franca et al. \cite{LaFranca05})
show that such simple models need considerable revision, with lower
absorbed AGN fraction at higher luminosities and luminosity dependent
density evolution of the hard X-ray AGN luminosity.

The fainter sources from deep surveys do not contribute much to the
final XRB intensity, and are often too faint optically and in X-rays
to be studied individually. Stacking of X-ray spectra is a valuable
tool in these circumstances (Worsley et al. \cite{Worsley06}, Civano,
Comastri \& Brusa \cite{Civano05}), but only average properties can be
studied in this fashion, missing all the diversity in the nature of
the sources necessary to account for the XRB. Wide shallow surveys
over a large portion of the sky (Schwope et al. \cite{Schwope00},
Della Ceca et al. \cite{DellaCeca04}, Ueda et al. \cite{Ueda03}) allow
detecting minority populations and studies of sources on an individual
basis, and serve as a framework against which to study and detect
evolution, but again with only a minor contribution to the XRB.

Medium depth wide-band surveys combining many sources with relatively wide sky
coverage ( Hasinger et al. \cite{Hasinger07}, Pierre et
al. \cite{Pierre04}, Barcons et al. \cite{Barcons02}, Baldi et
al. \cite{Baldi02}) have the potential to provide many of the missing pieces of
the XRB/AGN puzzle, since most (30-50\%) of the XRB below 10~keV comes from
sources at intermediate fluxes (Fabian \& Barcons \cite{FabianBarcons92},
Barcons, Carrera \& Ceballos \cite{Barcons06a}), and the optical and X-ray
spectra of many sources can be studied individually (e.g. Mateos et
al. \cite{Mateos05}). Furthermore, extensive \XMM{} source catalogues are
available (1XMM\footnote{The first \XMM{} Serendipitous Source Catalogue
(1XMM), released on 2003, contains source detections drawn from 585 \XMM{} EPIC
observations, and a total of $\sim$30~000 individual X-ray sources. The median
0.2-12~keV flux of the catalogue sources is $\sim3\times10^{-14}$~cgs, with
$\sim$12\% of them having fluxes below $10^{-14}$~cgs.}  -SSC \cite{SSC03}-),
and even larger ones will be available in the near future (2XMM,$\ldots$),
which, since they are constructed from serendipitous sources in typical public
\XMM{} observations, are themselves medium flux surveys. The exploitation of
the full potential of these catalogues depends on the results from
``conventional'' smaller scale medium flux surveys such as those above and the
one presented here.

Since AGN are the dominant population at medium and low X-ray fluxes,
and X-ray selected AGN are known to cluster strongly (Yang et
al. \cite{Yang06}, Gilli et al. \cite{Gilli05}, Mullis et
al. \cite{Mullis04}, Carrera et al. \cite{Carrera98}), it is
interesting to investigate the angular clustering of X-ray sources,
which can be performed without resorting to expensive ground-based
spectroscopy. Angular correlation can be translated into spatial
clustering via Limber's equation (Peebles \cite{Peebles80}), if the
luminosity function of the populations involved is known from other
surveys. Several studies have investigated the angular clustering in
the soft and hard bands in scales of tens-hundreds of arcsec to
degrees, with varying success in the detection of signal. Angular
clustering has been detected in the soft band (Gandhi et
al. \cite{Gandhi06}, Basilakos et al. \cite{Basilakos05}, Akylas et
al. \cite{Akylas00}, Vikhlinin \& Forman \cite{Vikhlinin95}) with
different values of the clustering strength. Angular clustering in the
hard band has evaded detection in some cases (Gandhi et
al. \cite{Gandhi06}, Puccetti et al. \cite{Puccetti06}), but not in
others (Yang et al. \cite{Yang03}, Basilakos et
al. \cite{Basilakos04}), in the latter cases finding clustering
strengths marginally compatible with the observed spatial clustering of
optical and X-ray AGN.

We present here a medium X-ray survey which includes 1433
serendipitous sources over 4.8~deg$^2$: AXIS (An X-ray International
Survey). This survey was undertaken under the auspices of the \XMM{}
Survey Science Centre\footnote{The \XMM{} Survey Science Centre is an
international collaboration involving a consortium of 10 institutions
appointed by ESA to help the SOC in developing the software analysis
system, to pipeline process all the \XMM{} data, and to exploit the
\XMM{} serendipitous detections, see {\tt
http://xmmssc-www.star.le.ac.uk}} (SSC). In this paper we present the
source catalogue (Section
\ref{XData}), overall X-ray spectral properties (Section
\ref{SecXSp}), number counts (Section
\ref{SeclogNlogS}, where some other samples have also been used),
fractional contribution to the XRB (Section \ref{ContributionXRB}),
and angular clustering properties (Section
\ref{SecAngularClustering}). Overall results are summarised in Section
\ref{SecConclusions}. In a forthcoming paper (Barcons et al. \cite{Barcons06b}) we
present the optical identifications of a subsample of AXIS, which we
have called the \XMM{} Medium Survey (XMS). In this paper we will use
cgs as a shorthand for the cgs system units for the flux:
erg~cm$^{-2}$~s$^{-1}$.

\section{The X-ray data}
\label{XData}

\begin{sidewaystable*}
   \caption[]{Basic data for the 36 fields in the AXIS survey:
including the target name, the \XMM{} OBS\_ID number of
the pn dataset used, the R.A. and
Dec. of the field centre, the ``clean'' exposure time and the filter
used for the pn camera, the Galactic column density in the direction
of the field (Dickey \& Lockman 1990), and, when applicable, the centre
and radius of the
circular target exclusion area, the width of the OOT exclusion area,
and the centre and radius of an additional exclusion area around bright/extended sources.}
\label{fields}

   \begin{tabular}{lcrrrcrrrrrrrr}
       \hline
       \noalign{\smallskip}

Target name & OBS\_ID & R.A. & Dec & Texp & Filter & $ N_{\rm H,Gal}$ & R.A. & Dec. & R & OOT & R.A. & Dec. & R' \\
 &  &(J2000) & (J2000) & (s) & & ($ 10^{20}$ cm$^{-2}$) & (J2000) & (J2000) & ($"$) & ($"$) & (J2000) & (J2000) & ($"$) \\
       \noalign{\smallskip}
       \hline
      \noalign{\smallskip}
  A2690          & 0125310101       & 00:00:30.30 & $-$25:07:30.00 &  21586.7 & Medium &  1.84  &  00:00:21.2 & $-$25:08:12.18 &  20 &   0 &           - &            - &   0\\
  Cl0016+1609$^a$& 0111000101       & 00:18:33.00 & +16:26:18.00 &  29149.3 & Medium &  4.07  &  00:18:33.2 &  +16:26:07.97 & 148 &   0 &           - &            - &   0\\
  G133-69pos\_2$^a$& 0112650501     & 01:04:00.00 & $-$06:42:00.00 &  18080.0 & Thin   &  5.19  &           - &            - &   0 &   0 &           - &            - &   0\\
  G133-69Pos\_1$^a$& 0112650401     & 01:04:24.00 & $-$06:24:00.00 &  20000.0 & Thin   &  5.20  &           - &            - &   0 &   0 &           - &            - &   0\\
  PHL1092        & 0110890501       & 01:39:56.00 & +06:19:21.00 &  16180.0 & Medium &  4.12  &  01:39:55.8 &   +06:19:19.67 &  88 &  40 &  01:40:09.0 & +06:23:27.67 &  68\\
  SDS-1b$^{a,d}$     & 0112371001   & 02:18:00.00 & $-$05:00:00.00 &  43040.0 & Thin   &  2.47  &           - &            - &   0 &   0 &           - &            - &   0\\
  SDS-3$^{a,d}$      & 0112371501   & 02:18:48.00 & $-$04:39:00.00 &  14927.9 & Thin   &  2.54  &           - &            - &   0 &   0 &           - &            - &   0\\
  SDS-2$^{a,d}$      & 0112370301   & 02:19:36.00 & $-$05:00:00.00 &  40673.0 & Thin   &  2.54  &           - &            - &   0 &   0 &           - &            - &   0\\
  A399$^{a,c}$   & 0112260201       & 02:58:25.00 & +13:18:00.00 &  14298.7 & Thin   & 11.10  &  02:57:50.2 &  +13:03:20.88 & 272 &   0 &           - &            - &   0\\
  Mkn3$^{a,c}$   & 0111220201       & 06:15:36.30 & +71:02:04.90 &  44506.1 & Medium &  8.82  &  06:15:36.6 &  +71:02:15.95 &  76 &  32 &           - &            - &   0\\
  MS0737.9+7441$^{a,e}$& 0123100201 & 07:44:04.50 & +74:33:49.50 &  20209.3 & Thin   &  3.51  &  07:44:04.3 &  +74:33:54.56 & 120 &  40 &           - &            - &   0\\
  S5 0836+71$^a$ & 0112620101       & 08:41:24.00 & +70:53:40.70 &  25057.3 & Medium &  2.98  &  08:41:24.3 &  +70:53:41.06 & 160 &  52 &           - &            - &   0\\
  PG0844+349     & 0103660201       & 08:47:42.30 & +34:45:04.90 &   9783.5 & Medium &  3.28  &  08:47:42.9 &  +34:45:03.27 & 200 &  40 &           - &            - &   0\\
  Cl0939+4713$^a$& 0106460101       & 09:43:00.10 & +46:59:29.90 &  43690.0 & Thin   &  1.24  &  09:43:01.8 &  +46:59:44.37 & 160 &   0 &           - &            - &   0\\
  B21028+31$^a$  & 0102040301       & 10:30:59.10 & +31:02:56.00 &  23236.0 & Thin   &  1.94  &  10:30:59.3 &  +31:02:56.08 & 140 &  72 &           - &            - &   0\\
  B21128+31$^a$  & 0102040201       & 11:31:09.40 & +31:14:07.00 &  13799.8 & Thin   &  2.00  &  11:31:09.6 &  +31:14:06.02 & 140 &  44 &           - &            - &   0\\
  Mkn205$^{a,e}$     & 0124110101   & 12:21:44.00 & +75:18:37.00 &  17199.6 & Medium &  3.02  &  12:21:43.8 &  +75:18:39.08 & 140 &  36 &           - &            - &   0\\
  MS1229.2+6430$^{a,e}$& 0124900101 & 12:31:32.32 & +64:14:21.00 &  28700.0 & Thin   &  1.98  &  12:31:31.2 &  +64:14:18.06 & 140 &  40 &           - &            - &   0\\
  HD111812$^b$   & 0008220201       & 12:50:42.56 & +27:26:07.70 &  37338.8 & Thick  &  0.90  &  12:51:42.6 &  +27:32:23.27 & 168 &  40 &           - &            - &   0\\
  NGC4968        & 0002940101       & 13:07:06.10 & $-$23:40:43.00 &   4898.7 & Medium &  9.14  &  13:07:06.3 & $-$23:40:33.23 &  40 &   0 &           - &            - &   0\\
  NGC5044        & 0037950101       & 13:15:24.10 & $-$16:23:06.00 &  20030.0 & Medium &  5.03  &  13:15:24.2 & $-$16:23:08.53 & 340 &   0 &           - &            - &   0\\
  IC883          & 0093640401       & 13:20:35.51 & +34:08:20.50 &  15849.4 & Medium &  0.99  &  13:20:35.4 &  +34:08:21.37 &  48 &   0 &  13:20:54.4 & +33:55:17.26 & 104\\
  HD117555$^{a,b}$& 0100240201      & 13:30:47.10 & +24:13:58.00 &  33225.4 & Medium &  1.16  &  13:30:47.8 &  +24:13:51.07 & 160 &  40 &           - &            - &   0\\
  F278           & 0061940101       & 13:31:52.37 & +11:16:48.70 &   4648.3 & Thin   &  1.93  &  13:31:52.4 &  +11:16:43.88 &  48 &   0 &           - &            - &   0\\
  A1837$^a$      & 0109910101       & 14:01:34.68 & $-$11:07:37.20 &  45361.3 & Thin   &  4.38  &  14:01:36.5 & $-$11:07:43.14 & 440 &   0 &           - &            - &   0\\
  UZLib$^a$      & 0100240801       & 15:32:23.00 & $-$08:32:05.00 &  23391.2 & Medium &  8.97  &  15:32:23.4 & $-$08:32:05.32 & 140 &  40 &           - &            - &   0\\
  FieldVI        & 0067340601       & 16:07:13.50 & +08:04:42.00 &   9634.0 & Medium &  4.00  &           - &            - &   0 &   0 &           - &            - &   0\\
  PKS2126-15$^a$ & 0103060101       & 21:29:12.20 & $-$15:38:41.00 &  16150.0 & Medium &  5.00  &  21:29:12.1 & $-$15:38:40.44 & 120 &  40 &           - &            - &   0\\
  PKS2135-14$^{a,f}$ & 0092850201   & 21:37:45.45 & $-$14:32:55.40 &  28484.3 & Medium &  4.70  &  21:37:45.1 & $-$14:32:55.22 & 120 &  44 &           - &            - &   0\\
  MS2137.3-2353  & 0008830101       & 21:40:15.00 & $-$23:39:41.00 &   9880.0 & Thin   &  3.50  &  21:40:15.1 & $-$23:39:39.32 & 140 &  48 &           - &            - &   0\\
  PB5062$^a$     & 0012440301       & 22:05:09.90 & $-$01:55:18.10 &  28340.9 & Thin   &  6.17  &  22:05:10.3 & $-$01:55:20.38 & 140 &  40 &           - &            - &   0\\
  LBQS2212-1759$^a$& 0106660101     & 22:15:31.67 & $-$17:44:05.00 &  90892.5 & Thin   &  2.39  &           - &            - &   0 &   0 &           - &            - &   0\\
  PHL5200$^a$    & 0100440101       & 22:28:30.40 & $-$05:18:55.00 &  43278.5 & Thick  &  5.26  &  22:28:30.4 & $-$05:18:53.12 &  16 &   0 &           - &            - &   0\\
  IRAS22491-1808$^a$& 0081340901    & 22:51:49.49 & $-$17:52:23.20 &  19867.2 & Medium &  2.71  &  22:51:49.4 & $-$17:52:25.02 &  32 &   0 &           - &            - &   0\\
  EQPeg$^a$      & 0112880301       & 23:31:50.00 & +19:56:17.00 &  12200.0 & Thick  &  4.25  &  23:31:52.7 &  +19:56:18.46 & 160 &  48 &           - &            - &   0\\
  HD223460       & 0100241001       & 23:49:41.00 & +36:25:33.00 &   6699.0 & Thick  &  8.25  &  23:49:40.8 &  +36:25:32.56 & 172 &  40 &  23:50:02.0 & +36:25:36.36 & 116\\
       \noalign{\smallskip}
       \hline
   \end{tabular}
   \begin{list}{}{}
        \item[$^a$]Fields belonging to the XMS
        \item[$^b$]Fields excluded from the angular correlation studies
        \item[$^c$]{\tt eposcorr} failed in these fields
        \item[$^d$]In common with SXDS (Ueda et al. \cite{Ueda07})
        \item[$^e$]In common with HELLAS2XMM (Baldi et al. \cite{Baldi02})
        \item[$^f$]Same area as one ChaMP field (Kim et al. \cite{Kim04})
   \end{list}
\end{sidewaystable*}

\subsection{Selection of fields}

A total of 36 \XMM{} observations were selected for optical follow-up
of X-ray sources within the AXIS programme (see Table \ref{fields}),
preferring those that were public early on (mid 2000), or belonging to
the SSC Guaranteed Time. We selected
fields with $\mid b \mid >20^\circ$, total exposure time $>$15~ks, and
devoid of (optical and X-ray) bright or extended targets (except in
two cases: A1837 and A399). Furthermore, after discarding the
observing intervals with high background rates, a few of the fields
ended up with exposure times shorter than 15~ks. A few fields (some
with shorter exposure times) were only intended to expand the solid
angle for bright X-ray sources.  All of these fields have been used
for the study of the cosmic variance, the angular correlation function
and the \ns{} in different bands.

Of those 36 \XMM{} observations, 27 were selected for optical
follow-up of medium flux X-ray sources. Two of those 27 fields (A2690
and MS2137) were later dropped from the main identification effort,
because their Declination was too low to observe them from Calar Alto
(Spain) with airmass lower than 2. The sources in the remaining 25
fields were used to form flux-limited samples in the 0.5-2~keV,
2-10~keV, 0.5-4.5~keV band, and a non-flux-limited sample in the
4.5-7.5~keV bands: the XMS (see Barcons et al. \cite{Barcons06b}). These 25 fields
are marked in Table
\ref{fields}.

\subsection{Data processing and relation to 1XMM}

The data used in our earlier follow-up efforts (see previous Section)
had been reduced with very different versions of the SAS (
Science Analysis System,
Gabriel et
al. \cite{Gabriel04}).  The reprocessing for the 1XMM catalogue (SSC \cite{SSC03})
allowed us to have a much more homogeneous set of X-ray data.  The
Observation Data Files (ODF)
were processed in the SSC Pipeline Processing System (PPS) facilities
at Leicester with the same SAS version used for 1XMM (very similar to
SAS version 5.3.3, see below for the difference), except for Mkn205,
which was processed with SAS version 5.3.3.
PHL1092
is a especial case, since the original \XMM{} observation was never
re-processed, instead we took a newer set of data from the \XMM{}
Science Archive, which was processed later with a different
SAS version (5.4.0).

The main difference between the versions of the SAS source detection task
({\tt emldetect}) used for Mkn205 and the rest of the fields is the
inclusion of the possibility of sources with negative count rates in
the latter. Negative count rates in individual energy bands were
allowed to avoid a bias in the total count rates of sources that were
undetected in one or more energy bands. However, since this option
caused numerical problems in some cases, the count rates were limited
to values $\geq 0$ in later versions of the pipeline. Since negative
count rates are meaningless, we have set all the negative count rates to
zero in what follows. The corresponding detection likelihoods have
also been set to zero, meaning that the presence of that source in
that band does not improve the fit (see below for details).

The standard SAS products include X-ray source lists (created by {\tt
emldetect}) for each of the three EPIC cameras (MOS1, MOS2 -Turner et
al. \cite{Turner01}- and pn -Str\"uder et al. \cite{Struder01}-), run in five
independent bands (1: 0.2-0.5 keV, 2: 0.5-2.0 keV, 3: 2.0-4.5 keV, 4:
4.5-7.5 keV, and 5: 7.5-12.0 keV). In addition, the source detection
algorithm was run in the 0.5-4.5 keV (band 9), which we will call the
XID band. Since EPIC pn has the highest sensitivity of the three
cameras, we have used only pn source lists. We have not allowed for
source extent when detecting the sources, and therefore all the
sources in our survey are treated as pointlike.

%
%

The internal calibration of the source positions from the SAS is quite
good (1.5~arcsec, Ehle et al. \cite{Ehle05}), but there may be some systematic
differences between the absolute X-ray source positions and optical
reference frames. We have registered the X-ray source positions to the
USNO-A2 reference frame field by field, using the SAS task {\tt
eposcorr}. This task shifts the X-ray reference frame to minimise the
differences between the X-ray source positions and the positions of
their optical counterparts in a given reference astrometric catalogue
(USNO-A2 in our case). These corrected positions are given in Table
\ref{XSouTable1}. The average shift in absolute value 
in R.A. (Dec.) was 1.3 (0.9) arcsec, and in all cases the shifts
were under 3~arcsec. The average number of X-ray-USNO-A2 matches used
to calculate the shifts was 28 (the minimum was 16 and the maximum
82). There were two fields for which this procedure failed (Mkn3 and
A399), probably because the first one had one quarter of the pn chips
in counting mode, and the second has two moderately strong extended
sources in opposite corners of the field. We therefore used the
original X-ray source positions for these two fields.

\begin{sidewaystable}
   \caption[]{Basic X-ray data for the AXIS X-ray sources. Column 1
   gives the IAU-style source name, columns 2 and 3 its sky position, column 4
  the error on that position, columns 5 to 14 give the
   count rates and likelihoods in bands 2 to 5 and XID, respectively. Sample table only, the full table will be made
   available in electronic format.}
\label{XSouTable1}

   \begin{tabular}{lcclclclclclclcccccccccc}
       \hline
       \noalign{\smallskip}

Name & R.A. & Dec. & $r_{90}$\footnote{90\% statistical error circle in the
position}
                                & CR2 & ML2 & CR3 & ML3 & CR4 & ML4 & CR5 & ML5 & CRXID & MLXID \\
  &(J2000) & (J2000) & ($''$)   & (cts/s) & & (cts/s) & & (cts/s) & & (cts/s) & &(cts/s)&       \\
(1)  & (2) & (3)   & (4)        & (5) & (6) & (7) & (8) & (9) & (10)& (11)& (12)& (13)  & (14)  \\ 
       \noalign{\smallskip}
       \hline
      \noalign{\smallskip}
  \object{XMM~U000002.7-251137} & 00:00:02.73 & $-$25:11:37.37 & 0.54 & 0.0089$\pm$0.0010 & 123.0 & 0.0028$\pm$0.0006 & 24.5 & 0.0043$\pm$0.0007 & 25.78 & 0.0008$\pm$0.0004 & 0.6 & 0.0110$\pm$0.0011 & 147.2 \\
       \noalign{\smallskip}
       \hline
   \end{tabular}
\end{sidewaystable}

\begin{sidewaystable}
   \caption[]{X-ray spectral fit data for the AXIS X-ray sources. Column 1
   gives the IAU-style source name (same as in Table~\ref{XSouTable1}), column 2
   gives the 0.5-4.5~keV spectral slope, column 3 the soft flux,
   column 4 the 2-12~keV spectral slope, column 5 the hard band
   flux, column 6 the XID flux, column 7 the ultra-hard flux,
   column 8 the 0.5-12~keV spectral slope and column 9 the
   0.5-10~keV flux. The last column (10) is a bit-coded number indicating
   which of our samples each source belongs to, in which a 1 indicates that it belongs,
   and a 0 that it does not, in the usual soft-hard-XID-ultra-hard order.
   The soft and XID fluxes are obtained using the 0.5-4.5~keV
   spectral slope, while hard and ultra-hard fluxes are obtained using the
   2-12~keV spectral slope. Sample table only, the full table will be made
   available in electronic format.}
\label{XSouTable2}

   \begin{tabular}{lccccccccccccccccc}
       \hline
       \noalign{\smallskip}

&\multicolumn{2}{c}{Soft}&\multicolumn{2}{c}{Hard}&XID&Ultra-hard&\multicolumn{2}{c}{Total}& Sample\\
& $\Gamma$ & $S$   & $\Gamma$ & $S$   & $S$ & $S$ & $\Gamma$ & $S$ \\
&         & ($10^{-14}$cgs) &          & ($10^{-14}$cgs) &  ($10^{-14}$cgs)&($10^{-14}$cgs)&  & ($10^{-14}$cgs) &  \\
(1)      & (2)   & (3)      & (4)   & (5) & (6) & (7)      & (8) & (9)\\ 
       \noalign{\smallskip}
       \hline
      \noalign{\smallskip}
  \object{XMM~U000002.7-251137} & $1.50^{+0.27}_{-0.19}$ & $1.45\pm0.08$ & $2.07^{+0.57}_{-0.56}$ & $2.71\pm0.41$ & $2.90\pm0.20$ & $0.85\pm0.22$ & $1.71^{+0.15}_{-0.17}$ & $4.31\pm0.40$ & 1110 \\
       \noalign{\smallskip}
       \hline
   \end{tabular}
\end{sidewaystable}


\subsection{Source selection}

In each detection band, the source detection algorithm fits a portion
of the image around the 
position of the candidate source trying to match the Point Spread
Function (PSF) shape to the photon distribution. In this process it uses the
background map, and the exposure map to fit a source count rate and 2D
position. A typical size of this region is the 80\% encircled energy radius,
which corresponds to about 5~pixels on-axis and 7~pixels at 15~arcmin off-axis
(we have used throughout 4~arcsec pixels). Sources closer than this distance to
pn chip edges, could have a worse determination of their positions and/or count
rates. We have therefore excluded all sources closer than the 80\% encircled
energy radius (taking into account the off-axis angle) to any pn chip edges. A
region ($\sim$12 pixels wide) on the readout (outer) edge of the pn chips is
masked out on board the satellite (Ehle et al. \cite{Ehle05}). We have
considered this ``effective'' edge as the outer edge of each chip when defining
our excluded regions.

We have visually inspected all X-ray images, excluding circular regions around
bright targets, and other bright/extended sources in the images (see Table
\ref{fields} for the sizes and positions of these regions). In a few images,
a bright band extending from the target to the pn chip reading edge
was visible, due to the photons arriving at the detector while
it was being read (called Out Of Time -OOT- region). We have excluded
a rectangular region around the OOT region, whose width is also given
in Table \ref{fields}. In addition, sources affected by bright
pixels/segments have been excluded, as well as those which were
obviously affected by the presence of nearby bright sources, or split
by the chip gaps, and were not picked up by the above procedures.

In addition, there were two sets of fields which partially
overlapped. G133-69pos\_2 and G133-69Pos\_1 on the one hand, and
SDS-1b, SDS-2 and SDS-3 on the other. We have dealt with this by
masking out the portion of the second (and third) field which overlap
with the first field, in the order given above.

We have used the following bands in the analysis performed in this paper:

\begin{itemize}

\item soft: 0.5-2 keV, identical to the standard SAS band 2

\item hard: combining standard SAS bands 3, 4 and
5, and hence corresponds to counts in the 2 to $\sim$12~keV
range. However, we have calculated (and will quote) all hard fluxes in
2-10 keV.

\item XID: 0.5-4.5 keV, identical to the SAS band 9

\item ultra-hard: 4.5-7.5 keV, identical to the standard SAS band 4

\end{itemize}

The detection likelihood in the hard band ($L_{345}$) was obtained
from the detection likelihoods in bands 3, 4 and 5 ($L_3$, $L_4$, and
$L_5$ respectively) using $L_{345}=-\log(1-Q(5/2,L'_3+L'_4+L'_5))$,
where $L'_i$ can be obtained from $L_i=-\log(1-Q(3/2,L'_i))$ and
$Q(a,x)$ is the incomplete gamma function (see manual for {\tt
emldetect}).

There were a total of 2560 accepted sources with a {\tt emldetect} detection
likelihood $\geq 10$ (the default value) in at least one band, which are listed
in Table \ref{XSouTable1}, with their corrected X-ray
positions, and count rates in the standard SAS and XID
bands.

The original unfiltered source lists are identical to the ones used by
Mateos et al. (\cite{Mateos05}) in their study of the detailed spectral
properties of medium flux X-ray sources, for the common fields. They
also used similar criteria for excluding sources close to the pn chip
edges, except for the readout edge. The differences in the final
accepted source lists arose from several reasons: Mateos et al. used
sources close to pn chip gaps (which we have excluded) if they were
far from chip gaps in the MOS detectors. They excluded from their
spectral analysis the sources with low number of counts. Finally, we
have treated slightly differently the sources close to the exclusion
zone boundaries. These differences are at the 10\% level: out of the
final 1137 accepted sources in the Mateos et al. sample, only 119
would have been excluded by our criteria.

\subsection{Sensitivity maps}
\label{SensMap}

The value of the sensitivity map at a given point is the minimum
count rate that a source should have to be detected with the desired
likelihood at that point.  As explained above, the detection
likelihood assigned by {\tt emldetect} takes into account the number
of counts in the detection box, how well do they fit the PSF shape,
and the variation of the exposure map over the detection box. The
likelihood is therefore in principle not trivially related to the Poisson
probability of an excess in the number of detected counts over the
expected background in the detection box.

However, we have found (see Appendix \ref{AppSensMap}) that the
count rate assigned by the software to the source (the ``observed''
count rate) is proportional to the count rate expected from a Poisson
distribution for the same likelihood (for detection likelihoods
between 8 and 20), with proportionality constant $\sim 0.9-1.1$,
depending on the band. It appears that the non-Poisson characteristics
taken into account by {\tt emldetect} have a relatively small
influence in the determination of the count rate. For a given
likelihood, radius of the detection region, total value of the
background map, and average value of the exposure map in that region,
we calculate the expected Poisson count rate at each point of the
detector and, using the proportionality constants, the corresponding
``observed'' count rate, i.e., the value of the sensitivity map (see
Appendix \ref{AppSensMap} for further details).

We have used the above procedure to create sensitivity maps for each
field in each band, in image $(X,Y)$ coordinates, which have 4~arcsec
pixels, have sky North to the top and East to the left, and are
centred approximately in the optical axis of the X-ray telescope. We
have also taken into account for each field the areas excluded close
to the detector edges, and around the bright sources and OOT regions,
excluding those regions from the sensitivity maps as well.

The final selection of sources in each band was done using their detection
likelihood and the corresponding sensitivity map at their sky position (to
ensure the validity of the sky areas calculated from the sensitivity maps, see
Section \ref{SkyAreas}).  We have chosen a detection confidence limit of
5-$\sigma$, which corresponds to $L=15$ in the band under consideration. We have
also imposed that the source has a count rate equal to or larger than the value
of the sensitivity map at their sky position, to ensure that the source
detection is reliable (this excluded less than 5\% of the $L\geq 15$ sources in
the soft band, $\sim$10\% of the sources in the XID and ultra-hard bands, and
$\sim$20\% of the sources in the hard band).  The number of sources
excluded by this criterion is much larger in the hard band than in the other
bands. This is somewhat contradictory with the proportionality between the
detected count rate and the Poisson count rate for this band being in a different
direction than in the other bands ($\sim$10\% smaller rather than $\sim$10\%
larger, see Table~\ref{crcrpoisTable}), since the sensitivity map will then be
relatively lower, and therefore would tend to include more sources, rather than
exclude them. In any case, the hard band is the widest, and the only one of our
bands which is not one of the bands used for source searching in the SAS, being
instead a composite of three default bands. In principle, all this makes the
hard band the more complex to deal with, and for which the uncertainties
associated with our empirical method to calculate the sensitivity maps might be
the highest.

The total number of distinct selected sources (i.e., fulfilling the above
criteria in at least one of the bands) is 1433.  We give in Table \ref{nXSou}
the total number of selected sources in each of the above bands. Unless
explicitly stated otherwise, for the \ns{} and the angular clustering analyses
below we have only used sources detected in the corresponding bands.  However,
it is important to emphasise that the fact that, if a source has been detected
in at least one band, it can be considered as a real source, and therefore its
counts in other bands can be used (e.g., for spectral fitting or hardness ratio
calculations), even if the source has a count rate smaller than the detection
threshold in those bands.

To estimate the number of spurious sources in our survey we first need
to calculate the number of independent source detection cells. {\tt
emldetect} uses an input list from {\tt eboxdetect} (which is a simple
sliding-box algorithm which uses a square $5\times 5$ pixels detection
cell), so the individual detection cell is $5\times 5\times
(4'')^2=400$~arcsec$ ^2$. The total area of our survey is
$\sim$4.8~deg$^2$ (see Section \ref{SkyAreas}), or about 155520
independent detection cells. Since the probability of a false
detection at the 5-$\sigma$ level is 0.000057, this corresponds to
about 9 spurious sources in each of our detection bands. This is less
than 1\% in the soft and XID bands, about 2\% in the hard band, and
almost 10\% in the ultra-hard band. The latter fraction could explain
in part the discrepant point in the ultra-hard \ns{} at the lowest
fluxes (see Section~\ref{MLfit}), since it is there where the
contribution from spurious sources is expected to be highest.


\begin{table*}
   \caption[]{ Number of sources selected in different bands $N$,
   weighted average slopes $\langle \Gamma\rangle$ and errors (taking
   into account both the error bars in the individual $\Gamma$ and the
   dispersion around the mean), and number of observed sources used in the
   average $N_{\rm ave}$. The soft, hard, XID and
   ultra-hard bands are as defined in the text. ``Soft and hard''
   refers to sources selected simultaneously in the soft and hard
   bands, ``Only soft'' refers to sources selected in the soft band
   but not in the hard band, and ``Only hard'' refers to sources
   selected in the hard band and not in the soft band} \label{nXSou}

   \begin{tabular}{lrcccccc}
       \hline
       \noalign{\smallskip}
    &    & \multicolumn{2}{c}{0.5-2~keV} && \multicolumn{2}{c}{2-10~keV} \\
\cline{3-4}\cline{6-7}
Selection & $N$ & $N_{\rm ave}$ & $\langle \Gamma\rangle$ && 
             $N_{\rm ave}$ & $\langle \Gamma\rangle$ \\
       \noalign{\smallskip}
       \hline
      \noalign{\smallskip}
Soft          & 1267 & 1239 & $ 1.811\pm0.015$ && 1145 & $1.53\pm0.02$  \\
Hard          &  397 &  397 & $ 1.76 \pm0.03 $ &&  394 & $1.55\pm0.03$  \\
XID           & 1359 & 1335 & $ 1.773\pm0.016$ && 1244 & $1.47\pm0.02$  \\
Ultra-hard    &   91 &   91 & $ 1.80 \pm0.06 $ &&   91 & $1.50\pm0.07$  \\
Soft and hard &  345 &  345 & $ 1.79 \pm0.03 $ &&  342 & $1.67\pm0.03$  \\
Only soft     &  922 &  894 & $ 1.878\pm0.017$ &&  803 & $1.04\pm0.04$  \\
Only hard     &   52 &   52 & $-0.29 \pm0.08 $ &&   52 & $0.53\pm0.12$  \\      
       \noalign{\smallskip}
       \hline
   \end{tabular}
\end{table*}



\section{X-ray properties of the sources}
\label{SecXSp}

We have studied the broad spectral characteristics of our sources by
fitting their count rates in several bands, to those expected from
power-law spectra with photon index $\Gamma$ and
Galactic Hydrogen column densities ($N_H$) from 21cm radio
measurements (Dickey \& Lockman \cite{Dickey90}, see Table \ref{fields}). The
spectra of a few sources will obviously not be well represented by a
power-law (e.g. stars or clusters showing thermal spectra), but for the
purpose of calculating fluxes in bands in which the power-law was fit,
a power-law is a reasonably accurate and very simple approximation.

The expected count rates for different values of $\Gamma$ (from -10 to 10 in
steps of 0.5, interpolating linearly for intermediate values), and the
Galactic $N_H$ values of each field were calculated with {\tt xspec} (Arnaud
\cite{Arnaud96}), using the ``canned'' on-axis redistribution matrix files, and
on-axis effective areas for each field, created with the SAS task {\tt
arfgen}. The inaccuracies from using standard response matrices instead of
source specific matrices (as generated by {\tt rmfgen}) are expected to be
small, since we only use broad bands, much broader than the spectral resolution
of the EPIC pn camera. Since the count rates from {\tt emldetect} are corrected
for the exposure map (which includes vignetting, and bad pixel corrections) and
the PSF enclosed energy fraction, the effective areas were generated disabling
the vignetting and PSF corrections as indicated in the {\tt arfgen} manual.

The corresponding fluxes in the bands 2 to 5 (in the case of band 5, the flux
was calculated in the 7.5 to 10 keV band) were also calculated using the
spectral model for the same values of $\Gamma$, setting $N_H=0$. We have
checked that the relatively coarse sampling does not introduce any biases in
our spectral fits by repeating the spectral fits for one field with a step in
$\Gamma$ of 0.001. The results were practically identical, any differences in
the spectral slopes being much smaller than their uncertainties.

We have performed spectral fits in bands 2 and 3 (for the soft and XID fluxes),
and bands 3, 4 and 5 (for the hard and ultra-hard fluxes). The average count
rates of our selected sources in the XID and hard bands are 0.0095 and 0.0032,
respectively, which for a typical exposure time of 15~ks, give an average of
more than 10 counts per bin, which ensures that Gaussian statistics is a good
approximation. However, they are not sufficient in many cases to warrant a
detailed spectral fitting with {\tt xspec}. The best fit $\Gamma$ and flux were
calculated by minimising the $\chi^2$ between the observed and expected count
rates. The minimisation was actually done in $\Gamma$, setting the flux from
the normalisation that minimised the $\chi^2$ in the corresponding band. One
sigma error bars for the photon index and flux were obtained from the values
which produced $\Delta\chi^2=1$ from the minimum. These error bars were
asymmetric in most cases, but we have used a symmetric error bar (the
arithmetic average of the upper and lower error bars) for the weighted averages
of the spectral slopes and the \ns.  In a few cases the fitted photon indices
are very steep $\mid\Gamma\mid\sim 10$, in all cases this corresponds to
sources with positive count rates in only one of the fitted bands, forcing the
power law to the steepest allowed slope.  The number of these pathological
cases in each band can be obtained from the difference between the $N$ and
$N_{\rm ave}$ columns in Table
\ref{nXSou} (typically $<10$\%).

The photon
indices and fluxes in each of the fitted bands, as well as their error
bars are given for each source in Table \ref{XSouTable2}. We show in Table \ref{nXSou}
the weighted average photon indices of the detected sources in each of
the fitted bands, as well as the number of sources used in the
averages, excluding in all cases sources with $\mid\Gamma\mid> 9$, to
avoid biasing the averages with a few outliers. The impact from these
sources would be small for the weighted averages used here.

We have checked the reliability of our simple spectral fit method (see
Appendix~\ref{CheckSpec}), performing both internal tests with simulations, and
external checks with respect to the full spectral fits of Mateos et
al. (\cite{Mateos05}) to a set of sources with a large overlap. We conclude
that our 2-3 band spectral slopes are good when taken individually for the
purposes of calculating fluxes in different bands. However, we have found that
there are significant systematic biases in the average photon indices (when
averaged over large samples) which are larger than the statistical
errors. These systematic biases are nonetheless small in both absolute
($\Delta\Gamma\sim0.12$) and relative ($\Delta\Gamma/\Gamma=0.2$) terms.  Our
method is therefore adequate for extracting broad spectral information from
medium flux X-ray sources, such as those in the 1XMM and 2XMM catalogues.

In broad terms, the soft/XID band slopes are $\sim$1.8 for sources detected in
all bands, being slightly softer for sources detected only in the soft band,
and much harder for sources detected only in the hard band (see Table
\ref{nXSou}). The hard band slopes are flatter, closer to 1.5-1.6, but
slightly steeper for sources detected both in the soft and hard bands
($\sim$1.7). The average hard slope of the sources only detected in the soft
band is quite flat $\langle\Gamma\rangle\sim 1$, but their un-weighted
average is $\langle\Gamma\rangle= 1.53 \pm 0.08$, much closer to the other
average slopes in that band. The origin of this difference is that ``Only
soft'' sources with steep spectra in the hard band tend to have larger errors
on the hard spectral slope (because they have few counts in \XMM{}
bands 4 and 5, and so the slope is not well constrained), and therefore they
have a very small weight in the weighted average.

The difference between the spectral slopes of sources only detected in
in the soft band and those only detected in the hard band is partly
due to a known bias in likelihood limited surveys (such as ours), that
occurs because source detection is done in photons rather than in flux
(Zamorani et al. \cite{Zamorani88}, Della Ceca et
al. \cite{DellaCeca99}), so that for a given flux, softer sources will
be much easier to detect in the soft band, while the inverse would be
true for harder sources. This is compounded by the fact that \XMM{} is
much more sensitive to soft photons.

   \begin{figure}
   \centering
   \includegraphics[width=6.5cm,angle=270.0]{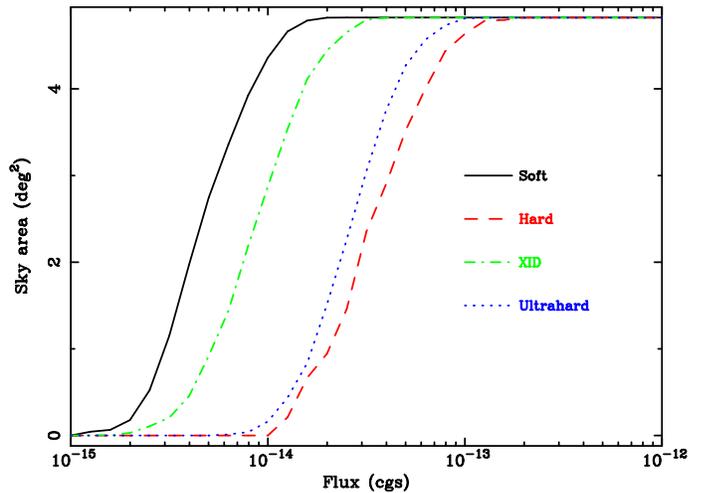}
      \caption{Sky area as a function of flux for different bands
              }
         \label{skyareafig}
   \end{figure}

\section{The \ns}
\label{SeclogNlogS}

We have first studied the sky density of the detected sources as a function of
flux (known as the \ns) in the soft, hard, XID and ultra-hard bands.

\subsection{Sky areas}
\label{SkyAreas}

The sky area over which we are sensitive to a given flux is easy to calculate
from the sensitivity maps. In principle, we just need to sum the sky area over
which the sensitivity maps have values below the desired flux. However, the
conversion between flux and count rate depends on the assumed spectral
shape. In each band $i$, we have calculated the sky areas at each flux for each
assumed spectral slope $\Omega_i(S,\Gamma)$ with the following procedure: in
each field, we have converted $S$ to count rate using $\Gamma$ and the response
matrices, and summed the area over which the sensitivity map is below that
count rate. The total sky area is the sum of the areas found for each field.

The sky area in each band for each source $j$ is then
$\Omega_{j,i}=\Omega_i(S_{j,i},\Gamma_{j,i})$, where the source's flux
$S_{j,i}$ and spectral slope $\Gamma_{j,i}$ are described in Section
\ref{SecXSp}.

Sky areas independent of any assumed spectrum can also be obtained by
weighting $\Omega_i(S,\Gamma)$ with the number of detected sources in
each $(S,\Gamma)$ bin. We will call these spectrally averaged sky
areas $\Omega_i(S)$. They are shown in Fig. \ref{skyareafig}. The
maximum value of all curves is 4.8~deg$^2$, which is the geometric
area covered by our survey, taking into account the excluded areas.

We have also obtained spectrally averaged sky
areas for each field $\Omega_{i,k}(S)$ ($k=1\ldots$number of fields)
from the sensitivity maps for each field $k$ in each band $i$, using
$S(\Gamma)$ for that field, and weighting with the number of sources
in the total sample with the corresponding $S,\Gamma$, assuming
implicitly that all fields have the same $S(\Gamma)$ distribution.

\subsection{Data from other surveys}
\label{DataOtherSurveys}

Our survey consists of \XMM{} exposures with a typical exposure time of about
15~ks, and it is hence a medium survey. Shallower wider area surveys are
required to obtain significant numbers of bright sources, while deeper
pencil-beam surveys will probe fainter fluxes. We have combined our
survey with both shallower and deeper surveys to obtain a
wide coverage in flux ($\sim$2-4 orders of magnitude):

\begin{itemize}

\item BSS: Della Ceca et al. (\cite{DellaCeca04}) have constructed a bright sample of \XMM{}
sources down to a flux of $7\times10^{-14}$~cgs in the XID band, with
a uniform coverage of 28.1~deg$^2$ above that flux. This sample is
ideally complementary to our XID, being both wider and shallower, and
selected exactly in the same band with the same observatory (but with
a different detector: MOS2 instead of pn). We have used a total of 389
sources from that survey, including all sources identified as stars,
which have been excluded from the \ns{} analysis of Della Ceca et
al. (\cite{DellaCeca04})

\item HBS: In the same paper, Della Ceca et al. (\cite{DellaCeca04}) also define a sample of
sources detected in the ultra-hard band, down to the same flux limit,
and with a uniform coverage of 25.17~deg$^2$, again both shallower and
wider than ours and with the same observatory. Results from a
subsample are presented by Caccianiga et al. (\cite{HBS28}). The source counts
are also discussed by Della Ceca et al. (\cite{DellaCeca04}) and compared to
previous results from {\sl BeppoSAX}. We have used a total of 65
sources from this survey

\item CDF: The deepest survey in the soft and hard bands so far is
the {\sl Chandra} Deep Field North (CDF-N) Bauer et al.  (\cite{Bauer04}),
obtained from a total exposure of 2~Ms in a location in the Northern
hemisphere. In a complementary effort in the South, the {\sl Chandra} Deep
Field South (Giacconi et al. \cite{Giacconi01}, Rosati et al. \cite{Rosati02})
gathered a total of 1~Ms. Source counts from both samples have been discussed
in Bauer et al. (\cite{Bauer04}). There are a total of 442 soft and 313 hard
sources in the CDF-N, and 282 soft and 186 hard sources in the CDF-S.  Recent
internal {\sl Chandra} calibrations have increased the estimate of the ACIS
effective area above 2~keV. We have used the applications on the {\sl Chandra}
calibration database to compare data from Cycles 8 and 5: above 2~keV the ratio
between the effective areas is reasonably flat, and well approximated by a
constant increase factor of about 12\%. The increase in the 2-8 keV fluxes is
unlikely to be due to the increasing contamination from the optical blocking
filter, since the latter only manifests itself below ~1 keV. Therefore, all
CDF-N and CDF-S hard fluxes (and their corresponding errors) have been
decreased by this factor. Furthermore, when comparing the sky areas calculated
individually for each source with those calculated interpolating from the sky
area of the full survey, the faintest soft sources in both CDF surveys showed
large (up to several orders of magnitude) discrepancies. Since our maximum
likelihood \ns{} fit method requires the use of a model for the sky area of the
full survey (see Section \ref{MLfit}), we have not used soft sources fainter
than 3 and 7$\times 10^{-17}$~cgs in the CDF-N and CDF-S respectively in the
\ns{} fit.

\item AMSS: The {\it {\sl ASCA} Medium Sensitivity Survey} (Ueda et al.
\cite{Ueda05}) is one of the largest high Galactic latitude broad-band X-ray
surveys to date, including 606 sources over 278~deg$^2$ with hard band
fluxes between $\sim 10^{-13}$ and $\sim 10^{-11}$ cgs.

\end{itemize}

The average sky area as a function of flux for different bands for the
CDF and AMSS are shown in Bauer et al. (\cite{Bauer04}) and Ueda et
al. (\cite{Ueda05}), respectively. We obtained them from the respective
first authors, and we have used them  for the Maximum
Likelihood fit (see Section~\ref{MLfit}). The {\sl Chandra}
re-calibration has also been applied to the CDF effective area fluxes.

\begin{table*}
   \caption[]{Maximum likelihood fit results to the \ns{} in different
bands and using different samples: the first column is the band used,
the second is the power-law slope above the flux break, the third is
the slope below that break, the fourth is the flux break, the fifth is
the normalisation, the last six columns indicate the number of
sources from each sample used in the fit.}
\label{MLNSfitTable}

   \begin{tabular}{lcccccccccc}
       \hline
       \noalign{\smallskip}
           &                        &                           &                             &                              &\multicolumn{6}{c}{$N_{\rm used}/N_{\rm tot}$}\\
Band       & $\Gamma_u$             & $\Gamma_d$                & $S_b$                       & $K$                          & AXIS      & BSS & HBS &CDF-N & CDF-S & AMSS \\
           &                        &                           & ($10^{-14}$~cgs)            & (deg$^{-2}$)                 &           &     &     &      &       &      \\
       \noalign{\smallskip}
       \hline
      \noalign{\smallskip}

soft$^g$       & 2.40$^{+0.06}_{-0.07}$ &    1.74$^{+0.06}_{-0.07}$ &      1.15$^{+0.18}_{-0.19}$ &    120.0$^{+22  }_{-18  }$   & 1267/1267 &  &  &  &   \\    
soft           & 2.39$^{+0.06}_{-0.06}$ &    1.69$^{+0.07}_{-0.06}$ &      1.15$^{+0.16}_{-0.13}$ &    123.4$^{+18.8}_{-17.1}$   & 1267/1267 &  &  &  &   \\    
soft$^f$       & 2.38$^{+0.14}_{-0.09}$ &    1.56$^{+0.01}_{-0.01}$ &      1.02$^{+0.14}_{-0.17}$ &    141.4$^{+42.1}_{-7.3}$    & 1267/1267 &  &  & 429/442$^a$ & 269/282$^b$ \\
       \noalign{\smallskip}
       \hline
      \noalign{\smallskip}
hard$^g$       & 2.74$^{+0.08}_{-0.07}$ &    -                      &      1.00                   &    735  $^{+92  }_{-76  }$   & 348/397$^c$ &  &  &  &  \\     
hard           & 2.72$^{+0.07}_{-0.08}$ &    -                      &      1.00                   &    684.4$^{+74.1}_{-84.0}$   & 348/397$^c$ &  &  &  &  \\     
hard           & 2.03$^{+0.12}_{-0.11}$ &    1.00$^{+0.11}_{-0.12}$  &      0.30$^{+0.07}_{-0.05}$ &   1086.6$^{+134.3}_{-146.5}$ &             &  &  & 313/313 &         \\
hard           & 2.51$^{+0.50}_{-0.28}$ &    0.89$^{+0.12}_{-0.12}$  &      0.72$^{+0.11}_{-0.10}$ &    743.7$^{+109.6}_{-105.6}$  &             &  &  &         & 186/186 \\
hard           & 2.12$^{+0.13}_{-0.01}$ &    1.10$^{+0.01}_{-0.01}$  &      0.44$^{+0.04}_{-0.01}$ &    799.1$^{+226.8}_{-8.4}$    &             &  &  & 313/313 & 186/186 \\
hard           & 2.66$^{+0.08}_{-0.05}$ &    1.20$^{+0.01}_{-0.01}$  &      1.00$^{+0.08}_{-0.01}$ &    611.5$^{+49.1}_{-34.3}$   & 397/397     &  &  & 313/313 & 186/186 \\
hard           & 2.58$^{+0.02}_{-0.02}$ &    -                       &      1.00                   &    606.5$^{+46.8}_{-46.3}$   & 348/397$^c$ &  &  &         &         & 606/606\\     
hard           & 2.53$^{+0.25}_{-0.18}$ &    1.18$^{+0.14}_{-0.08}$  &      0.92$^{+0.66}_{-0.19}$ &    607.8$^{+366.4}_{-208.1}$ &             &  &  & 313/313 & 186/186 & 606/606\\     
hard$^f$       & 2.58$^{+0.17}_{-0.02}$ &    1.30$^{+0.01}_{-0.01}$  &      1.17$^{+0.01}_{-0.05}$ &    485.3$^{+10.1}_{-24.3}$   & 397/397     &  &  & 313/313 & 186/186 & 606/606\\     
       \noalign{\smallskip}
       \hline
      \noalign{\smallskip}
XID$^g$        & 2.39$^{+0.05}_{-0.20}$ &    1.37$^{+0.09}_{-0.32}$ &      1.08$^{+0.07}_{-0.48}$ &    265  $^{+214 }_{-19  }$   & 1359/1359 &  &  &  &  \\
XID            & 2.46$^{+0.11}_{-0.07}$ &    1.29$^{+0.09}_{-0.18}$ &      1.45$^{+0.16}_{-0.26}$ &    212.2$^{+47.4}_{-22.8}$   & 1359/1359 &  &  &  &  \\
XID$^f$        & 2.54$^{+0.03}_{-0.04}$ &    1.35$^{+0.06}_{-0.25}$ &      1.64$^{+0.13}_{-0.28}$ &    193.0$^{+33.1}_{-15.8}$   & 1359/1359 & 389/389 &  &  &  \\
       \noalign{\smallskip}
       \hline
      \noalign{\smallskip}
ultra-hard$^g$ & 2.63$^{+0.15}_{-0.15}$ &    -                      &      1.00                   &     102 $^{+23  }_{-20  }$   & 84/89$^d$ &  &  &  &  \\
ultra-hard     & 2.59$^{+0.09}_{-0.05}$ &    -                      &      1.00                   &     95.0$^{+11.1}_{-12.8}$   & 84/89$^d$ &  &  &  &  \\
ultra-hard$^f$ & 2.62$^{+0.10}_{-0.10}$ &    -                      &      1.00                   &    102.2$^{+20.0}_{-21.1}$   & 84/89$^d$ &  & 58/65$^e$ &  &  \\
       \noalign{\smallskip}
       \hline
   \end{tabular}
   \begin{list}{}{}
        \item[$^a$]$S_{\rm min,CDF-N}=3\times10^{-17}$~cgs
        \item[$^b$]$S_{\rm min,CDF-S}=6\times10^{-17}$~cgs
        \item[$^c$]$S_{\rm min,AXIS}=1.5\times10^{-14}$~cgs
        \item[$^d$]$S_ {\rm min,AXIS}=10^{-14}$~cgs
        \item[$^e$]$S_{\rm max,HBS}=2.5\times10^{-13}$~cgs
        \item[$^f$]Best fit used for contribution to XRB, and in Figs. \ref{logNlogSFig} and \ref{ratiologNlogSFig}.
	\item[$^g$]Using fixed photon indices of 1.8 in the soft and XID bands, and 1.7 in the hard and ultra-hard bands.
   \end{list}
\end{table*}

\begin{figure*}
\hbox{
\includegraphics[width=6.5cm,angle=270.0]{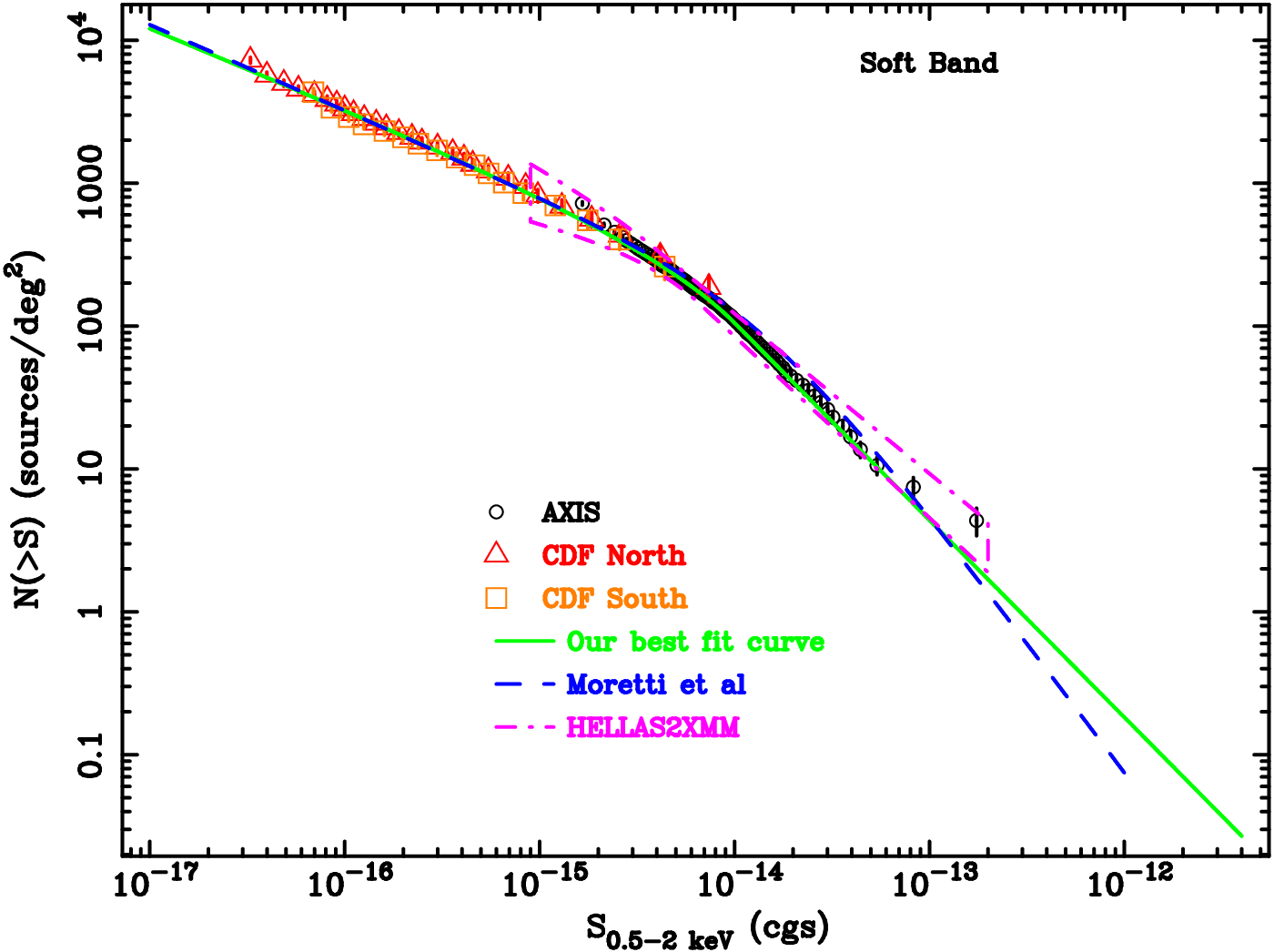}
\includegraphics[width=6.5cm,angle=270.0]{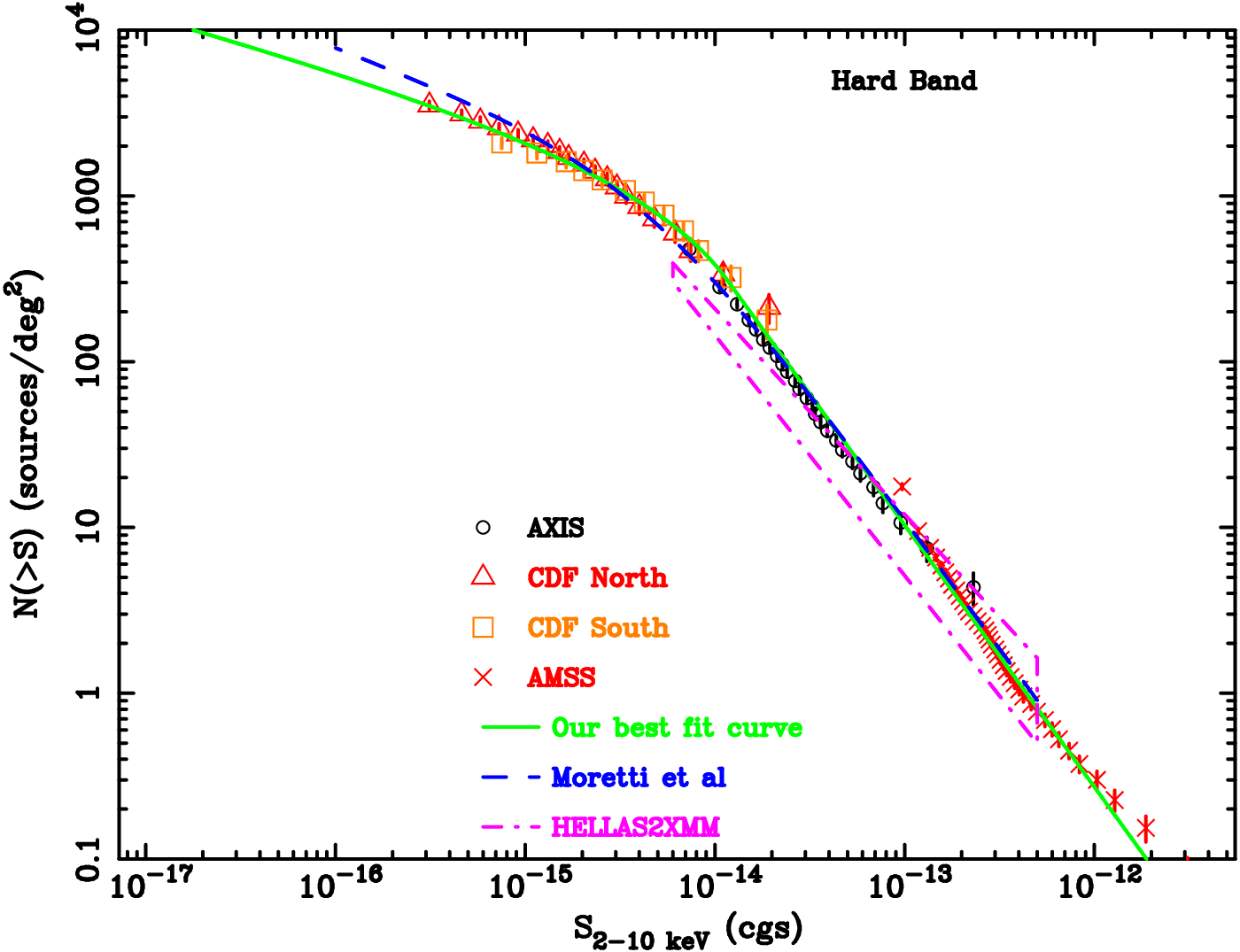}
}
\hbox{
\includegraphics[width=6.5cm,angle=270.0]{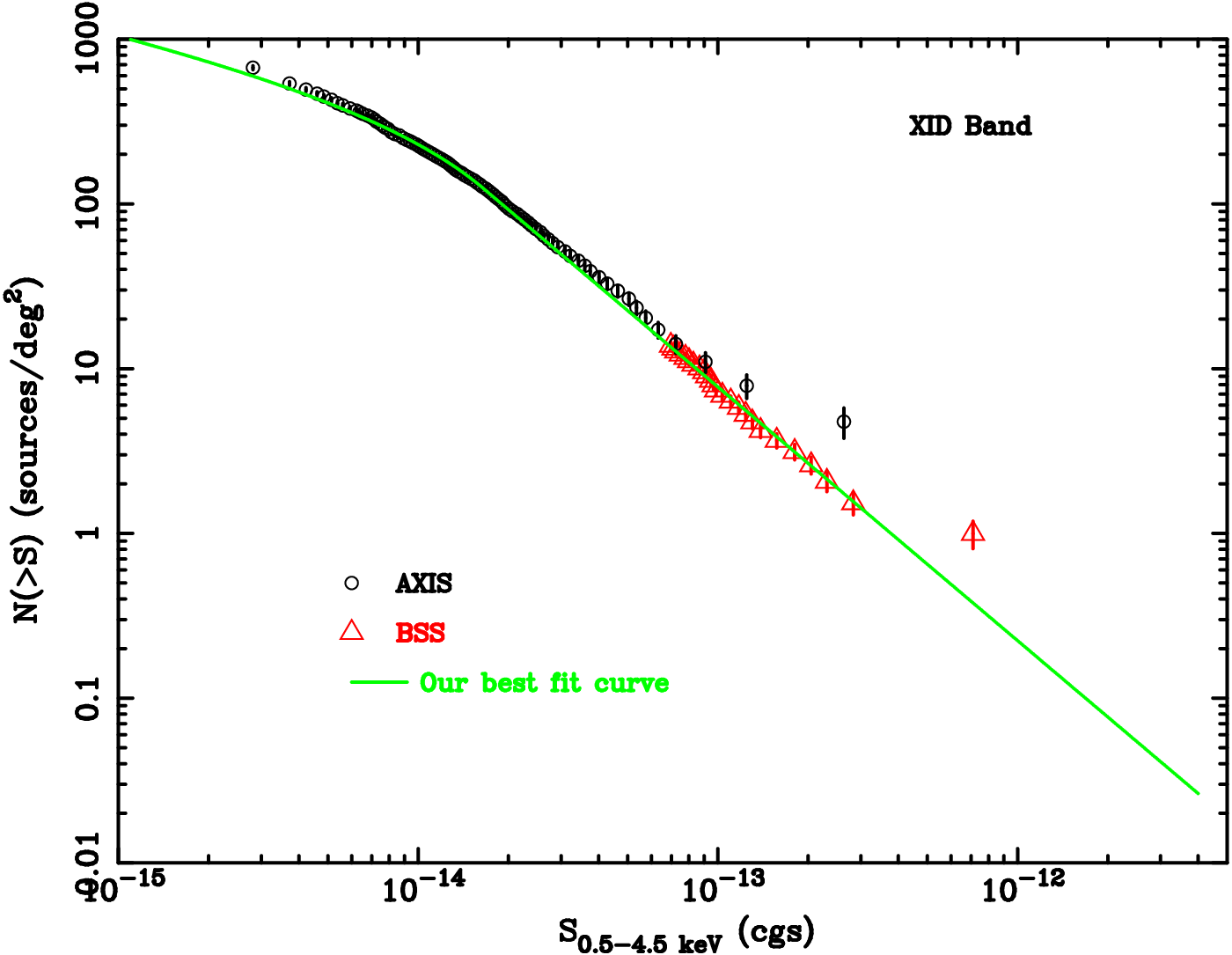}
\includegraphics[width=6.5cm,angle=270.0]{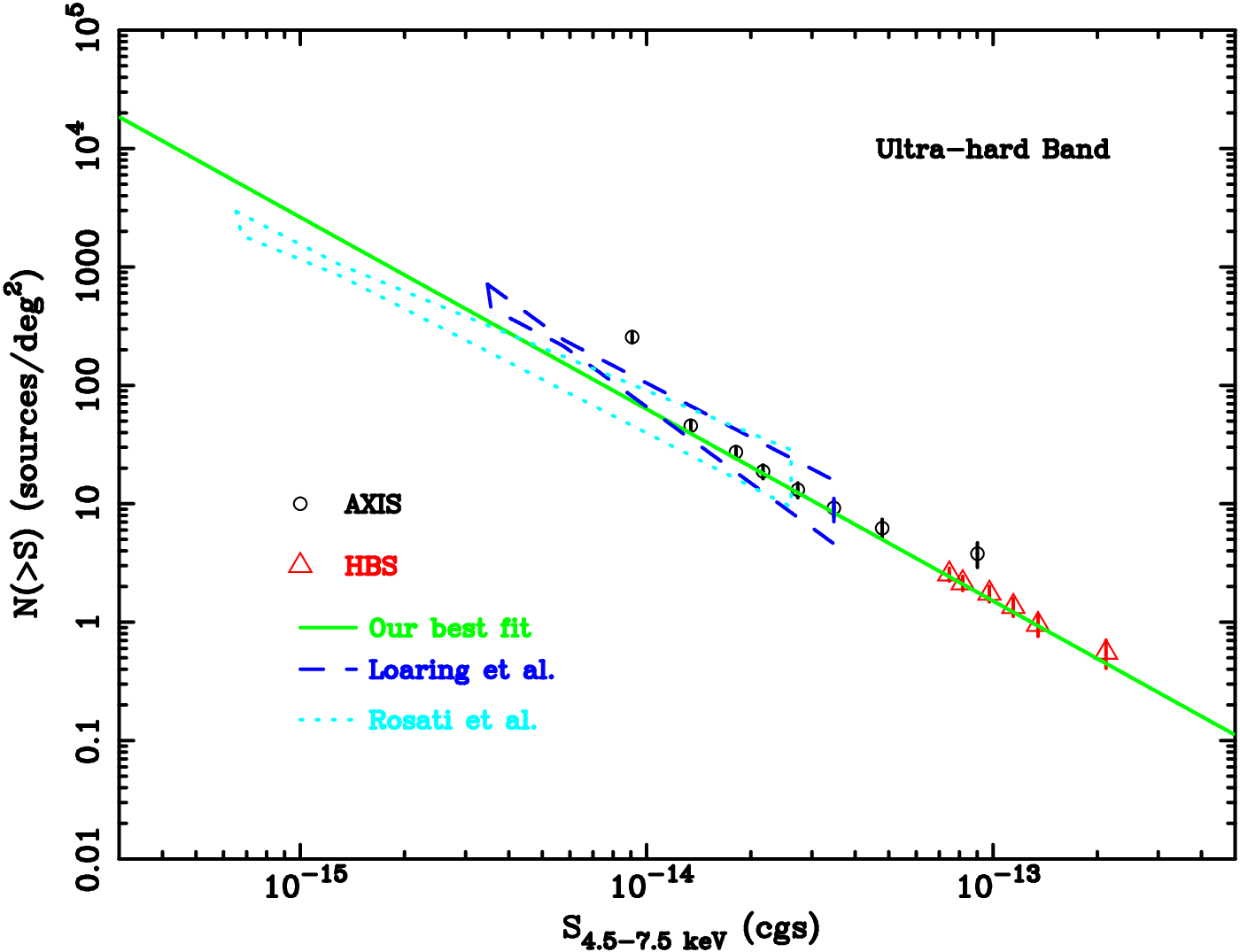}
}
\caption{Integral \ns{} in different bands, along with our best
fits and some previous results}
\label{logNlogSFig}
\end{figure*}

\begin{figure*}
\hbox{
\includegraphics[width=6.5cm,angle=270.0]{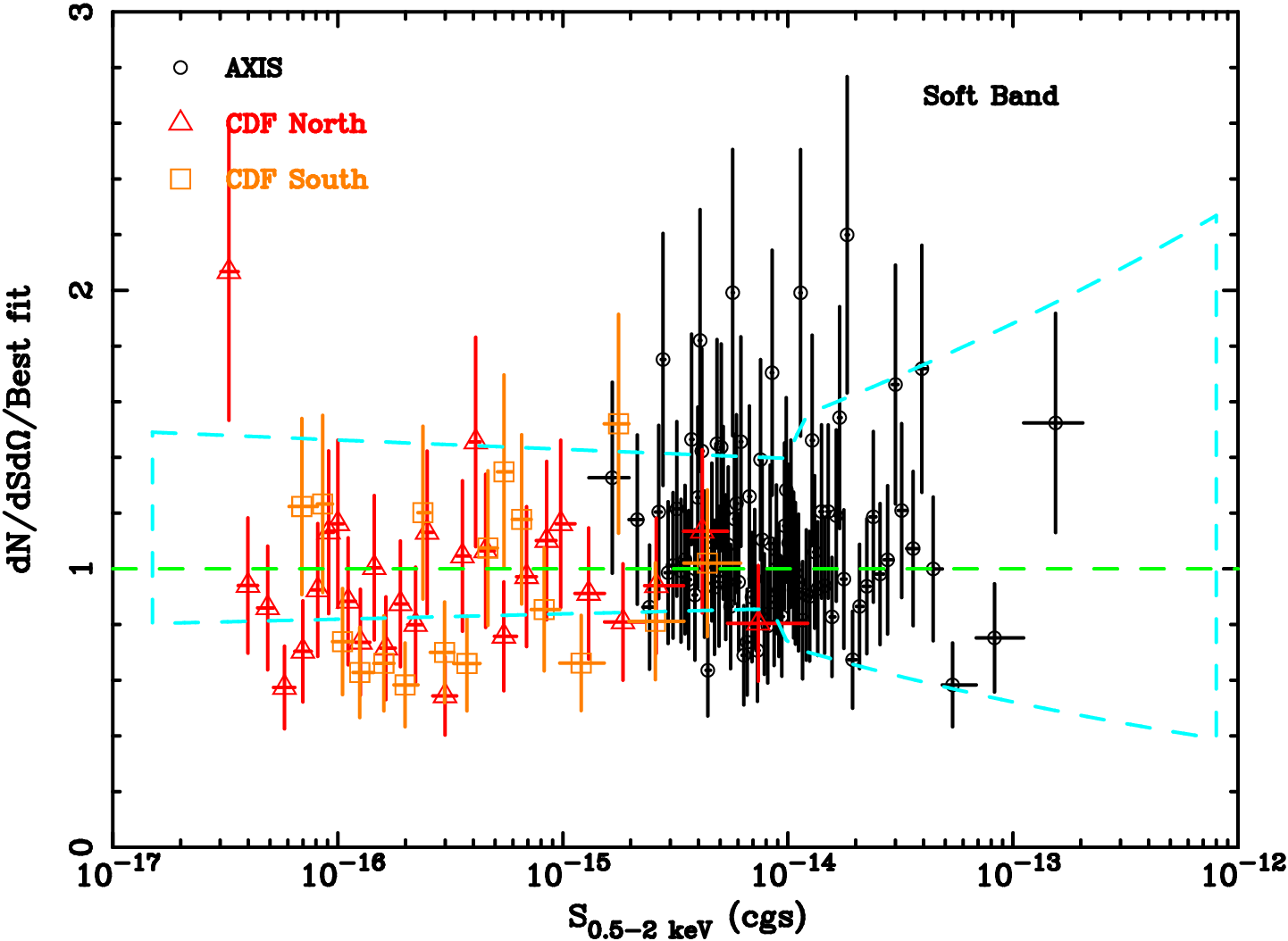}
\includegraphics[width=6.5cm,angle=270.0]{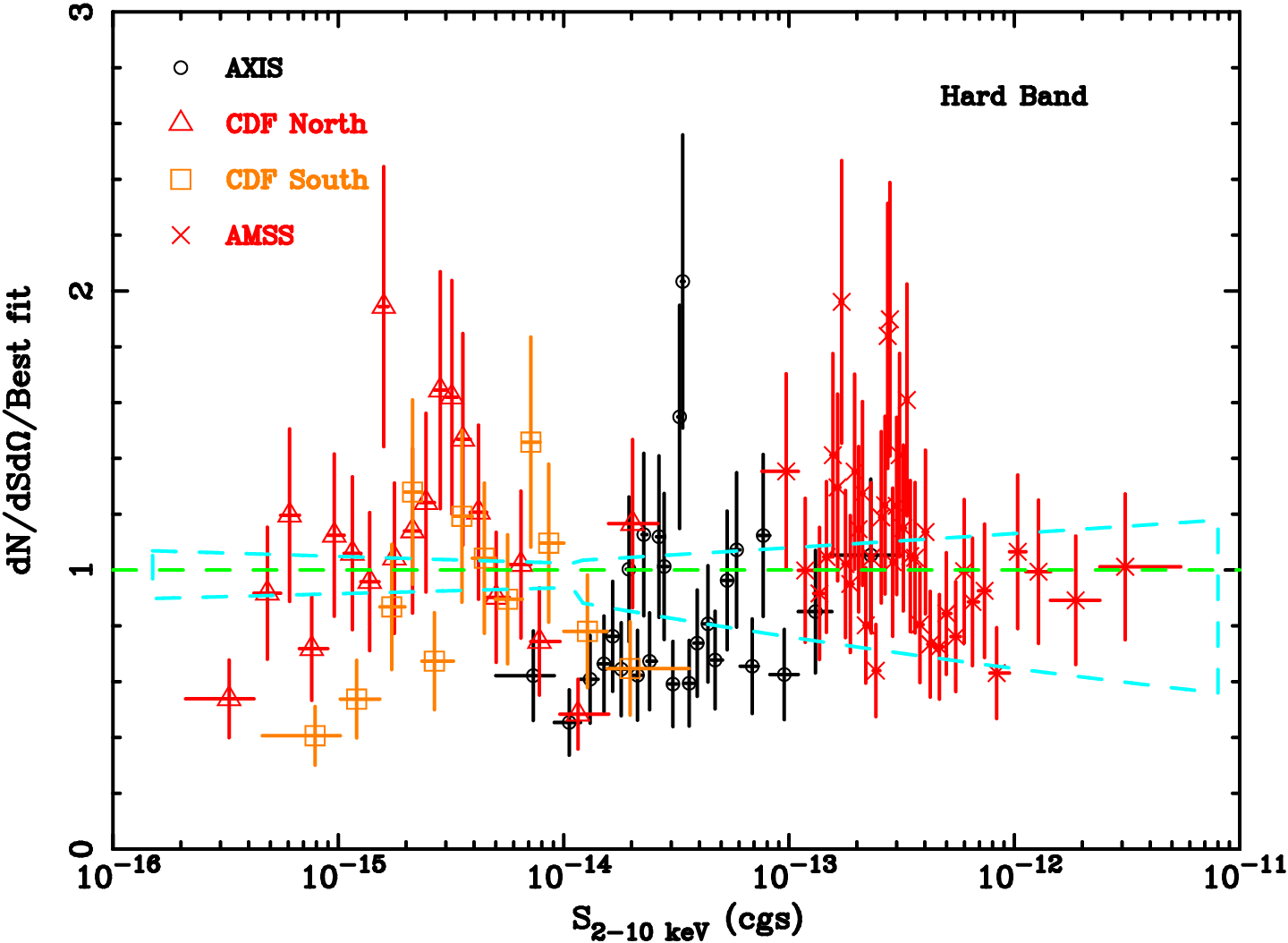}
}
\hbox{
\includegraphics[width=6.5cm,angle=270.0]{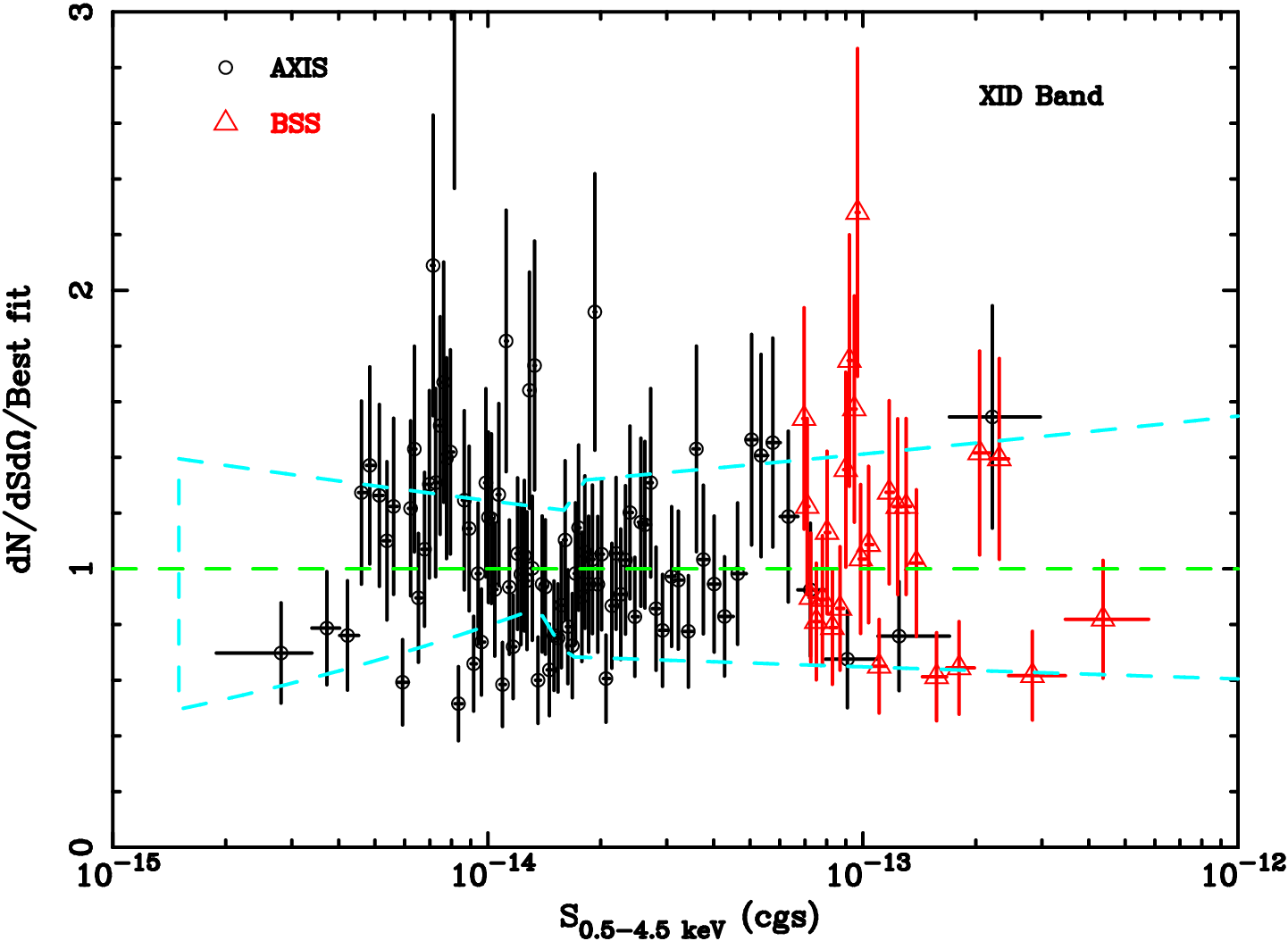}
\includegraphics[width=6.5cm,angle=270.0]{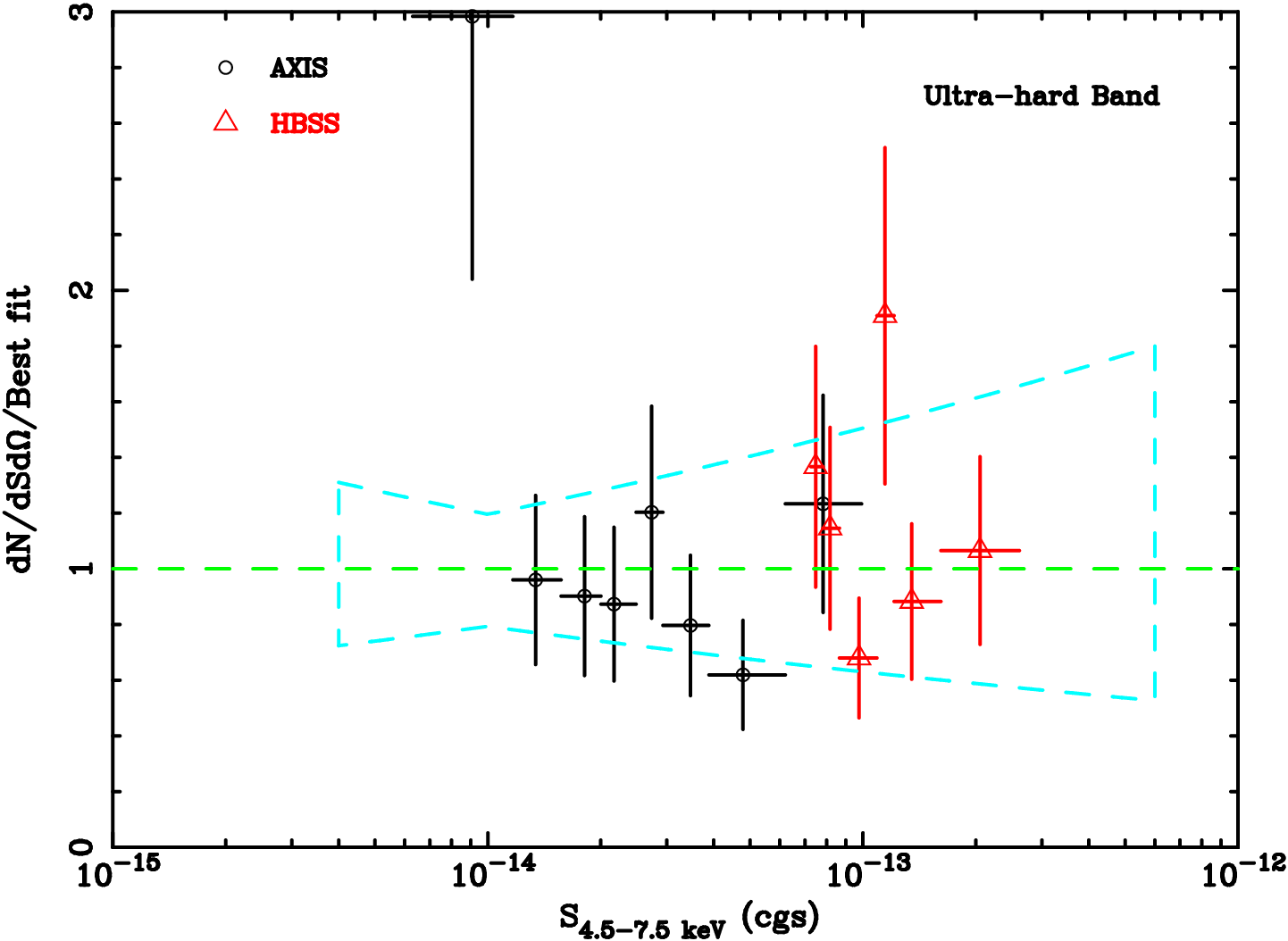}
}
\caption{Ratio between the binned differential \ns{} and the best
fit model in different bands. The 1-$\sigma$ uncertainty
interval on the \ns{} is also shown as bow-tie-shaped dashed lines above
and below unity.}
\label{ratiologNlogSFig}
\end{figure*}

\subsection{Construction of the binned \ns{}}

The binned differential \ns{}
(number
of sources per unit flux and unit sky area at a given flux $S$:  $dN/dSd\Omega$)
have been constructed
summing the inverse of the $\Omega_{j,i}$ for the sources in each flux
bin, and dividing that number by the width of the bin. The errors are
calculated dividing the $dN/dSd\Omega$ by the square root of the
sources in each bin. Since we have chosen to have a minimum number of
sources per bin (see below), their widths are all in principle
different, determined by the flux of the first source in the bin and
the flux of the first source in the next bin going up in flux. With
this definition, the width of the last bin was left undefined. Since
we have a large number of sources, we have dropped the brightest
source in each sample (only when dealing with binned \ns), defining
the upper limit in the last bin to be the flux of this brightest
source.

The integral \ns{} 
(number of sources per unit sky area with fluxes higher than $S$:
$N(>S)$)
from our sample and other samples are shown in
Fig.~\ref{logNlogSFig}. We have chosen to plot the integral \ns{} in bins
containing 15 sources each (except the last one, which can contain up to 29
sources), to avoid apparent features due to fluctuations of a few sources,
specially at the brighter ends of each sample, were the number of sources is
low. We have simply added the inverse of the sky areas for each source
$\Omega_{j,i}$, for all sources with fluxes above the lower limit of each bin.
The error bars are calculated dividing the $N(>S)$ by the square root of the
total number of sources with flux equal or greater than the lower limit of the
bin.

Sources just below the faint flux limit of a survey could experience
statistical fluctuations of their fluxes, promoting them into the survey, the
fainter sources just above that limit could drop from the survey for the same
reason, but since fainter sources are much more abundant (because the source
counts are steep), the net effect is to increase artificially the observed
source counts close to the fainter survey limits. This is known as the
Eddington bias (Eddington \cite{Eddington13}), and sometimes causes a
re-steepening of the \ns{} at the faintest fluxes, as observed in the AXIS
ultra-hard source counts, for example.

\subsection{\ns{} model}
\label{nsmodel}

Previous X-ray source count results (e.g. Bauer et al. \cite{Bauer04},
Ueda et al. \cite{Ueda03}, Moretti et al. \cite{Moretti03}, Baldi et
al. \cite{Baldi02}, Hasinger et al. \cite{Hasinger98}, Cagnoni, Della
Ceca \& Maccacaro \cite{Cagnoni98}) have shown that the \ns{} is well
approximated by a steep power law at bright fluxes, flattening at
lower fluxes. We have hence adopted the following model for the
differential \ns{}:

\begin{displaymath}
{dN\over dSd\Omega}(S)=\left\{
\begin{array}{ll}
{K\over S_b}\left({S\over S_b}\right)^{-\Gamma_d}& ,S\leq S_b\\
{K\over S_b}\left({S\over S_b}\right)^{-\Gamma_u}& ,S>    S_b\\
\end{array}
\right\}
\end{displaymath}

The above model has four independent parameters: the break flux $S_b$,
the normalisation $K$, the slope at high fluxes $\Gamma_u$, and the
slope at low fluxes $\Gamma_d$. If the change in the slope
of the \ns{} is not significant, we fixed $S_b\equiv
10^{-14}$~cgs and $\Gamma_u=\Gamma_d$, leaving only two independent
variables: $K$ and $\Gamma_u$.

The integral \ns is therefore:

$$N(>S)=\int_S^\infty dS {dN\over dSd\Omega}(S)$$

Finally, assuming a given \ns{}, the number of sources with fluxes
between $S_{\rm min}$ and $S_{\rm max}$ is

$$N(S_{\rm min}\leq S \le S_{\rm max})=\int_{S_{\rm min}}^{S_{\rm max}} dS
{dN\over dSd\Omega}(S)\Omega(S)$$

This number is the total expected number of sources in our survey
$\lambda$ with fluxes in band $i$ in that interval if
$\Omega(S)=\Omega_i(S)$ defined above, or it can be $\lambda_k$ the
expected number of sources in that flux interval and band in a given
field $k$ if $\Omega(S)=\Omega_{i,k}(S)$.

The contribution to the intensity of the XRB from the interval
$(S_{\rm min},S_{\rm max})$ is

\begin{equation}
I(S_{\rm min}\leq S \leq S_{\rm max})=\int_{S_{\rm min}}^{S_{\rm max}} dS\,
S{dN\over dSd\Omega}(S)
\label{ISminSmax}
\end{equation}

\subsection{Maximum likelihood fit method and results}
\label{MLfit}

We have fitted the \ns{} using a Maximum Likelihood method which takes
into account the uncertainties on the fluxes of the sources, as well
as the changing shape of the sky areas with flux.

Specifically, we have minimised the following expression: 
$$L_{\rm tot}=\sum_{\rm sample}L_{\rm sample}$$ 
where $L_{\rm sample}$ is given by

\begin{equation}
L_{\rm sample}=-2\sum_{j=1}^N \log(P(S_j))-2\log(P_\lambda(N))
\label{lsample}
\end{equation}

\noindent where the sum
is over all the $N$ sources in the corresponding sample, and $P(S_j)$
is the probability of finding a source of flux $S_j$ in that sample
(see below). The second term  is
$P_\lambda(N)=e^{-\lambda}\lambda^N/N!$ the Poisson probability of
finding $N$ sources in that sample when the expected number is
$\lambda$ (see Section~\ref{nsmodel}).

We have defined $P(S_j)$ as
$$P(S_j)={{\int_{S'_{\rm min}}^{S'_{\rm max}}dS {dN\over
dSd\Omega}(S)\Omega(S){\exp\left(-(S_j-S)^2\over
2\sigma_j^2\right)\over\sqrt{2\pi}\sigma_j}}\over \lambda}$$

This expression takes into account both the variation of the sky area with flux
(through $\Omega(S)$), and the uncertainty in the flux of the source $\sigma_j$
(Page et al. \cite{Page00}), through the last  exponential term in the
numerator, in which we have assumed a Gaussian distribution of the fluxes
around the measured value. To speed up the numerical calculation of the
integral in the numerator above, we have defined $S'_{\rm min}={\rm
max}(S_j-4\sigma_j,S_{\rm min})$ and $S'_{\rm max}={\rm
min}(S_j+4\sigma_j,S_{\rm max})$, since the tails of the Gaussian distribution
decrease very quickly.  The normalisation of the \ns{} ($K$) appears both in
the numerator and the denominator of $P(S_j)$ and it is unconstrained. This is
why we have introduced the second term in Eq. \ref{lsample}.

The 1-$\sigma$ uncertainties in the \ns{} parameters are estimated from the
range of each parameter around the minimum which makes $\Delta L_{\rm
tot}=1$. For each parameter, this is done by fixing the parameter of interest
to a value close to the best fit value, and varying the rest of the parameters
until a new minimum for the likelihood is found, this is repeated for several
values of the parameter until this new minimum equals $L_{\rm tot,min}+1$.

The results of the Maximum Likelihood fits to various (single and combined)
samples and bands are given in Table~\ref{MLNSfitTable}. Except when stated
otherwise, the flux interval used in the fit is $S_{\rm min}=10^{-17}$~cgs and
$S_{\rm max}=10^{-12}$~cgs. These numbers have been chosen to span the observed
fluxes of the sources. The final results are not very dependent on them, since
the sky area falls very quickly at low fluxes, and the sky density of bright
sources is very small. The initial values for the numerical search for the best
fit have been obtained from a $\chi^2$ fit to the total binned differential
\ns. The best overall fits are marked in the first column of Table~\ref{MLNSfitTable}
and also shown in Fig.~\ref{logNlogSFig}.

The first two rows in Table~\ref{MLNSfitTable} for each band allow
comparing the results of the fit to the AXIS sources using a fixed spectral
slope to calculate fluxes and sky areas ($\Gamma=1.8$ in the soft and XID
bands, $\Gamma=1.7$ in the hard and ultra-hard bands, first row), to the
results using the best fit spectral slope for each source (second row). The
results are mutually compatible in all bands, but the error bars are noticeably
larger in the XID and ultra-hard band when using a fixed spectral slope.
Since by fitting the spectra of the sources we are ``forcing'' them to
have a power law spectral shape, the uncertainties in the fitted fluxes are
smaller than those in the fluxes with fixed spectral slopes. This might explain at
least in part the smaller uncertainties in the \ns{} fitted parameters in the
latter case.
In
what follows, we have hence used the slopes from spectral fits to calculate
fluxes from count rates for all the AXIS sources, since this approach seems to
produce smaller uncertainties in the fitted parameters.

There is general agreement between data and model fits, but
it is difficult to quantify this statement, since, unlike the $\chi^2$
statistic, the absolute value of $L_{\rm tot}$ is not an indicator of
the goodness of the fit. Each panel in Fig. \ref{logNlogSFig} covers
several orders of magnitude, and hence a detailed visual comparison is
also difficult. We have therefore plotted in
Fig. \ref{ratiologNlogSFig} the ratio between the binned differential
\ns{} and the best fit model. Systematic deviations from unity in those
plots would reveal differences between the data and the best fit
model. In that Figure we also show the 1-$\sigma$ uncertainty interval
on the best fit, estimated in a conservative way: for each flux, we
have calculated the differential \ns{} in the 16 corners of the
hypercube defined by the 1-$\sigma$ uncertainty intervals in the best
fit \ns{} parameters, and taken the maximum and minimum values.

As expected, the relative agreement of the AXIS and BSS/HBS source
counts is quite good, merging with each other well, and following the
same \ns{} shape. This confirms the good relative calibration of the two
EPIC cameras on board \XMM{} (pn for AXIS and MOS2 for BSS/HBS).
The XMM-COSMOS \ns{} results (Cappelluti et al. \cite{Cappelluti07}), are also consistent with ours within 1 to 2-$\sigma$ in
the soft and hard bands, but the uncertainties in our best fit parameters are
smaller, probably due to our much higher number of
sources and wider flux coverage in those bands.

The agreement with the CDF samples is very good. In the soft
band, our joint AXIS-CDF fit is in excellent agreement with the CDF
results of Bauer et al. (\cite{Bauer04}) ($\Gamma_d=1.55\pm0.03$), as expected
since virtually all sources at low fluxes are the same in both
samples. Our AXIS-only fit prefers a steeper slope below the break,
but still compatible with the joint AXIS-CDF fit at
$<2\sigma$. The agreement with Moretti et al.  (\cite{Moretti03}) is very good
below the break ($\Gamma_d=1.60^{+0.02}_{-0.03}$, hereafter we have
translated all integral \ns{} slopes to our differential slopes),
but their break happens at higher fluxes ($S_b=(1.5\pm0.3)\times
10^{-14}$~cgs), resulting in a steeper slope at high fluxes
($\Gamma_u=2.82^{+0.07}_{-0.09}$). Our results are fully compatible
with those of the HELLAS2XMM (Baldi et al. \cite{Baldi02}) sample at their
low flux end, but our source counts are systematically at their lower envelope
above the break.

In the hard band, the CDF, AXIS and AMSS \ns{} match well in their overlapping
regions. However, the ratio in the top-right panel of
Fig. \ref{ratiologNlogSFig} is rather wiggly and shows clear differences in
detail. 
The AXIS data are mostly below the model, perhaps because (see
Section~\ref{SensMap}) the high fraction of sources excluded for being fainter
than the sensitivity map, and the relatively lower value of the sensitivity map
with respect to the Poisson count rate (which would in turn increase the
effective area).  There are two ``bumps'' with the data systematically above
the model at about $3\times10^{-15}$ and $3\times10^{-13}$ cgs (although in
this last one the discrepancy is well inside the uncertainties on the \ns{}
fit). These bumps probably arise from fitting a broken power-law with a sharp
break to the \ns{}, which is really a smooth distribution coming from the
superposition of different populations at different redshifts.  This is
compounded by possible calibration uncertainties: the hard band calibration of
{\sl Chandra} has changed by 12\% in the last 3 years (see Section
\ref{DataOtherSurveys}).

A break at $3\times10^{-15}$~cgs is actually present in the joint fit to the
CDF-N and CDF-S hard (all-CDF) samples (see Table~\ref{MLNSfitTable}), with the
low flux slope compatible with the value from Bauer et al. \cite{Bauer04}
($\Gamma_d=1.56\pm0.14$) within about 1-$\sigma$, and the high flux slope
flatter than in the global fit. The low flux slopes for the two single CDF
samples are much flatter than and incompatible with any previously reported
values ($\sim 1$). We believe that the change in the slope of the source counts
between AXIS and the CDF forces the global break flux to the overlapping fluxes
between those samples ($\sim 10^{-14}$~cgs), and hence the fit below this flux
somehow averages the two slopes present in the CDF, forming a bump around the
CDF break, rising and dropping above and below this flux, respectively.
The overall behaviour of the \ns{} can be obtained by a fit to the CDF and AMSS
data alone, and, despite the lower AXIS source counts, the best fit values
between the CDF+AMSS and the CDF+AMSS+AXIS are mutually compatible within
$<$1-$\sigma$, the main effect being pushing the break flux to higher fluxes
and steepening the low flux slope. Additionally, the uncertainties in all
parameters are significantly reduced by the inclusion of the AXIS data, by
factors between about 2 and 20.

The best fit curve by Moretti et al. (\cite{Moretti03}) is again very similar
to our all-sample fit at the highest ($\Gamma_u=2.57^{+0.10}_{-0.08}$) and
lowest fluxes ($\Gamma_d=1.44^{+0.12}_{-0.13}$), where it is however above the
CDF data (perhaps because of the change in the hard band calibration of {\sl
Chandra}).  Their transition flux occurs at lower fluxes
($S_b=(4.5^{+3.7}_{-1.7})\times 10^{-15}$~cgs), compatible with the CDF-N and
all-CDF fits, perhaps because their functional form is smoother than a broken
power-law and cannot accommodate the relatively strong break between the
AMSS/AXIS steep slope and the CDF. The HELLAS2XMM confidence interval overlaps
with ours, but has a flatter slope, and falls clearly below both the AXIS and
the CDF points below about $2\times 10^{-14}$~ cgs,  perhaps suggesting
incompleteness at their lower fluxes.

Our fit to the joint AXIS-BSS source counts in the XID band looks
visually quite good, passing through the middle of the AXIS and BSS
data points (see Fig.~\ref{logNlogSFig}). The XID band source counts
from the BSS (Della Ceca et al. \cite{DellaCeca04}) require a steeper
slope ($\Gamma_u=2.80\pm 0.11$) than both our AXIS-only and AXIS+BSS
fits. The origin of this apparent discrepancy is probably that we have
used all the selected sources in both samples, while they have
excluded the stars, which contribute more at higher fluxes, and hence
would produce a steeper source count distribution, as observed.

The AXIS-only ultra-hard source counts merge smoothly with the HBS, if
we ignore our higher flux point (which could be affected by the
exclusion of the bright targets in our survey). We have not used AXIS
sources with fluxes below $10^{-14}$~cgs in the fit because the \ns{}
steepens suddenly at those fluxes, perhaps because the Eddington bias
is most important in this band where the number of sources is
smallest,
and/or because of the contribution from spurious sources.
The best fit values including these lowest flux sources are
very similar to the ones excluding them, but the errors on the values
are much larger.  Since the ML fit is dominated by the bulk of the
sources, rather than by a few discrepant points, and given the low
number of sources, the AXIS-only and AXIS+HBS source counts are very
similar and mutually compatible. They are also compatible as well with
the HBS-only value ($\Gamma_u=2.64^{+0.25}_{-0.23}$), despite the fact
that this \ns{} again only includes the extragalactic sources. The
reason why the discrepancy is almost unnoticeable in the ultra-hard
band is probably because the stars contribute negligibly to the high
Galactic latitude source counts in this energy band.

We have compared our ultra-hard (4.5-7.5~keV) source counts to some previous
results in the 5-10~keV band, assuming a photon index of 1.7 for the flux
conversion between these bands. Our source count slope is slightly steeper than
in Baldi et al. (\cite{Baldi02}, $\Gamma_u=2.54^{+0.25}_{-0.19}$) and flatter
than in Loaring et al. (\cite{Loaring05}, $\Gamma_u=2.80^{+0.67}_{-0.55}$), but
well within the 1-$\sigma$ limits in both cases. The source counts from Rosati
et al. (\cite{Rosati02}, $\Gamma_u=2.35\pm 0.15$) are much flatter, but still
compatible with ours within less than 2-$\sigma$. They also find an indication
of a break in the source counts at a 5-10~keV flux of $4\times 10^{-15}$~cgs
(or about $3\times 10^{-15}$~cgs in the ultra-hard band). Since they also reach
fainter fluxes ($S_{\rm ultra-hard}\sim 1.6\times 10^{-15}$~cgs), their flatter
source count slope probably arises from a combination of an Euclidean slope at
brighter fluxes and an even flatter slope at their faintest fluxes. Our
absolute source counts agree well with both Loaring et al. (\cite{Loaring05})
and Rosati et al. (\cite{Rosati02}) at their brightest flux limits, and this
agreement is maintained for our best fit \ns{} over the whole flux interval in
the former case, while the latter is clearly flatter and below our best fit,
indicating a flattening of the ultra-hard source counts below our flux limit as
discussed above.
The XMM-COSMOS ultra-hard number counts coincide with ours at the
lowest fluxes ($\sim 10^{-14}$~cgs), but they are above ours at higher
fluxes, although within their relatively large error bars.

\begin{table*}
   \caption[]{Intensity in different bands from different origins and flux
intervals. The first column indicates the band, the second gives the origin of the
intensity, the third and fourth columns list the flux interval, the fifth
column total intensity from that interval, the sixth column shows the
fraction of the XRB intensity contributed by sources in that interval.}
\label{IntensityTab}

   \begin{tabular}{llrrcc}
       \hline
       \noalign{\smallskip}
Band       & Origin                & $S_{\rm min}$ & $S_{\rm max}$ & $I(S_{\rm min}\leq S\leq S_{\rm max})$ & $f_{\rm XRB}$\\
           &                       & (cgs)         & (cgs)         & ($10^{-12}$~cgs deg$^{-2}$)\\
       \noalign{\smallskip}
       \hline
      \noalign{\smallskip}

soft       & Best fit AXIS+CDF \ns      &                  0    & $3\times 10^{-17}$    & 0.25         & 0.03 \\
soft       & Best fit AXIS+CDF \ns      & $3\times 10^{-17}$    & $2\times 10^{-13}$    & 5.60         & 0.75 \\
soft       & Best fit AXIS+CDF \ns      & $3\times 10^{-17}$    &         $10^{-11}$    & 6.54         & 0.87 \\
soft       & RBS sources$^a$            &         $10^{-11}$    &                       & 0.20         & 0.03 \\
soft       & Total resolved             & $3\times 10^{-17}$    &                       & 6.55$^e$     & 0.87 \\
soft       & XRB$^b$                    &                       &                       & $7.5\pm 0.4$ \\
       \noalign{\smallskip}
       \hline
      \noalign{\smallskip}
hard       & Best fit AXIS+CDF+AMSS \ns &                  0    & $3\times 10^{-16}$    &  0.62        & 0.03 \\
hard       & Best fit AXIS+CDF+AMSS \ns & $3\times 10^{-16}$    & $1\times 10^{-11}$    & 17.08        & 0.85 \\
hard       & Best fit AXIS+CDF+AMSS \ns & $3\times 10^{-16}$    & $3.1\times 10^{-11}$  & 17.18        & 0.85 \\
hard       & HEAO-1 A2 sources$^c$      & $3.1\times 10^{-11}$  &                       &  0.43        & 0.02 \\
hard       & Total resolved             & $3\times 10^{-16}$    &                       & 17.25$^e$    & 0.85 \\
hard       & XRB$^b$                    &                       &                       & $20.2\pm 1.1$ \\
       \noalign{\smallskip}
       \hline
      \noalign{\smallskip}
XID        & Best fit AXIS+BSS \ns      & $3\times 10^{-15}$    & $10^{-12}$            &  8.48        & 0.56 \\ 
XID        & Best fit AXIS+BSS \ns      & $3\times 10^{-15}$    & $2.38\times 10^{-11}$ &  9.00        & 0.59 \\ 
XID        & Bright sources$^d$         & $2.38\times 10^{-11}$ &                       &  0.39        & 0.02 \\ 
XID        & Total resolved             & $3\times 10^{-15}$    &                       &  9.20$^e$    & 0.60 \\ 
XID        & XRB$^d$                    &                       &                       & $15.3\pm 0.6$ \\ 
       \noalign{\smallskip}
       \hline
      \noalign{\smallskip}
ultra-hard & Best fit AXIS+HBS \ns      & $9\times 10^{-15}$    & $2\times 10^{-13}$    & 1.50         & 0.21 \\
ultra-hard & Best fit AXIS+HBS \ns      & $9\times 10^{-15}$    & $1.05\times 10^{-11}$ & 1.74         & 0.24 \\
ultra-hard & Bright sources$^d$         & $1.05\times 10^{-11}$ &                       & 0.15         & 0.02 \\
ultra-hard & Total resolved             & $9\times 10^{-15}$    &                       & 1.79$^e$     & 0.25 \\
ultra-hard & XRB$^d$                    &                       &                       & $7.2\pm 0.4$ \\ 
       \noalign{\smallskip}
       \hline
   \end{tabular}
   \begin{list}{}{}
        \item[$^a$]Schwope et al. (\cite{Schwope00})
        \item[$^b$]Moretti et al. (\cite{Moretti03})
        \item[$^c$]Piccinotti et al. (\cite{Piccinotti82})
        \item[$^d$]See text
        \item[$^e$]After subtracting the stellar contribution (see Section \ref{IntensityBrightSources})
   \end{list}
\end{table*}

\section{Contribution to the X-ray background}
\label{ContributionXRB}

Using the best fit parameters from Table~\ref{MLNSfitTable}, we can
estimate the intensity contributed by sources in different flux
intervals (Eq.~\ref{ISminSmax}). In order to compare the total
intensity contributed by resolved sources with the total XRB
intensity, we need to estimate the intensity from sources brighter
than our survey limit (see Section~\ref{IntensityBrightSources}), and
to adopt a total XRB intensity (see Section~\ref{TotalXRBIntensity}).

\subsection{Intensity from bright sources and stars}
\label{IntensityBrightSources}

The contribution from bright sources is straightforward to obtain for
the ``traditional'' soft and hard bands. For the soft band we have
followed the same method as Moretti et al. (\cite{Moretti03}) summing
the fluxes of the sources in the {\it Rosat Bright Survey} (RBS,
Schwope et al. \cite{Schwope00}) with soft flux higher than
$10^{-11}$~cgs, and dividing by the area covered by the RBS. For the
hard band we have used the sources in the {\sl HEAO-1}~A2 survey
(Piccinotti et al. \cite{Piccinotti82}), which is complete down to
$3.1\times 10^{-11}$~cgs. We have estimated the contribution of bright
sources for the ultra-hard band from the hard band values, converting
both the flux limit and the intensity to the ultra-hard band using a
power-law with slope $\Gamma=1.7$. The XID band strides the hard and
soft bands and it is necessary to combine measurements taken in
different bands and with different instruments. We have estimated the
2-4.5~keV flux limit and intensity from the hard band values using
again $\Gamma=1.7$, and added both of them to the soft band values.

Since the estimates of the XRB intensity discussed below are for the
extragalactic XRB, we need to estimate the contribution from Galactic
stars down to our fainter flux limits, in order to subtract it from
our estimated source intensities and get the resolved fraction of the
extragalactic XRB. We take the estimates of Bauer et
al. (\cite{Bauer04}) from the fractions in their Table 2 and their
assumed total XRB intensities: $0.19^{+0.05}_{-0.04}\times10^{-12}$
and $0.36^{+0.34}_{-0.19}\times10^{-12}$~cgs~deg$^{-2}$ from stars in
the soft and hard bands respectively. Since most of the contribution
from stars comes from high fluxes (Bauer et al. \cite{Bauer04}), and
the stars at those fluxes have mainly low temperature thermal spectra
(Della Ceca et al. \cite{DellaCeca04}), we have estimated the stellar
contribution in the XID band from the soft band contribution assuming
a 0.5~MK {\tt mekal} model under {\tt xspec}, 
obtaining an XID-to-soft flux ratio of 1.  Our estimate of the XID band
intensity from Galactic stars is thus
$0.19^{+0.05}_{-0.04}\times10^{-12}$~cgs~deg$^{-2}$. This is a conservative
estimate, since the source counts in the XID band are about one order of
magnitude shallower than in the soft band.
Using that same spectral model, the stellar contribution in the 1-2~keV band is
$(0.036\pm 0.010)\times10^{-12}$~cgs~deg$^{-2}$.
 Given
the even higher flux limit in the ultra-hard band, and the strong
dependence of the thermal spectrum with energy, we have instead
estimated the stellar contribution to the ultra-hard band intensity by
summing the fluxes of the two sources identified as stars in the HBS,
and dividing it by the sky coverage of that survey, obtaining
$(0.099\pm0.005)\times10^{-12}$~cgs~deg$^{-2}$. No
sources have been identified as stars among the 60 identified XMS
deeper ultra-hard survey sources (Barcons et al. \cite{Barcons06b}),
and all of the 10 remaining unidentified objects in that sample are
extended. This lower fraction of stellar identifications of X-ray
sources as the flux decreases and the band hardens, is consistent with
the average soft thermal X-ray spectra of stars and their flat source
counts (Bauer et al. \cite{Bauer04}).

\begin{figure}
\includegraphics[width=6.5cm,angle=270.0]{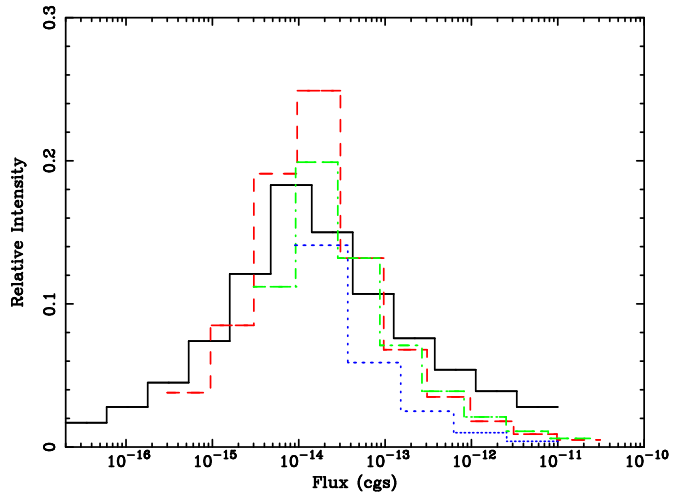}
\caption{
Relative contribution to the ``total'' intensity from equally spaced
logarithmic flux intervals, using the best fit
models in each band (see text): soft (solid line), hard (dashed), XID
(dot-dashed) and ultra-hard (dotted).}
\label{XRBContribFig}
\end{figure}

\subsection{Total X-ray background intensity}
\label{TotalXRBIntensity} 

The total extragalactic XRB intensity measured with different
instruments produces different results (Barcons et
al. \cite{Barcons00}), and not all the differences are attributable to
cosmic variance. Moretti et al. (\cite{Moretti03}) have 
averaged several measurements available in the literature in the 1-2~keV and
hard bands, obtaining $4.54\pm 0.21$ and $20.2\pm 1.1$ (in units of
$10^{-12}$~cgs~deg$^{-2}$), respectively.
We have used those values to estimate the XRB intensity in our bands assuming a
power-law with $\Gamma=1.4$, which is an adequate model for the extragalactic
XRB spectrum above 2~keV (Lumb et al. \cite{Lumb02}). The extrapolation of the
value of Moretti et al. (\cite{Moretti03}) in the hard band to the 1-2~keV band
using that spectral shape produces a value similar to the one obtained directly
by them in that band, this justifies extrapolating again the same spectral
shape down to 0.5~keV, although the shape of the XRB is poorly known below
1~keV. The results are given in Table~\ref{IntensityTab}.

Our adopted hard band XRB intensity (from Moretti et al.) is in agreement with
the estimate from Lumb et al. (\cite{Lumb02}, $21.5\pm 2.6$ in the above
units), and compatible with the result of De Luca \& Molendi (\cite{DeLuca04},
$22.4\pm 1.6$) within about 1-$\sigma$. Extrapolating those intensities to the
soft band using $\Gamma=1.4$ we obtain $7.46\pm 0.09$ and $7.78\pm 0.06$,
respectively, again fully compatible with our adopted value of $7.5\pm 0.4$.

Recently, Hickox \& Markevitch (\cite{HM06}, henceforth HM06) have
re-estimated the total XRB intensity in the 1-2~keV and 2-8~keV bands,
using {\sl Chandra} CDF data, and contributions from brighter
sources. They have analysed thoroughly the non-cosmic background in
{\sl Chandra}, and isolated unresolved components of the XRB in those
bands, which are only about $\sim$20\% and $\sim$4\% of the non-cosmic
background contributions in those bands, respectively. 
The origin of this unresolved components is unknown. One possibility is stray
light from sources outside the field of view. The HM06 total XRB intensity
estimates in those bands are obtained by summing to these unresolved components
the contributions from the resolved sources in the CDF data, and from brighter
sources using the \ns{} of Vikhlinin et al. (\cite{Vikhlinin95a}), with
different spectral slopes at different fluxes.  Their 1-2~keV intensity is
$4.6\pm0.3$ (in the above units), very similar to our adopted value.
With their estimates of the total XRB intensity, the total resolved fractions
of the XRB are only $77\pm 3$\% and $80\pm8$\% in the 1-2~keV and 2-8~keV
bands, respectively, lower than previous estimates, in particular in the softer
band. We will discuss the origin of these differences in
Section~\ref{SectionHM06}

\subsection{Contribution of different flux intervals to the X-ray background}

X-ray intensities are shown in Table~\ref{IntensityTab} for the
observed flux intervals in the samples used here, and for flux
intervals with higher maximum fluxes, chosen to ``join'' the observed
intervals with the contribution from bright sources, which are also
given in that table, as well as the XRB intensities from Moretti et
al. (\cite{Moretti03}).

We have plotted in Fig.~\ref{XRBContribFig} the relative contribution of two
flux intervals per decade to the total XRB intensity for the four default
bands, assuming our best fit \ns. It is clear that the maximum contribution
comes from fluxes around the break flux $\sim10^{-14}$~cgs. The contribution
from the bins in one decade around that value are close to 50\% of the total in
the soft and hard bands. In the ultra-hard band we do not reach deep enough to
detect the break.  From the 5-10~keV band source counts of Hasinger et
al. (\cite{Hasinger01}) and Rosati et al. (\cite{Rosati02}), there could be a
break just below $\sim 10^{-14}$~cgs.

The extrapolation the soft and hard \ns{} to zero flux using our best
fit model (see Table~\ref{IntensityTab}) does not saturate the XRB
intensity (although in the soft band total XRB intensity is within the
intensities spanned by the uncertainties in the best fit
\ns{} parameters). This suggests that the possibility of a new
dominant population at lower fluxes is still open, mainly in the hard
band. Under the assumption that the fraction of truly diffuse XRB is
negligible (and the fluctuation analysis of Miyaji \& Griffiths
\cite{Miyaji02} shows that the source counts continue growing in the
soft band down to at least $7\times10^{-18}$~cgs), it is possible to
estimate the minimum source counts slope necessary to just saturate
the XRB at zero flux, if the source counts steepened just below the
minimum flux studied here. Those slopes are 1.85 and 1.84 in the soft
and hard bands, respectively. Comparing these values to the observed
slopes of the separate AGN and galaxy source counts in the CDF (Bauer
et al. \cite{Bauer04}), only the absorbed AGN (slope $\sim 1.62$,
versus 1.2-1.5 for the rest of the AGN estimates) come close to be
able to saturate the soft XRB, while galaxies can do it with just
about any estimate for their source counts slope (2.1-2.7), if it
keeps growing at the same rate below the resolved fluxes. The
situation in the hard band is again the same, with the absorbed AGN
(source counts slope 1.95, versus 1.4-1.5) being the only AGN
population able to contribute the rest of the unresolved intensity,
while the galaxy source counts estimates (slope 3-3.5) could easily
fulfil this role.
A similar conclusion is reached if a higher intensity for the XRB is adopted
(HM06), since this would steepen the faint source counts slope required to
saturate the XRB.

Even if the source counts re-steepened as discussed in the previous
paragraph, it is clear that the maximum contribution to the XRB in the
soft and hard bands (and probably also in the XID band, and perhaps
also in the ultra-hard band) comes from sources with fluxes within a
decade $\sim10^{-14}$~cgs, where most of the AXIS sources in those
bands lie. 
This is also true if the XRB intensity is higher than the value used here.
Medium depth surveys with limiting fluxes close to that value are therefore
crucial to understand the evolution of X-rays in the Universe, at least in the
(relatively) soft bands considered here. Sources at lower fluxes and/or heavily
obscured are of course much more important for the overall energy content of
the XRB (Gilli et al. \cite{Gilli01}, Fabian \& Iwasawa
\cite{FabianIwasawa99}), since most of it resides in harder X-rays, where the
resolved fraction is much smaller (Worsley et al. \cite{Worsley04}).

\subsection{Resolved and unresolved components of the X-ray background}
\label{SectionHM06}

%
%
\begin{table}
   \caption[]{Comparison between the estimated X-ray intensities from HM06 and
   this and previous works, originating from different origins and flux
   intervals. The table is divided in
three sections: the top one is for the soft band, the middle one for the
1-2~keV band and the bottom one for the hard band. In the middle section the
conversion between 0.5-2~keV and 1-2~keV have been done in each flux interval
using the photon indices given at the beginning of Section~\ref
{SectionHM06}. The first two columns indicate the flux limits between which has
been estimated the intensity, if the first one is missing the intensity is
calculated from zero flux, while if the second one is missing, the intensity is
calculated to infinity. The third column is the intensity from HM06, and the
fourth column the intensity from this work (or previous ones as indicated). The
last three rows in each table section give the total intensity (as estimated
directly by the sum of the values in the third column by HM06, and as estimated
from previous works in the fourth column), the total intensity resolved into
sources (with the stellar contribution subtracted, see
Section~\ref{IntensityBrightSources}), and the fraction of the total intensity
resolved into sources.}
\label{TableHM06}

   \begin{tabular}{rrrl}
       \hline
       \noalign{\smallskip}
\multicolumn{4}{c}{0.5-2~keV}\\
       \noalign{\smallskip}
       \hline
      \noalign{\smallskip}
\multicolumn{2}{c}{Flux limits}&\multicolumn{2}{c}{Intensity}\\
\multicolumn{2}{c}{(cgs)}&\multicolumn{2}{c}{($10^{-12}$~cgs~deg$^{-2}$)}\\
                        && HM06 & Here \\
       \noalign{\smallskip}
       \hline
      \noalign{\smallskip}
                    & $2.5\times10^{-17}$ & $1.8\pm0.3$   &  $0.23\pm0.09$$^a$ \\
$2.5\times10^{-17}$ & $5.0\times10^{-15}$ & $2.4\pm0.4$   &  $2.2\pm0.4$$^a$ \\
$5.0\times10^{-15}$ & $1.0\times10^{-11}$ & $4.2\pm0.3$   &  $4.4\pm1.4$ \\
$1.0\times10^{-11}$ &                     &  -            &   0.2$^b$ \\
\multicolumn{2}{c}{Total}                 & $8.4\pm0.6$   &  $7.5\pm0.4$$^c$ \\
\multicolumn{2}{c}{Total resolved}        & $6.6\pm0.5$   &  $6.6\pm1.7$$^e$ \\ 
\multicolumn{2}{c}{Fraction resolved}     & $0.79\pm0.07$ &  $0.88\pm0.23$ \\
       \noalign{\smallskip}
       \hline
      \noalign{\smallskip}

\multicolumn{4}{c}{1-2~keV}\\
       \noalign{\smallskip}
       \hline
      \noalign{\smallskip}
\multicolumn{2}{c}{Flux limits}&\multicolumn{2}{c}{Intensity}\\
\multicolumn{2}{c}{(cgs)}&\multicolumn{2}{c}{($10^{-12}$~cgs~deg$^{-2}$)}\\
                        && HM06 & Here \\
       \noalign{\smallskip}
       \hline
      \noalign{\smallskip}
                    & $1.5\times10^{-17}$ & $1.0\pm0.1$   &  $0.13\pm0.05$$^a$ \\
$1.5\times10^{-17}$ & $3.0\times10^{-15}$ & $1.5\pm0.3$   &  $1.3\pm0.2$$^a$ \\
$3.0\times10^{-15}$ & $0.5\times10^{-11}$ & $2.1\pm0.1$   &  $2.2\pm0.7$ \\
$0.5\times10^{-11}$ &                     &  -            &   0.1$^b$ \\
\multicolumn{2}{c}{Total}                 & $4.6\pm0.3$   &  $4.5\pm0.2$$^c$ \\
\multicolumn{2}{c}{Total resolved}        & $3.6\pm0.2$   &  $3.7\pm0.7$$^e$ \\ 
\multicolumn{2}{c}{Fraction resolved}     & $0.78\pm0.07$ &  $0.81\pm0.16$ \\
       \noalign{\smallskip}
       \hline
      \noalign{\smallskip}

       \noalign{\smallskip}
\multicolumn{4}{c}{2-10~keV}\\
       \noalign{\smallskip}
       \hline
      \noalign{\smallskip}
\multicolumn{2}{c}{Flux limits}&\multicolumn{2}{c}{Intensity}\\
\multicolumn{2}{c}{(cgs)}&\multicolumn{2}{c}{($10^{-12}$~cgs~deg$^{-2}$)}\\
                        && HM06 & Here \\
       \noalign{\smallskip}
       \hline
      \noalign{\smallskip}

                    & $1.6\times10^{-16}$ & $ 4.2\pm2.1$   &   $0.40\pm0.04$$^a$ \\
$1.6\times10^{-16}$ & $1.6\times10^{-14}$ & $ 9.5\pm1.3$   &   $9.3\pm0.4$$^a$ \\
$1.6\times10^{-14}$ & $1.0\times10^{-11}$ & $ 7.0\pm0.4$   &   $7.9\pm1.5$ \\
$1.0\times10^{-11}$ & $3.1\times10^{-11}$ &   -            &   $0.09\pm0.06$ \\
$3.1\times10^{-11}$ &                     &   -            &    0.4$^d$ \\
\multicolumn{2}{c}{Total}                 & $20.7\pm2.5$   &  $20.2\pm1.1$$^c$ \\
\multicolumn{2}{c}{Total resolved}        & $16.5\pm1.3$   &  $17.3\pm1.7$$^e$ \\ 
\multicolumn{2}{c}{Fraction resolved}     & $0.80\pm0.11$  &  $0.86\pm0.10$ \\
       \noalign{\smallskip}
       \hline
      \noalign{\smallskip}

   \end{tabular}
   \begin{list}{}{}
        \item[$^a$]Extrapolating our best fit \ns{}
        \item[$^b$]Schwope et al. (\cite{Schwope00})
        \item[$^c$]Moretti et al. (\cite{Moretti03})
        \item[$^d$]Piccinotti et al. (\cite{Piccinotti82})
        \item[$^e$]After subtracting the stellar contribution
   \end{list}
\end{table}

Comparison between HM06 results and ours (and previous) results
involve an uncertainty concerning the different bands under
consideration (0.5-2~keV vs. 1-2~keV, and 2-8~keV vs. 2-10~keV). We
have assumed power-law photon indices of 1.5 for the unresolved
component (as fitted to the unresolved XRB spectrum by HM06), 1.43 for
the resolved faint sources (again as fitted by HM06 to the summed
spectrum of their resolved sources), and 2 for the resolved sources
brighter than $\sim 10^{-14}$~cgs (Mateos et al. \cite{Mateos05}),
for the conversions from the 1-2~keV to band the soft band and vice versa, as
well as the conversions from the 2-8~keV band to the 2-10~keV band.
 HM06 have excluded from their study
of the unresolved XRB areas around the sources in Alexander et
al. (\cite{Alexander03}), with 0.5-2~keV and 2-8~keV flux limits of
$2.5\times10^{-17}$ and $1.4\times10^{-16}$~cgs, respectively. Hence,
HM06 define as unresolved intensity that coming from sources below
those flux limits (or from a truly diffuse component).

An additional correction comes from the 2-8~keV band flux limits for
the resolved sources of HM06, which they take from Alexander et
al. (\cite{Alexander03}). This same sample was also the basis of Bauer
et al. (\cite{Bauer04}). We have already seen
(Section~\ref{DataOtherSurveys}) that those fluxes need to be
decreased by 12\% due to a change in the {\sl Chandra} calibration, in
addition to the correction due to the different bands, just
discussed. Assuming an spectral slope of 1.43 and taking into account
this flux correction, the flux limit in the 2-10~keV band becomes
$1.6\times10^{-16}$~cgs.

We compare in Table~\ref{TableHM06} the 0.5-2~keV, 1-2~keV and 2-10~keV X-ray
intensity (from different origins and flux intervals) from HM06 and
from the results from this and previous works. The error bars on our
estimates of the intensities have been estimated from the errors on
the \ns{} best fit parameters using the standard error propagation
rules (Wall \& Jenkins \cite{WJ03}). This has not been possible for
the extrapolation to zero flux, since the expressions involve the
natural logarithm of the lower limit of the interval. In this case we
have used the values spanned by the uncertainties in the best fit
\ns{} parameters, as explained in Section \ref{MLfit}.

The most  noticeable difference is between the total soft XRB intensities,
which are not compatible at the $\sim$1-$\sigma$ level, while the corresponding
1-2~keV intensities are compatible at 0.16-$\sigma$ (see
Section~\ref{TotalXRBIntensity}). This is because of the very different ways
they have been obtained: our adopted value is from an extrapolation of the
Moretti et al. (\cite{Moretti03}) 1-2~keV total XRB intensity assuming
$\Gamma=1.4$, while the HM06 estimate uses different contributions with
different values of the photon index. Since the contribution from the brightest
sources is almost half of the total, and they have the steepest spectra, the
``effective'' spectral slope in the conversion of the HM06 XRB intensity from
1-2~keV to 0.5-2~keV is $\Gamma\sim 1.72$, much steeper than our assumed
$\Gamma=1.4$, and hence with a much larger contribution from the 0.5-1~keV
interval. The HM06 resolved contribution also increases in the soft band with
respect to the 1-2~keV band, but the resolved fraction increases only slightly,
because of the very similar effective spectral shapes of the resolved component
and the XRB intensity. In contrast, our estimated resolved component is very
similar to HM06, but our assumed XRB intensity in the 0.5-2~keV has a much
flatter spectral shape, resulting in a smaller resolved fraction, but still
compatible within the errors.

We have also converted our different resolved contributions from the soft band
to 1-2~keV using different spectral slopes for different flux intervals (as
indicated at the beginning of this Section). The differences in the resolved
components and in the resolved fraction with respect to HM06 are very small (see
Table~\ref{TableHM06}) and well within the mutual uncertainties.

In summary, in the soft band the difference in the total background
intensity is  just above 1-$\sigma$, and it mostly arises because the
different effective spectral shape assumed for the total XRB intensity between
HM06 and most previous works. In contrast, the resolved intensities are very
similar in HM06 and in our work.

A further source of uncertainty in the comparison of both XRB intensities
is our extrapolation of the $\Gamma=1.4$ XRB spectral shape below 1~keV, where
its real shape is not known.  Assuming that the XRB spectrum steepens just
below 1~keV with a power-law shape, we have estimated that $\Gamma=2.2$ would be
needed to produce our estimate of the HM06 soft XRB intensity from their
1-2~keV XRB intensity, which is difficult to accommodate, given the average
slope of the faintest sources detected and of the unresolved component in the
soft band ($\Gamma\sim1.4-1.5$, HM06). Therefore, the uncertainty in the
$<$1~keV XRB spectral slope cannot fully account for the difference in the soft
XRB intensities discussed above.

%



In the hard band there is a small, statistically not significant,
discrepancy in the total background intensity, and we agree quite well
also in the resolved component, with us estimating a slightly higher
value, but well within 1-$\sigma$. Our estimate of the bright source
contribution in the hard band is fairly robust, since the AMSS sources
cover that part of the \ns.

Extrapolating the \ns{} to zero flux, we cannot
saturate the unresolved component, so either there is a diffuse
component, or the \ns{} has to steepen again somewhere below the
current flux limits (see Tables \ref{IntensityTab} and \ref{TableHM06}, and HM06).

\begin{table*}
   \caption[]{Information on the simulations for the angular correlation
   function, together with fit results: Band is the band where the sources were
   selected, $N_{\rm pool}$ is the number of sources in the pool used for the
   bootstrap simulation (see text), $N$ is number of sources selected in the
   real sample, $N_{\rm sim}$ is the number of simulations, $\chi^2_0$ is the
   value of the $\chi^2$ fit when using a ``null'' model, $\chi^2$ is the best
   fit value for a power-law model (see Eq.~\ref{WThetaEq}), with the best fit
   parameters and their 1-$\sigma$ uncertainty intervals given the the last two
   columns,  $N_{\rm bin}$ is the number of bins used in the fit, and $P(F)$ is
   the F-test probability that the improvement in the fit is significant (the
   smaller the $P(F)$ the higher the significance). The second row in each band
   corresponds to the best fit fixing $\gamma$ to the ``canonical'' value
   $0.8$.}
\label{WThetaResults}

   \begin{tabular}{lrrrrrrrcc}
       \hline
       \noalign{\smallskip}
Band  & $N_{\rm pool}$ &  $N$ & $N_{\rm sim}$ & $\chi^2_0$  & $\chi^2$ & $N_{\rm bin}$ & $P(F)$ & $\theta_0 (``)$ & $\gamma$ \\
       \noalign{\smallskip}
       \hline
      \noalign{\smallskip}
Soft  &   1177 &   1131 &  1000  & 17.10 &     5.10 &  10  &  0.0079 & $19^{+7}_{-8} $     &  $  1.2^{+0.3}_{-0.2}$ \\
      &        &        &        &       &     7.80 &  10  &  0.0096 & $6^{+2}_{-2}$        &  $\equiv 0.8$          \\
       \noalign{\smallskip}
       \hline
      \noalign{\smallskip}
Hard  &    415 &    351 &  2500  &  8.47 &     7.33 &  10  &  0.5622 & $12^{+20}_{-12}$    &  $  1.1^{+2.8}_{-2.3}$ \\
      &        &        &        &       &     7.47 &  10  &  0.3017 & $4^{+5}_{-4}$       &  $\equiv 0.8$          \\
       \noalign{\smallskip}
       \hline
      \noalign{\smallskip}
XID   &   1301 &   1218 &  1000  & 16.00 &     5.30 &  10  &  0.0120 & $19^{+7}_{-8} $     &  $  1.3^{+0.4}_{-0.3}$ \\
      &        &        &        &       &     8.80 &  10  &  0.0238 & $4^{+2}_{-2}$       &  $\equiv 0.8$          \\
       \noalign{\smallskip}
       \hline
      \noalign{\smallskip}
UH    &     88 &     77 & 10000  &  1.97 &     1.87 &  10  &  0.8119 & $0.7^a$             &  $  0.22^a$            \\
      &        &        &        &       &     1.90 &  10  &  0.5759 & $8^{+26}_{-8}$      &  $\equiv 0.8$          \\
       \noalign{\smallskip}
       \hline
      \noalign{\smallskip}
HR    &    250 &    225 & 10000  &  3.60 &     1.10 &  10  &  0.0087 & $42^{+8}_{-12}$     &  $  3.9^{+2.5}_{-1.3}$ \\
      &        &        &        &       &     3.22 &  10  &  0.3165 & $5^{+9}_{-5}$       &  $\equiv 0.8$          \\
       \noalign{\smallskip}
       \hline
      \noalign{\smallskip}
Soft/3&    392 &    380 &  2500  & 14.28 &    10.42 &  10  &  0.2835 & $35^{+11}_{-18}$    &  $  1.7^{+1.0}_{-0.6}$ \\
      &        &        &        &       &    12.53 &  10  &  0.2912 & $ 7^{+7}_{-6}$      &  $\equiv 0.8$          \\
       \noalign{\smallskip}
       \hline
      \noalign{\smallskip}
   \end{tabular}
   \begin{list}{}{}
	\item[$^a$] Parameter unconstrained by the fit
   \end{list}
\end{table*}

\begin{figure*}
\hbox{
\includegraphics[width=6.5cm,angle=270.0]{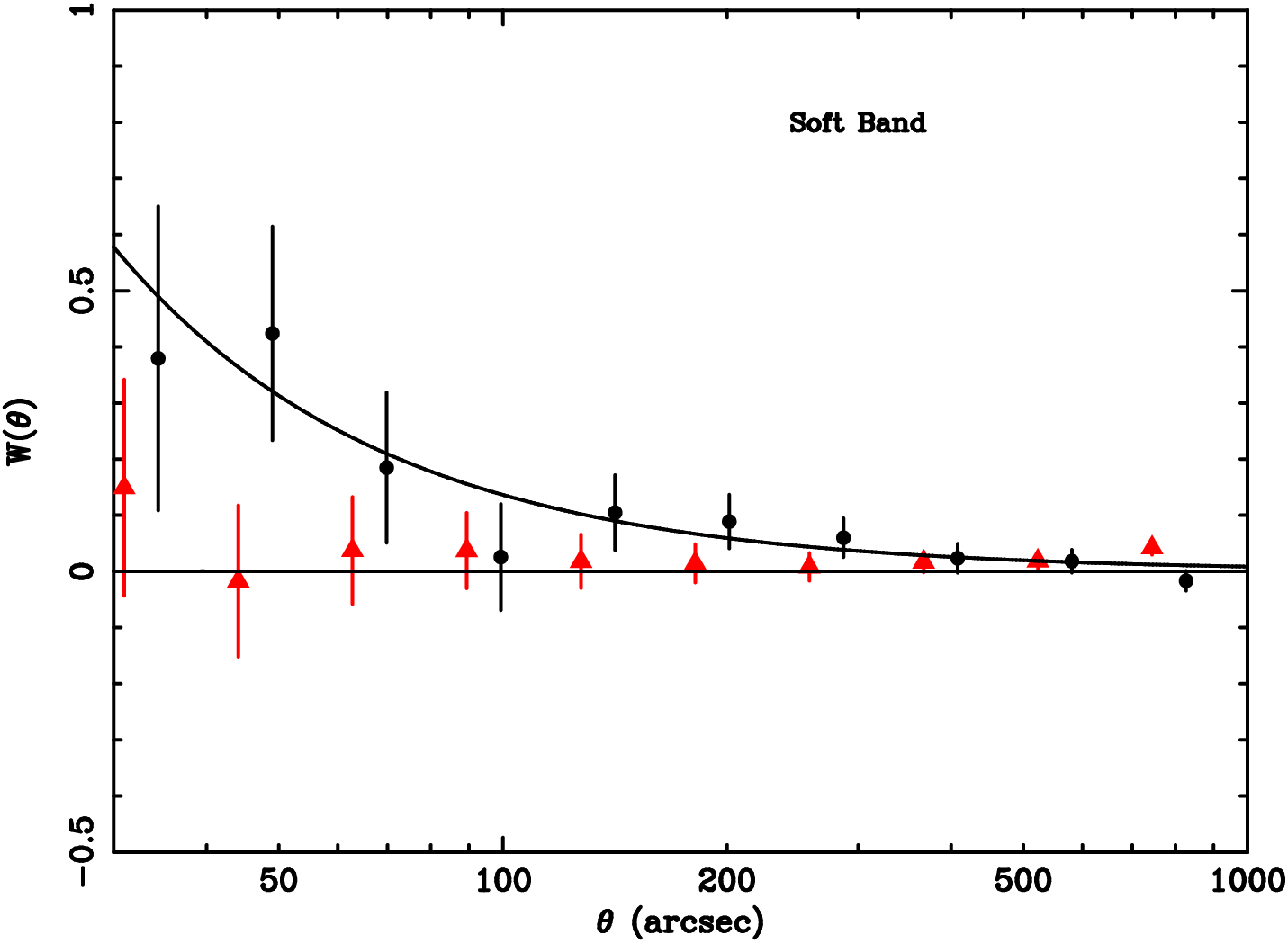}
\includegraphics[width=6.5cm,angle=270.0]{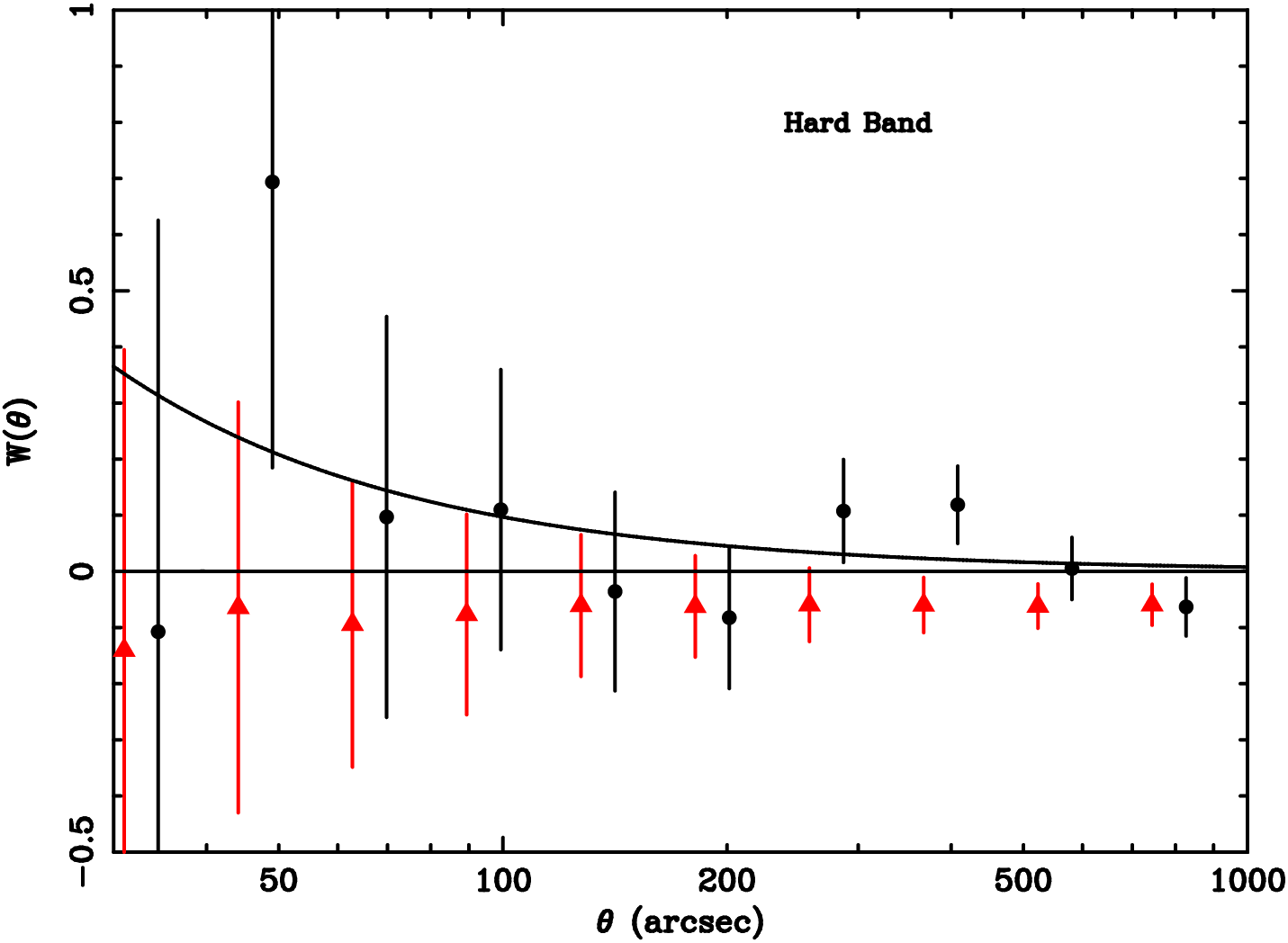}
}
\hbox{
\includegraphics[width=6.5cm,angle=270.0]{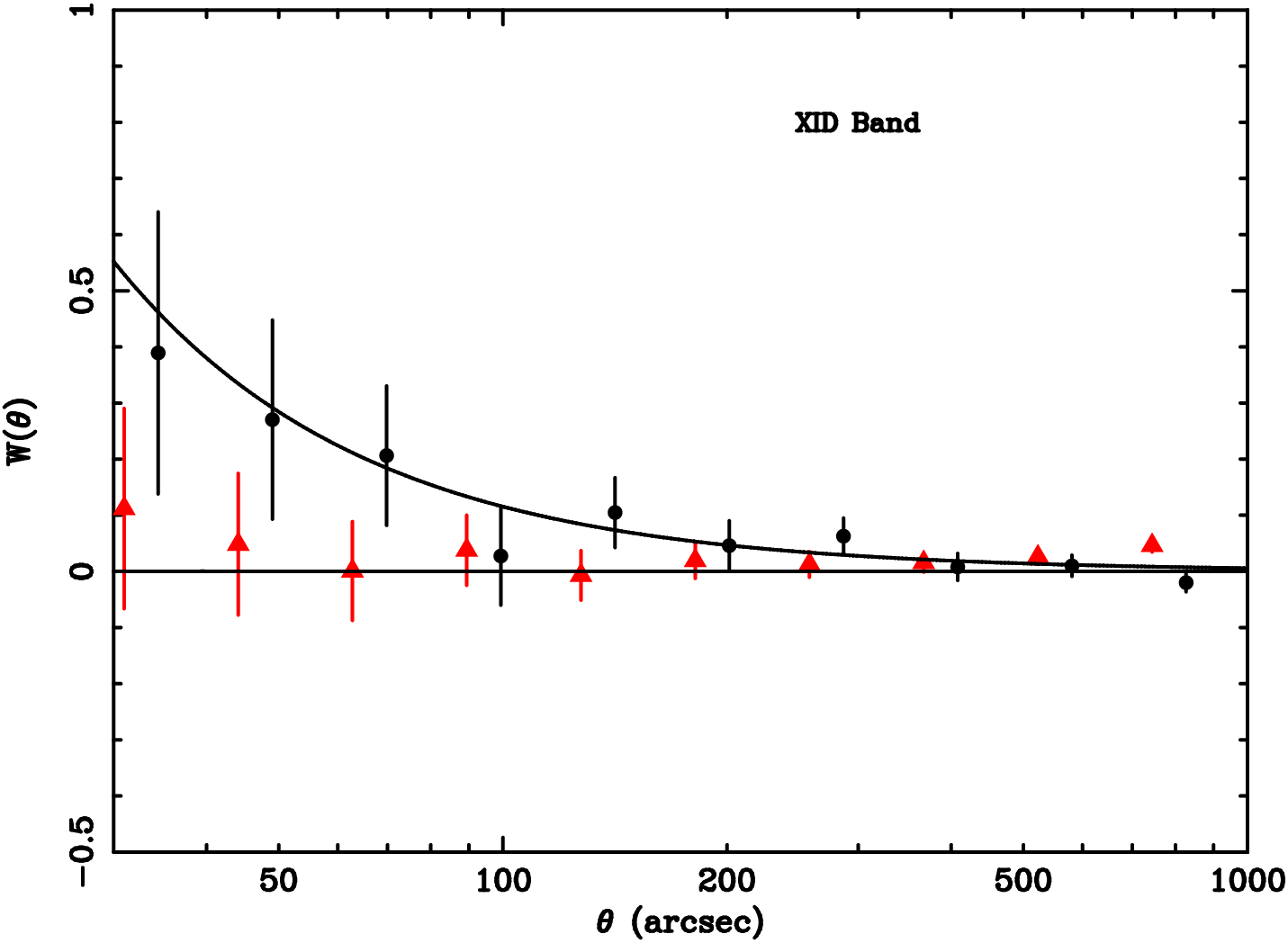}
\includegraphics[width=6.5cm,angle=270.0]{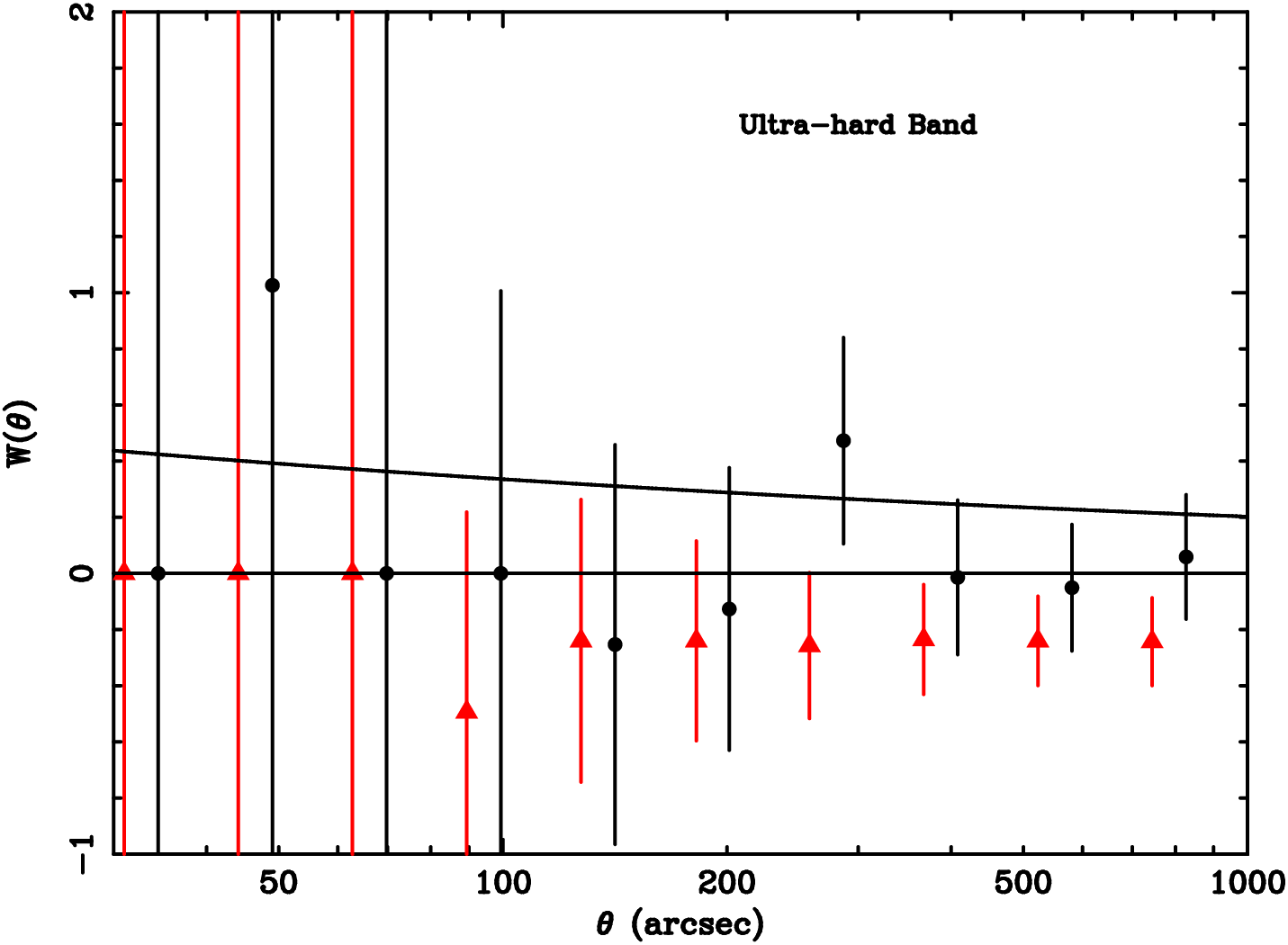}
}
\hbox{
\includegraphics[width=6.5cm,angle=270.0]{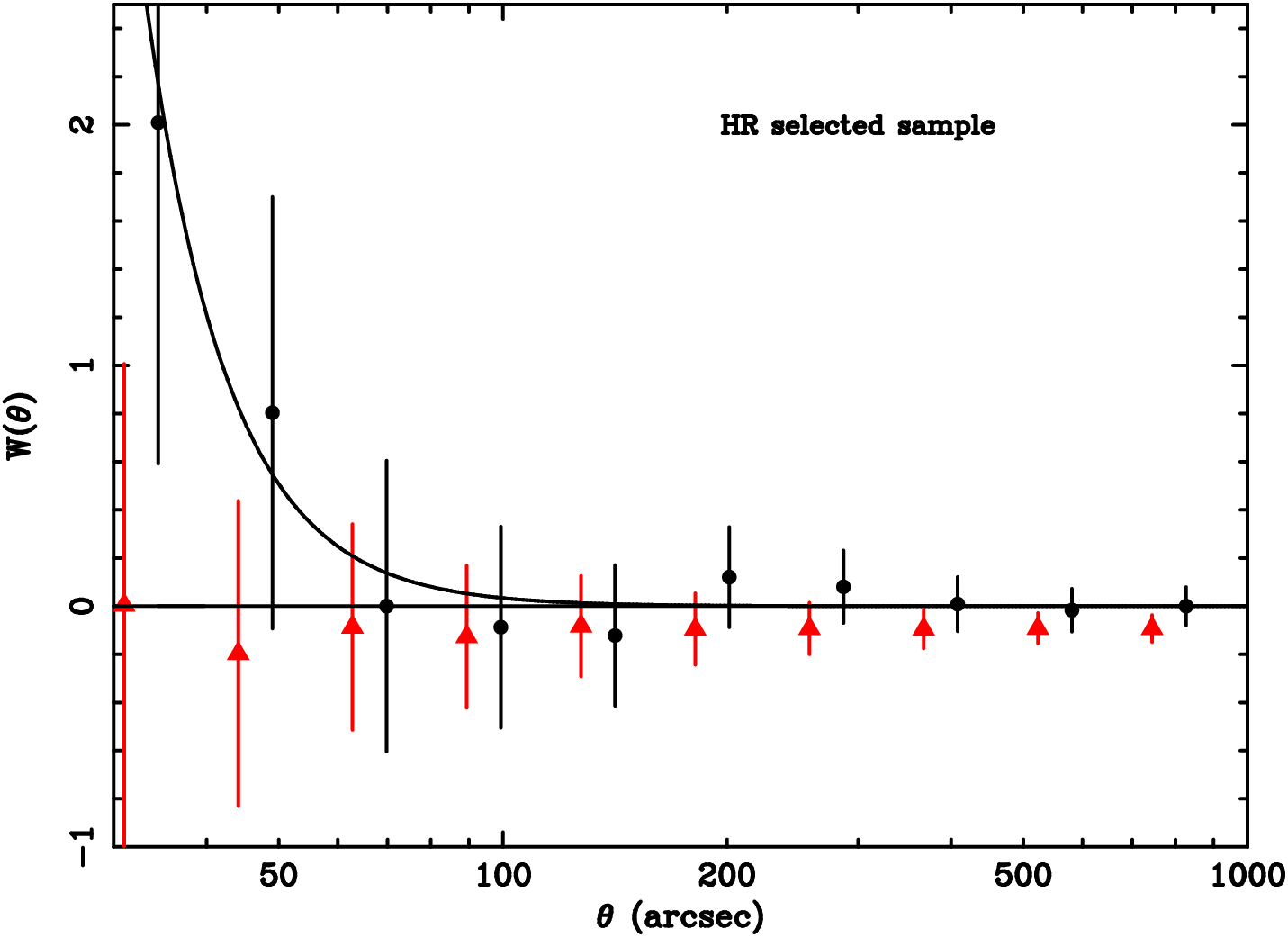}
\includegraphics[width=6.5cm,angle=270.0]{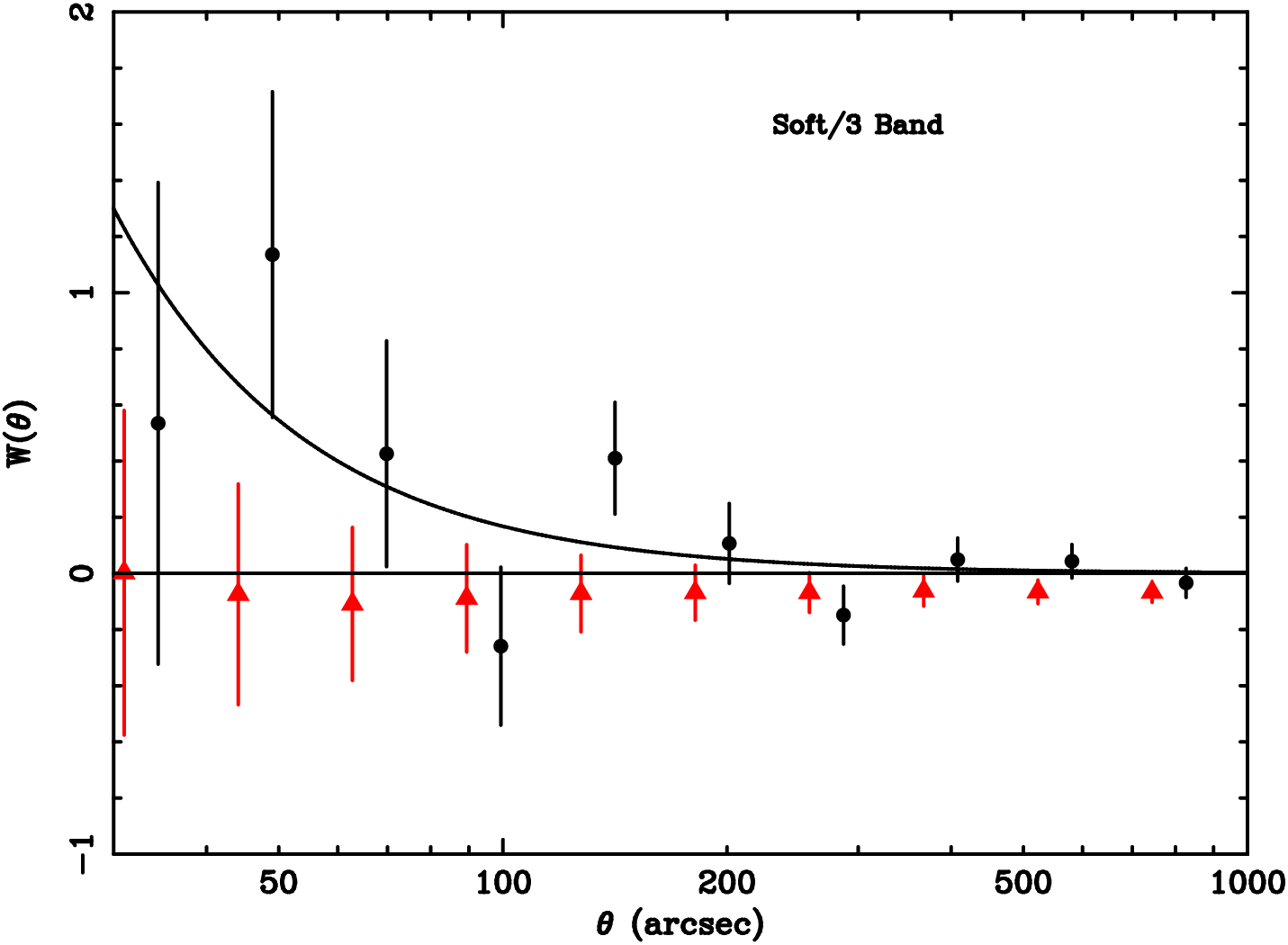}
}
\caption{$W(\theta)$ vs. $\theta$ for the soft (top-left), hard (top-right),
XID (middle-left), and ultra-hard (middle-right) samples, for a hardness-ratio selected
sample (bottom-left) and for a random selection of one third of the soft sources
(bottom-right). The solid dots are the integral-constraint-corrected observed values, the grey triangles
show the ``zero points'' from the integral constraint (displaced to the left
for clarity), and the solid lines are the best fit $\chi^2$ fits to
Eq.~\ref{WThetaEq}. The middle-right and two bottom panels have
different scales on their Y-axis.}
\label{WThetaFig}
\end{figure*}

\section{Looking for source clustering}
\label{SecAngularClustering}

A survey to detect source clustering requires both depth (to achieve
high angular density), and width (to minimise the possibility that a
single structure biases the overall average). In practice, a
compromise between those two conflicting requirements has to be
achieved. The AXIS sample comprises 36 fields outside the Galactic
plane, and a relatively high source density (about 
35 sources per field, at least in the soft and XID samples), and we have tested
what can it say about the angular distribution of sources in the sky.

\subsection{Cosmic variance}
\label{CosmicVariance}

The simplest test for source clustering is to compare the actual number of
sources detected in each field $N_k$ with the number expected $\lambda_k$ from
the best fit \ns{} and the sky area of each individual field $\Omega_k$ (this
is similar to the traditional counts-in-cells method). In principle, if just
one field (or a few fields) happens to look through strong cosmic structures,
its number of sources should be significantly different from the expected
number from a random uniform distribution, as measured by the overall \ns.

The statistics we have used to measure the deviation
from such a random uniform distribution are the cumulative Poisson
distributions:

\begin{equation}
\begin{array}{lcll}
P_{\lambda_k}(\geq N_k) & = & \sum_{l=N_k}^{\infty}P_{\lambda_k}(l) & {\rm if}\,N_k>\lambda_k \nonumber\\
P_{\lambda_k}(\leq N_k) & = & \sum_{l=0}^{N_k}     P_{\lambda_k}(l) & {\rm if}\,N_k<\lambda_k \\
\end{array}
\label{PgtN}
\end{equation}

\noindent where $P_\lambda(l)$ is the Poisson probability of detecting $l$
sources when the expected number of sources is $\lambda$. The cumulative
distributions above give the probability of finding $\geq N_k$ (first row) or
$\leq N_k$ (second row) sources when the expected number from the source counts
is $\lambda_k$. This method is similar to the one used in Carrera et
al. (\cite{Carrera98}), but in that work we used $P_\lambda(l)$ instead of the
cumulative probabilities. We believe that the approach used here is a more
conservative estimate of how likely is to find a number of sources in a field
which is different from the expected value.

The maximum likelihood statistics for the whole sample is then
$$L'=\sum_{k}P_{\lambda_k}(\geq N_k)+\sum_{k'}P_{\lambda_{k'}}(\leq N_{k'})$$,
where $k$ runs over the fields for which $N_k>\lambda_k$, and $k'$ over the
fields for which $N_{k'}<\lambda_{k'}$.

We have compared the observed $L'$ values in each band with 10000 simulated
values, using the values of $\lambda_k$ and Poisson statistics. The number of
simulations with likelihood values above the observed ones were 1388, 4580, 778
and 4958 for the soft, hard XID and ultra-hard bands, respectively.  Nothing
significant is found in the hard or ultra-hard samples, while some deviation
below or about at the 90\% significance level is found in the soft and XID
bands. These results, although formally at a low significance, have
encouraged us to try more elaborated tests for clustering.

\subsection{Angular correlation function}
\label{ACF}

If cosmic structure is present in all (or most) fields, a test for
significant deviations from the mean number of sources in each field
from some overall average will not give significant results. We should
look instead for evidence of sources tending to appear together in the
sky with respect to an unclustered source distribution. The classic
parametrisation for this effect is the angular correlation function
$W(\theta)$ which measures the excess probability of finding two
sources in the sky at an angular distance $\theta$ with respect to a
random uniform distribution (Peebles \cite{Peebles80}): 

\begin{equation}
\delta P=n^2\delta\Omega_1\delta\Omega_2[1+W(\theta)]
\label{WThetaDefinition}
\end{equation}

\noindent where $\delta P$ is the probability of finding two objects in two
small angular regions $\delta\Omega_1$ and $\delta\Omega_2$, separated
by an angle $\theta$, when the sky density of objects is $n$.

Since the angular separation is a projection in the sky of the real
spatial separations of the sources at different redshifts, the
underlying spatial clustering is somewhat blurred with this purely
angular measurement. Unfortunately, the more powerful spatial
clustering depends on having redshifts for a very high fraction of the
sources, or at least knowing precisely what is their redshift
distribution. Since none of these two conditions are fulfilled by any
of the AXIS samples, we have used the data
presently at hand to study the angular correlation function.

There are several proposed ways of measuring the angular correlation function,
most of which look for an excess number of source pairs at a given angular
separation $\theta$ with respect to a simulated random ``uniform'' sample (Landy \&
Szalay \cite{Landy93}, Efstathiou et al. \cite{Efstathiou91}). We have chosen the one proposed by
Efstathiou et al. (\cite{Efstathiou91}).

\begin{equation}
W(\theta)=f{DD \over DR}-1
\label{DDDR}
\end{equation}

\noindent where $DD$ is the number of actual pairs of sources with
angular separation $\theta$, and $DR$ is the number of pairs of one real and
one randomly placed simulated source (see below) at the same separation, while
$f$ is a normalisation constant to take into account the different number of
real and simulated sources. 
 The error bars around each point are
given by $\Delta W(\theta)=\sqrt{(1+W(\theta))/DR}$ (Peebles \cite{Peebles80}).

The ``randomly placed'' simulated sources have to follow as closely as
possible the real distribution of the source detection sensitivity of
the survey, giving a ``flat'' random sky against which to judge the
presence (or otherwise) of significant over-densities at different
angular separations. We have used bootstrap simulations by forming a
pool of real sources with detection likelihoods higher than 15 (our
standard value) in the band under study, irrespectively of whether
their count rates were above or below the sensitivity map of the
corresponding field at the source positions. Then, keeping the number
of sources simulated in each field equal to the real number of sources
$N_k$, we have extracted sources from this pool, keeping their count
rates, and their distances to the optical axis of the X-ray telescope,
but randomising their azimuthal angle around it. If the source had a
count rate above the sensitivity map of the field under consideration
at its ``new'' $(X,Y)$ position, the source was kept in the simulated
sample, otherwise a new one is extracted until $N_k$ valid simulated
sources are found. In this way, the angular distribution of the
simulated sources mimics the decline of the source detection
sensitivity with off-axis angle. The number of random samples $N_{\rm
sim}$ used in each band are shown in Table~\ref{WThetaResults}. They
have been chosen so as to give a total of about a million simulated
sources in each band.

With this recipe, the normalisation constant $f$ above is given by
$$f={ \sum_k N_k\left(N_k-1\right)\over2N_{\rm sim}\sum_kN_k^2}$$ 
\noindent $N_k$ being the number of sources in field $k$ and $N_{\rm sim}$ the
number of simulations.

If a positive correlation is present at angular scales comparable to
the individual field size, the estimate of the mean surface density of
objects from the survey is too high, and this causes a negative bias
in the angular correlation function known as the integral constraint
(Basilakos et al. \cite{Basilakos04}). Given the complicated dependence of the
sensitivity over the area of our survey, we have corrected for this
effect empirically, finding the angular correlation function that we
would have detected in the absence of correlation, via the average of
$N_{\rm sim}$ simulated realisations of $W(\theta)$, where the real data were
replaced by random samples, simulated independently following the
recipe in the previous paragraph. The triangles in
Fig.~\ref{WThetaFig} show these ``zero points'' at each angular scale,
which have been used to increase the corresponding observed
$W(\theta)$, convolving the error bars using Gaussian statistics.


Applying Eq.~\ref{DDDR} to the full sample results in a $\sim$3-$\sigma$
significant bump at about 200~arcsec, which was also present in the
simulations, but at lower significance. The bump turned out to be 
present only in two fields (HD111812 and HD117555), a search in the
literature revealed that both fields have stellar clusters in
their field of view (Eggen \& Iben \cite{Eggen89}). Since we are interested mainly in
extragalactic structure, we have opted for excluding these two fields
from all subsequent angular clustering analysis, leaving 34 fields.

The $W(\theta)$ obtained in this fashion are shown in
Fig.~\ref{WThetaFig}, as well as the corresponding best $\chi^2$ fits
to a power-law model

\begin{equation}
W(\theta)=(\theta/\theta_0)^{-\gamma}\label{WThetaEq}
\end{equation}

The best fit parameters for this model are given in
Table~\ref{WThetaResults}. In that Table we also give the significance of the
detection from an F-test comparing the $\chi^2$ value of the power-law model
with a simple $W(\theta)=0$ no-clustering model. The F-test suggests
significant correlation in the soft and XID bands at the $\sim$99\% level.
The lack of significant
detections in the two harder bands might be due to the lower number of
sources with respect to the soft and XID bands (see below).

Gandhi et al. (\cite{Gandhi06}) also found significant angular
correlation in the XMM-LSS sample in the soft band, with a similar
slope ($\gamma=1.2\pm 0.2$), and a lower correlation length
($\theta_0=7\pm3$~arcsec), but still compatible with ours within
$\sim$1-$\sigma$. Vikhlinin \& Forman (\cite{Vikhlinin95}) found
$\gamma=0.7\pm0.3$ and $\theta_0=4\pm3$~arcsec, again a lower
correlation length than us, but compatible within less than 1-$\sigma$
with our $\gamma\equiv0.8$ result. On the other hand, Basilakos
et al. (\cite{Basilakos05}) also found significant angular correlation
in this band, but with a higher correlation length
($\theta_0=10.4\pm1.9$~arcsec, with the canonical slope), just
compatible with our results at about the 2-$\sigma$ level. Basilakos et
al. (\cite{Basilakos05}) also used Limber's equation (Peebles
\cite{Peebles80}) to calculate the spatial correlation function from
the angular correlation function, assuming several different AGN X-ray
luminosity functions (since most X-ray sources at low flux levels are
expected to be AGN). Their high correlation lengths are only
compatible with AGN optical (e.g. Croom et al. \cite{Croom02}, Grazian
et al. \cite{Grazian04}) or X-ray correlation functions (Carrera et
al. \cite{Carrera98}, Akylas et al. \cite{Akylas00}, Mullis et
al. \cite{Mullis04}) if the clustering is constant in physical
coordinates, while optical QSO clustering seems instead to be constant
in comoving coordinates (Croom et al. \cite{Croom01}, but see also
Grazian et al. \cite{Grazian04}, Croom et al. \cite{Croom05}). Akylas
et al. (\cite{Akylas00}) also found significant angular correlation in
the soft band sources in the RASS-BSC, but at much higher angular
distances ($\sim 8^\circ$).

\begin{figure*}
\hbox{
\includegraphics[width=6.5cm,angle=270.0]{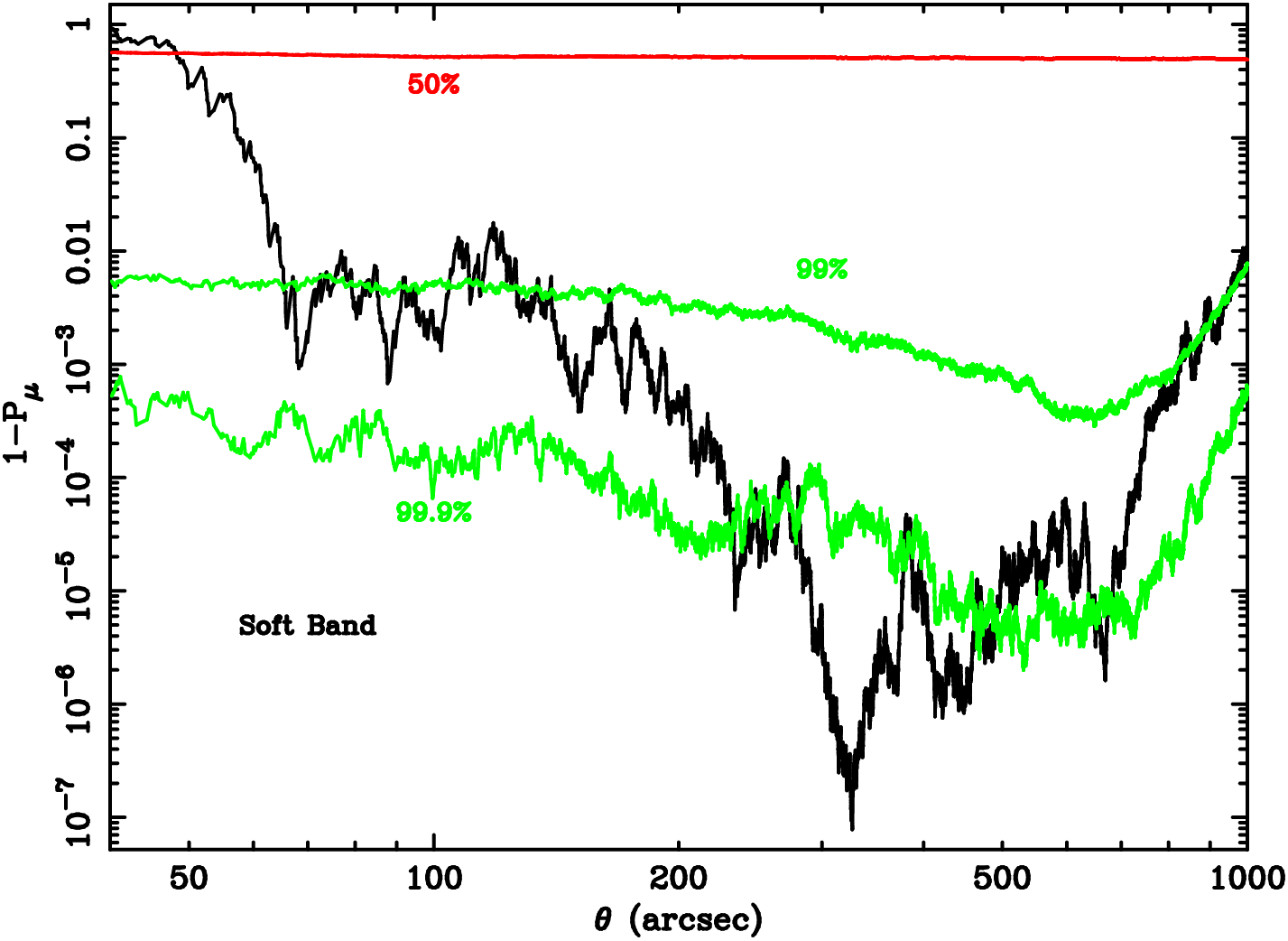}
\includegraphics[width=6.5cm,angle=270.0]{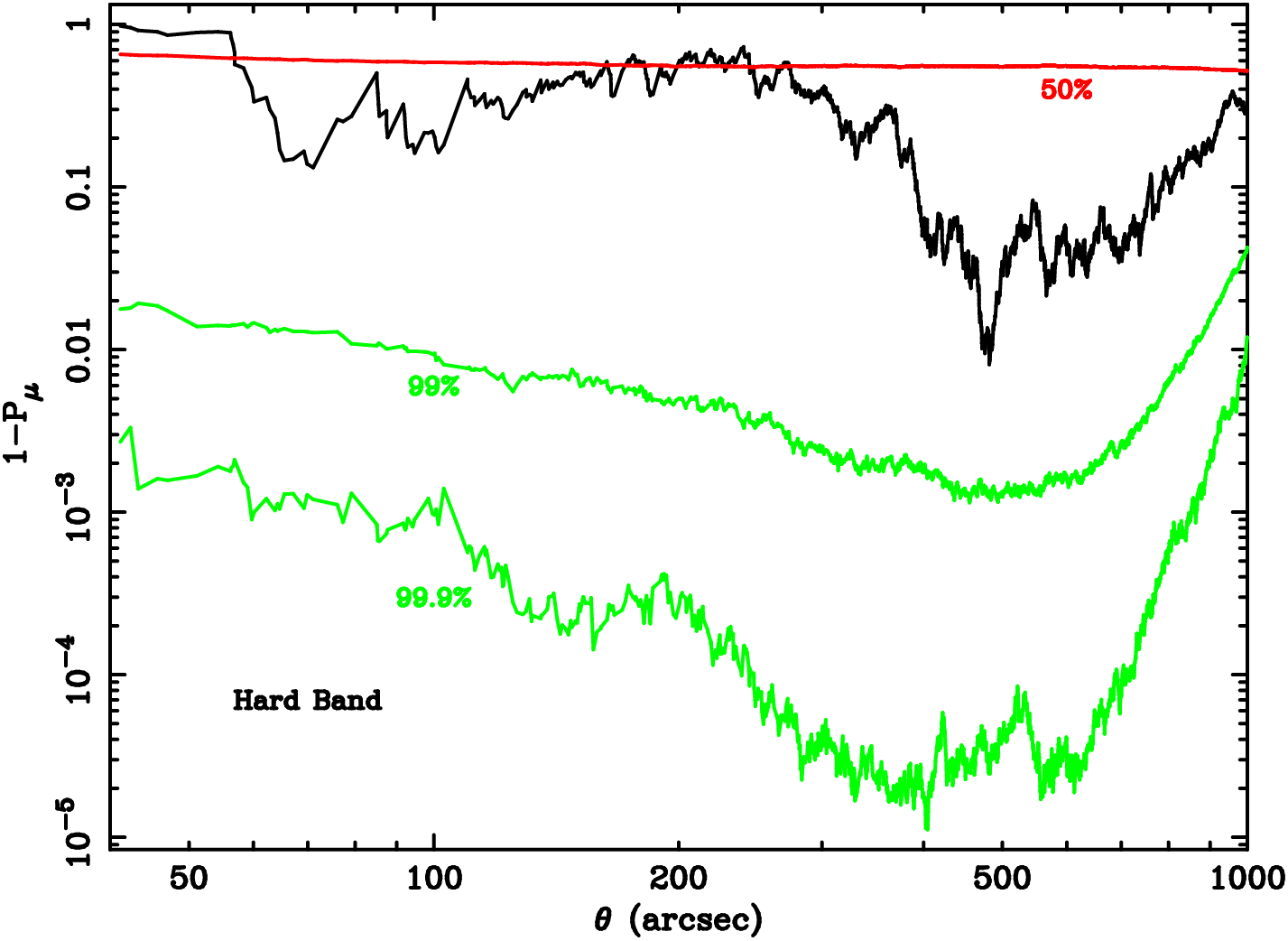}
}
\hbox{
\includegraphics[width=6.5cm,angle=270.0]{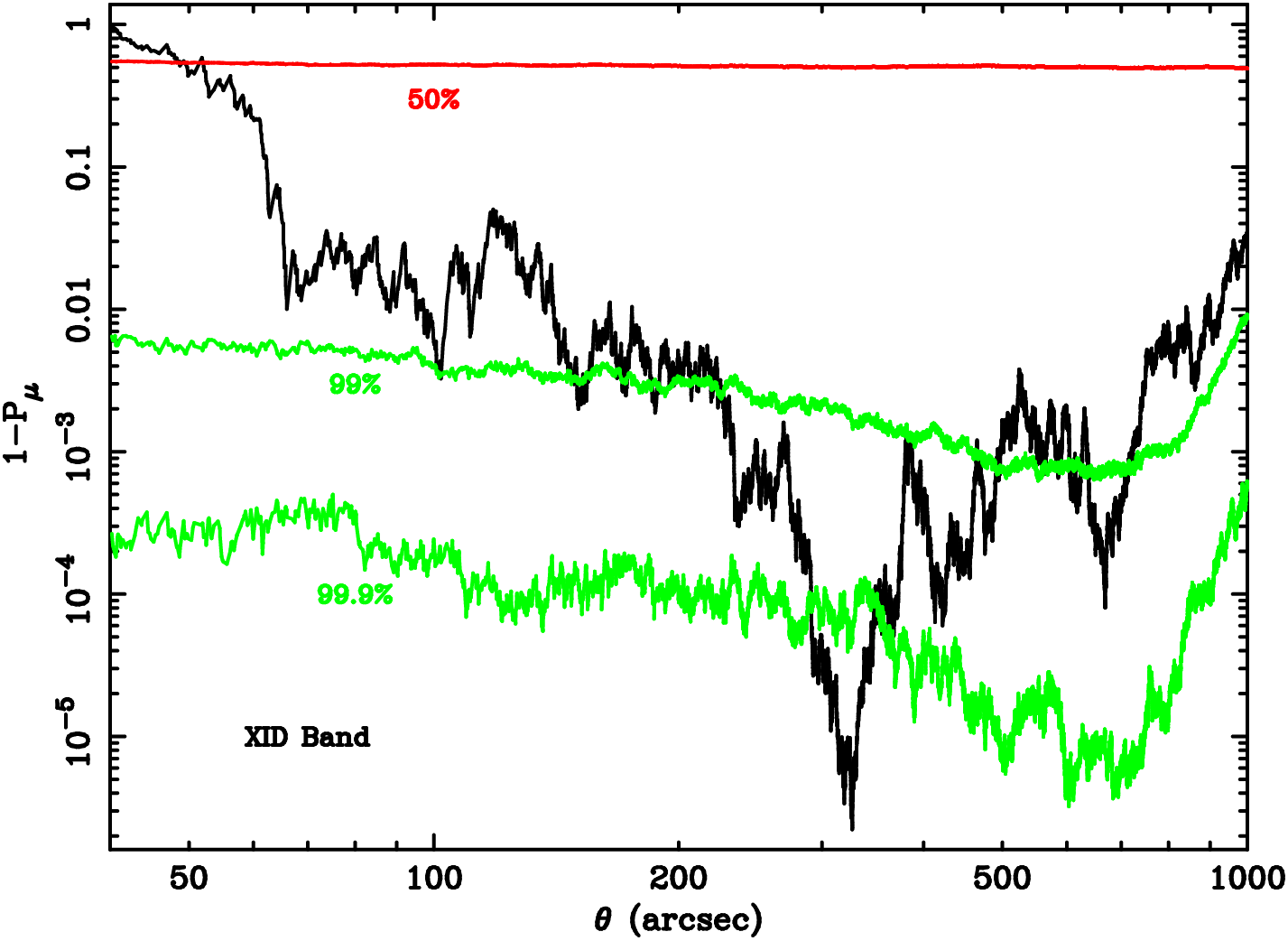}
\includegraphics[width=6.5cm,angle=270.0]{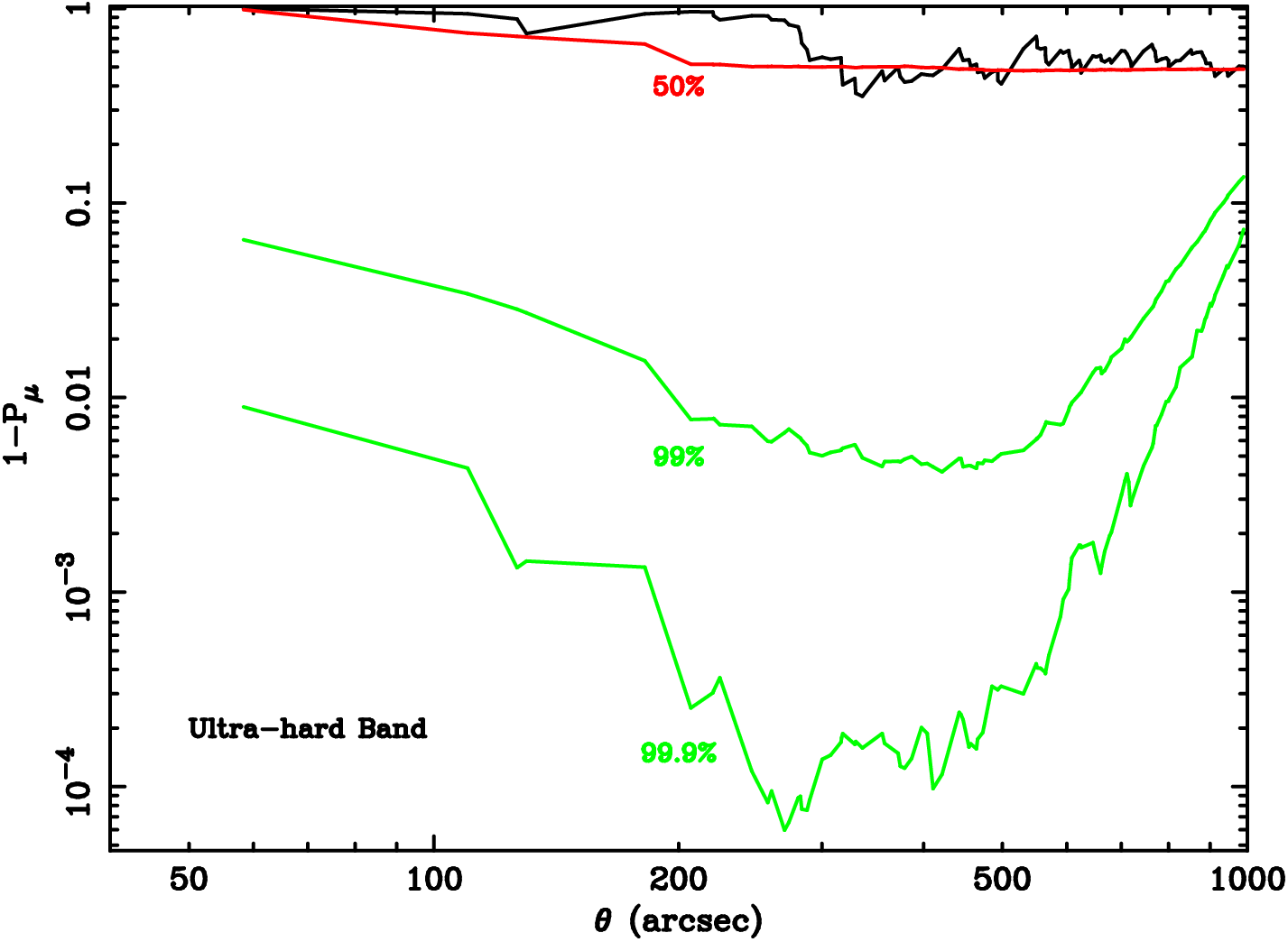}
}
\hbox{
\includegraphics[width=6.5cm,angle=270.0]{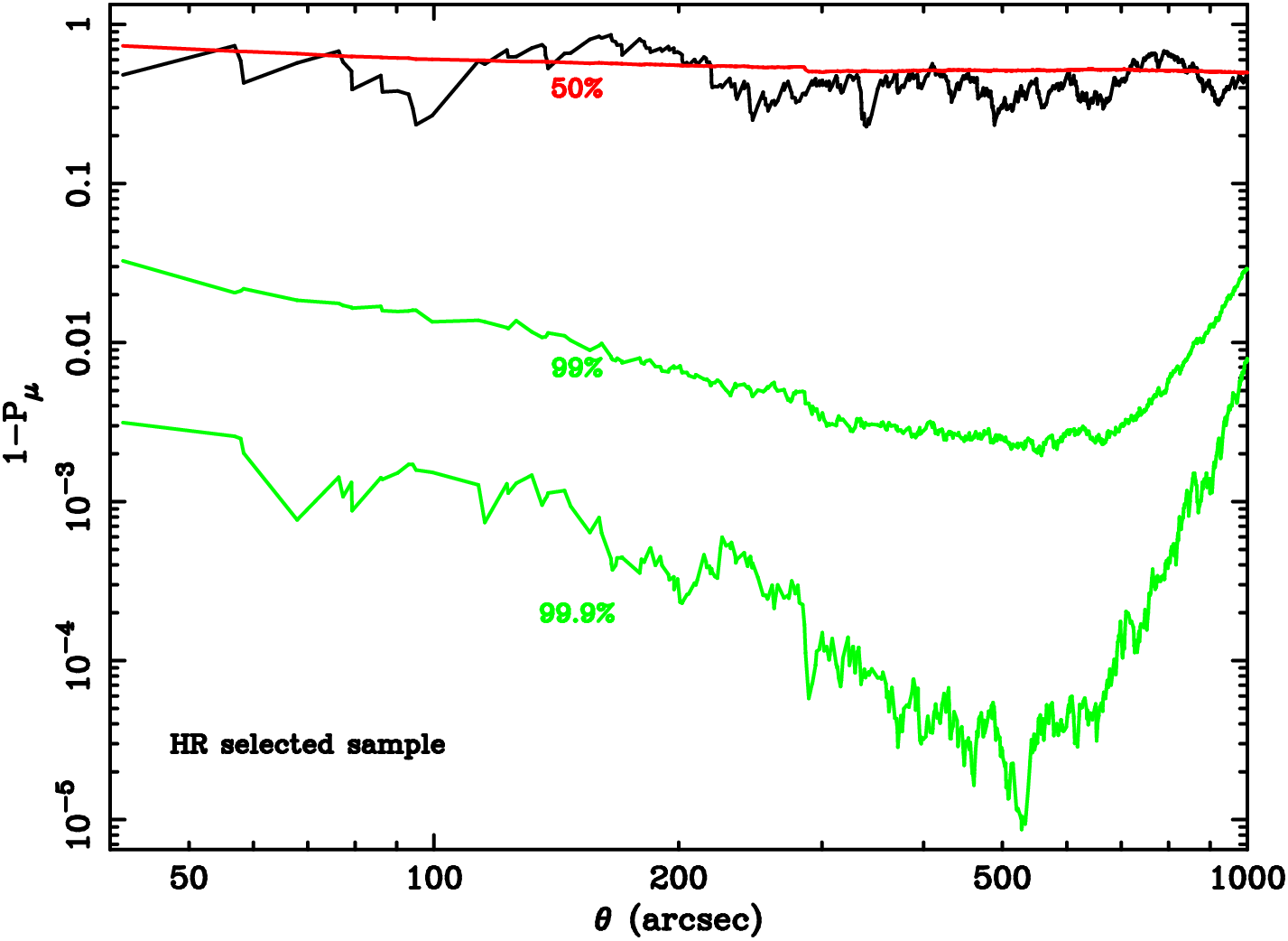}
\includegraphics[width=6.5cm,angle=270.0]{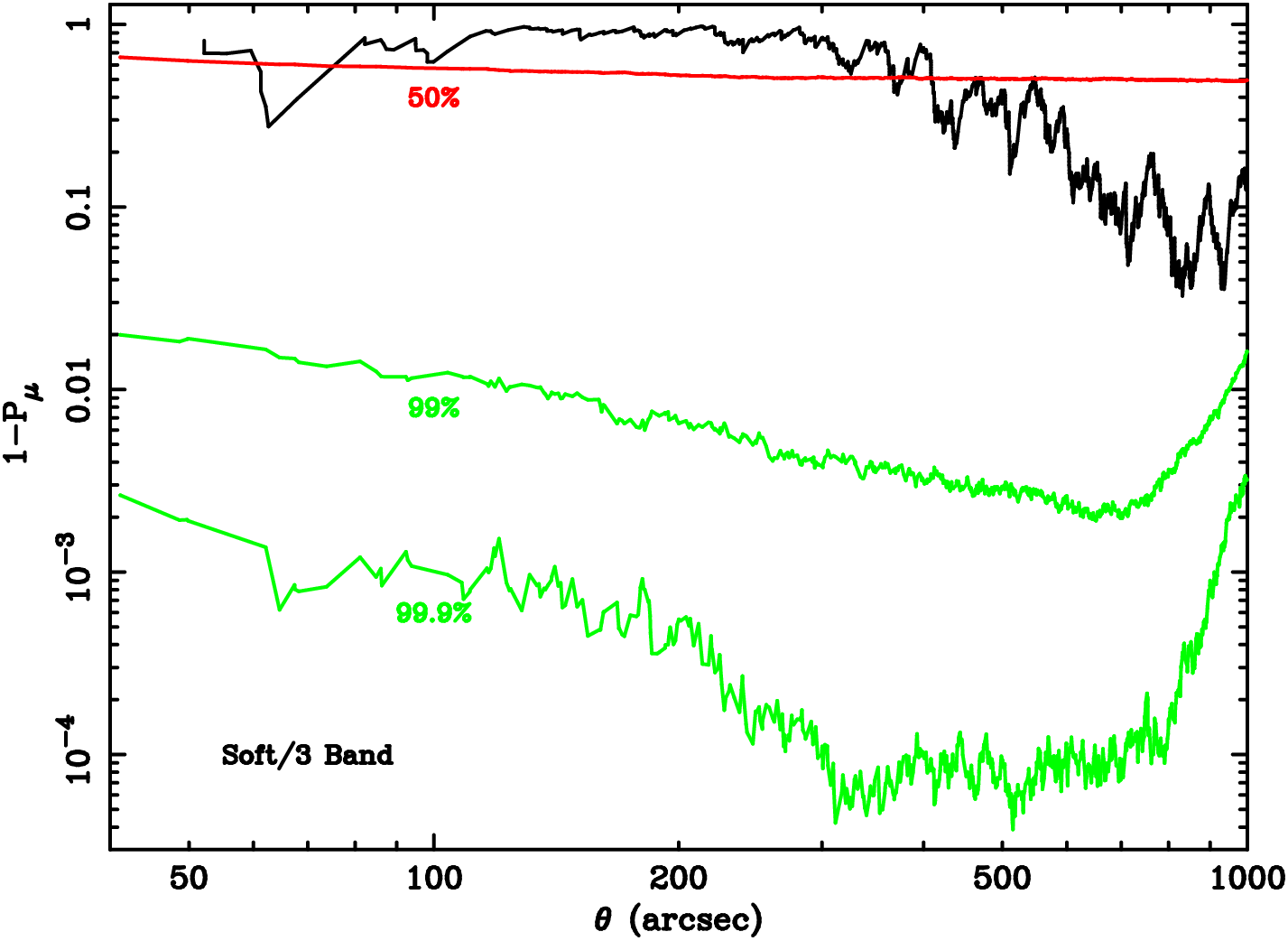}
}
\caption{
$1-P_\mu(N_\theta)$ vs. $\theta$ for the soft (top-left), hard
(top-right), XID (middle-left), ultra-hard (middle-right) selected
samples, for a hardness-ratio selected sample (bottom-left) and for a
random selection of one third of the soft sources (bottom-right). The
median (50\%), 99\% and 99.9\% levels from random simulations are also
shown (grey jagged lines). The Y-axis scaling is different in
different panels, and has been chosen for maximum clarity.}
\label{PMuThetaFig}
\end{figure*}

Correlation of hard band selected sources is not detected by Gandhi et
al. (\cite{Gandhi06}) (413 sources) and Puccetti et
al. (\cite{Puccetti06}) (205 sources). However, clustering is very
significantly detected ($>4\sigma$) by Basilakos et
al. (\cite{Basilakos04}) (171 sources) with $\gamma=1.2\pm 0.3$ and
$\theta_0=49^{+16}_{-25}$~arcsec. Clustering is also detected at
similar significance by Yang et al. (\cite{Yang03}), with
$\theta_0=40\pm11$~arcsec (for $\gamma\equiv0.8$) for 278 sources.
This is somewhat surprising since the Basilakos et al. sample has less
than half the number of sources than Gandhi et al., and a similar
number of sources to Puccetti et al., and the Yang et al. sample is
again smaller than the Gandhi et al. sample.  Our angular correlation
length is compatible with the Basilakos et al. (\cite{Basilakos04})
(and with the Puccetti et al.) result within 1-$\sigma$, but not with
the Yang et al. correlation length. However, the F-test on the
correlation function, and the superior $P_\lambda(N)$ test
(see below and Fig.~\ref{PMuThetaFig}) do not detect clustering.  Again, our sample
seems to give somewhat intermediate results in between those of
Basilakos et al. (\cite{Basilakos04}) and Yang et al. (\cite{Yang03}) on one
hand, and Gandhi et al. (\cite{Gandhi06}) and Puccetti et
al. (\cite{Puccetti06}) in the other.

\begin{figure*}
\hbox{
\includegraphics[width=6.5cm,angle=270.0]{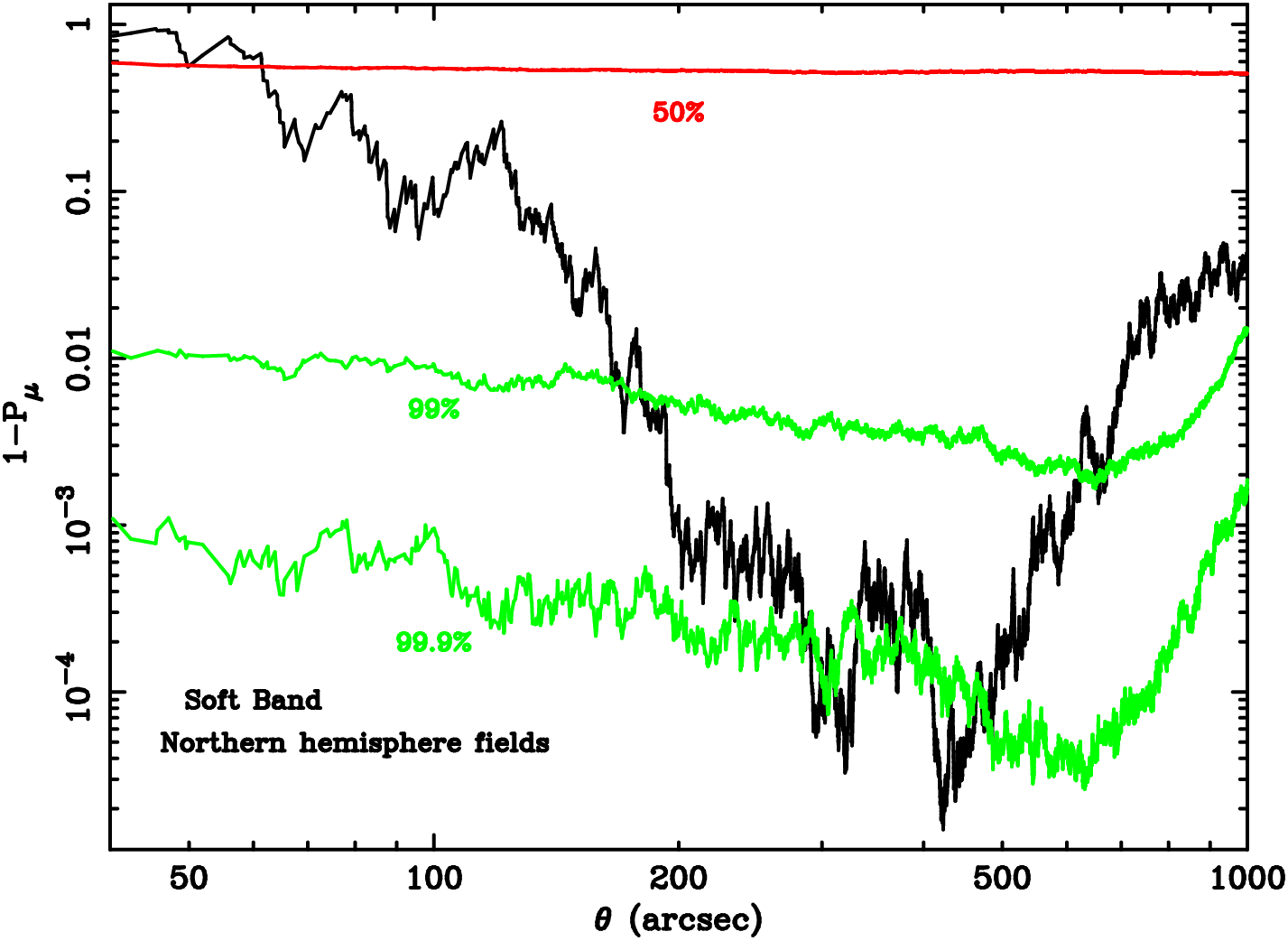}
\includegraphics[width=6.5cm,angle=270.0]{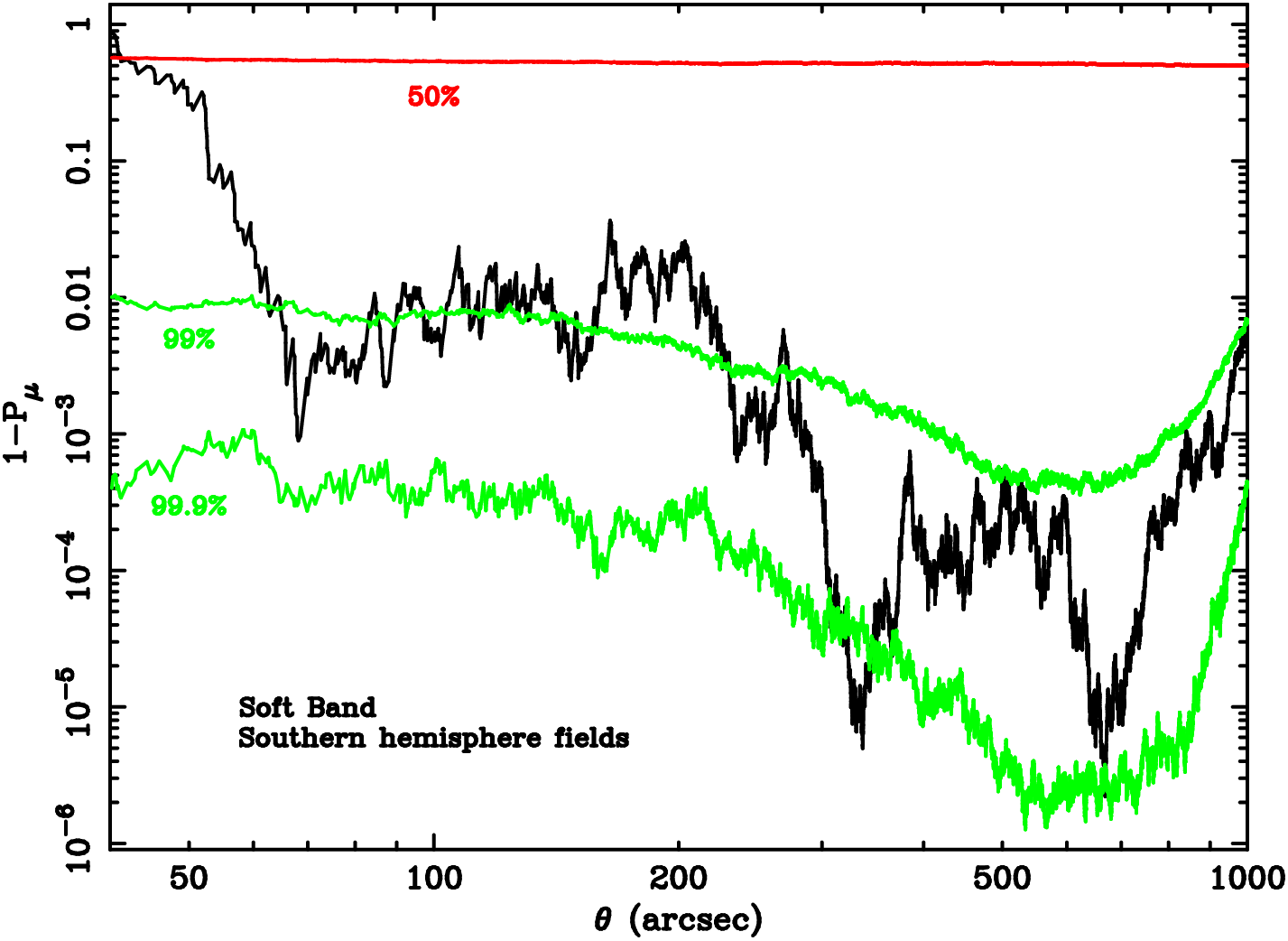}
}
\caption{
$1-P_\mu(N_\theta)$ vs. $\theta$ for the soft band and for the fields in
the northern (left) and southern (right) Galactic hemispheres. The
median (50\%), 99\% and 99.9\% levels from random simulations are also
shown (grey jagged lines).}
\label{PMuThetaFigSky}
\end{figure*}

Since each bin in $W(\theta)$ contains information from many pairs,
perhaps involving the same real and simulated sources many times, it
is at least debatable whether the hypothesis of independent bins
underlying the $\chi^2$ test applies in this case. To circumvent this
problem, we have applied a simple test for the significance of the
correlation,
using Poisson statistics to compare the expected number of pairs (at distances
smaller or equal than each individual observed distance) in absence of
correlation ($\mu=\sum_{\theta'<\theta} DR(\theta')/f$) with the number of
observed pairs ($N_\theta=\sum_{\theta'<\theta} DD(\theta')$).  The quantity
$1-P_\mu(N_\theta)$ versus the angular distance $\theta$ is shown in
Fig.~\ref{PMuThetaFig} for each band. We can estimate the ``signal'' in the
absence of correlation (similarly to what we have done to quantify the integral
constraint) repeating this exercise, but replacing the real data by randomly
placed sources (see above). For each sample we have performed 10000 simulations
along these lines. In this way, we have for each angular scale and sample the
distribution of the expected value of $1-P_\mu(N_\theta)$ if the sources are
not correlated.  We show in Fig.~\ref{PMuThetaFig} the median and the upper
99\% and 99.9\% percentiles of those distributions for each angular scale. The
simulations show clearly that the detection significance is actually lower than
the value given purely by Poisson statistics, when there are more than
$\sim$1000 sources, as in the soft and XID bands, where the 99\% percentile is
around the Poisson 3-$\sigma$ level. While in the hard and ultra-hard bands the
99\% percentile is around the ``right'' Poisson position, albeit with large
excursions.

Using the percentiles from the simulations, we have found a $>99$\%
(but mostly $<$99.9\%) significant detection of correlation in the
angular positions of in our soft sample at separations
200-700~arcsec. In the XID band the significance is lower, at just
about 99\% significance, with essentially one peak of higher
significance at $\sim$300~arcsec.  In contrast, no signal is found in
either the hard or ultra-hard bands.  The lack of significant
detections in the two harder bands might be due to the lower number of
sources with respect to the soft and XID bands.

%
%
%

We have performed a test for the influence of the number of sources in
the detection of clustering, since we only detect clustering in the
samples with more than 1000 sources. We have rejected at random two of
every three sources in the soft band sample, and repeated both tests
for correlation: the clustering signal disappears (see
Table~\ref{WThetaResults}, Figs. \ref{WThetaFig} and
\ref{PMuThetaFig}). We have also repeated this test keeping at
random one half and two thirds of the soft sources, only finding
correlation at the 99\% level in the last case. We conclude that
clustering in the hard and ultra-hard bands may be as strong or
stronger than clustering in the soft band, but our relatively low
number of sources prevents us from detecting it. This is compatible
with the results of Gilli et al. (\cite{Gilli05}), which do not find
any significant difference between the clustering properties of soft
or hard X-ray selected sources in the CDF.

The significant clustering found with the $P_\mu(N_\theta)$ of the soft
band corresponds to a broad bump between about 150 and 700~arcsec,
with ``peaks'' at about 300, 400 and 650~arcsec (the first two are
significant even at the 99.9\% level). These ``peaks'' also appear to
be present in the XID band. While the increase in the number of real
pairs giving rise to the broad bump is very gradual, the ``peaks''
must correspond to significant increases in relatively narrow ranges
of angular distances. To investigate if these ``peaks'' come from a
small number of fields in a particular region of the sky, we have
divided the sample between the fields
in the northern and southern Galactic
hemispheres.
The $1-P_\mu(N_\theta)$ for these regions are shown in
Fig. \ref{PMuThetaFigSky}. Despite the lower significances (expected
because of the lower numbers of sources), the broad bump and the peaks
are still present in the samples from the two hemispheres, the
$\sim$300~arcsec peak being common to both, with the one at
400~arcsec appearing predominantly in the northern hemisphere, and the
one at 650~arcsec in the southern hemisphere. We have also repeated
the analysis excluding the two fields with obvious cluster emission
(A1837 and A399) from the northern and southern Galactic hemisphere
sub-samples, respectively, finding no significant differences in the
strength or position of the ``peaks''.  As a further test, we have
looked for signal in the 7 deeper fields (Texp$>$40~ks, $\sim$400
sources) and in the remaining 27 fields ($\sim$800 sources), finding
it only in the latter, and hence the signal is not due to a few deep
fields.

A detailed discussion of the (cosmic or statistical) origin of these
structures is outside the scope of this paper, and will be discussed
with a larger sample elsewhere. However, whatever the origin of the
angular structure we have found, it must be relatively widespread over
the data and/or the sky. We take this to mean that, if it has a real
cosmic origin, it must come from sources at $z\leq 1.5$, which is the
peak of the redshift distribution of medium depth X-ray surveys
(Barcons et al. \cite{Barcons06b}).

%
%

\subsection{Angular correlation of a hardness-ratio-selected
sample of sources}

A low significance or absence of correlation in the hard band is
somewhat surprising, since most sources at the flux levels of our
sample are AGN (Barcons et al. \cite{Barcons06b}), which are known to
cluster strongly (both X-ray selected -Mullis et al. \cite{Mullis04},
Yang et al. \cite{Yang06}- and optically selected -Croom et
al. \cite{Croom05}-). Furthermore, obscured AGN appear more commonly
in hard X-ray selected samples (Della Ceca et al. \cite{DellaCeca06},
Caccianiga et al. \cite{HBS28}, Della Ceca et al. \cite{DellaCeca04},
Barcons et al. \cite{Barcons06b}).

Gandhi et al. (\cite{Gandhi06}) have found that, despite the absence of
significant angular correlation in their hard sample, if they select
only the hardest spectrum sources (using the soft-to-hard hardness
ratio), and they use lower significance sources, the significance of
the correlation increases to $\sim 2-3\sigma$. They lowered their significance
level in order to increase the number of sources in the tested
sample. This significant correlation only in the hardest of the
hard-X-ray-selected sources is a very intriguing and potentially
interesting result.

We have done a similar analysis, defining the same hardness ratio
$HR=(H-S)/(H+S)$ where $H$ and $S$ are the 2-12~keV and 0.5-2~keV
count rates, respectively. We have selected the sources detected in
either the soft or hard bands (in the sense explained in
Section~\ref{SensMap}) which have $HR\geq -0.2$ (Gandhi et al. \cite{Gandhi06}).

The F-test seems to indicate correlation at $>99$\% level (Table
\ref{WThetaResults}), but the superior Poisson method (Fig. \ref{WThetaFig})
does not detect clustering on any angular scales, with the data being
actually consistent with the median of the random simulations. We
cannot thus confirm the results of Gandhi et al. (\cite{Gandhi06}),
despite having a slightly larger sample of sources (250 versus their
209). This is compatible with the results of Yang et
al. (\cite{Yang06}) which do not find any significant differences
between the spatial clustering of sources with soft or hard spectra.


\section{Conclusions}
\label{SecConclusions}

We present the results from AXIS (An \XMM{} International Survey)
which comprises 1433 distinct serendipitous X-ray sources detected
with a likelihood of 15 ($\sim$5-$\sigma$) in \XMM{} EPIC pn
observations in 36 different \XMM{} observations at high Galactic
latitude. We have defined sub-samples in four bands: 0.5-2~keV(soft,
1267 sources), 2-10~keV (hard, 397 sources), 0.5-4.5~keV (XID, 1359
sources) and 4.5-7.5~keV (ultra-hard, 91 sources). The first two being
the ``standard'' X-ray bands, the third one chosen to span the energy
range with the best sensitivity of that camera, and the last one
taking advantage of the unprecedented hard X-ray sensitivity of
\XMM{} above 4~keV. Given the distribution of exposure times of the fields
selected for AXIS, it will serve as a pathfinder to the X-ray and
optical properties of the sources in large scale \XMM{}
catalogues (1XMM, 2XMM, $\ldots$).

Using count rates in the 5 standard \XMM{} bands as low-resolution
spectra we have fitted single power-laws to all sources in the XID and
hard bands independently, in order to calculate source fluxes without
assuming a constant spectrum (see Table~\ref{nXSou}). Our internal and
external checks make us confident on the results from this
technique. The average best fit power-law slope in the XID band is
$\langle\Gamma\rangle\sim 1.8$, independently of the sample to which
the sources belong, while the average is $\langle\Gamma\rangle\sim
1.5-1.6$ in the hard band.

We have constructed empirical sensitivity maps in the four bands, taking into
account the exposure maps, the background maps and the excluded regions (see
Table~\ref{fields}). We have derived sky coverages as a function of flux in the
four bands, taking into account the distribution of spectral slopes of the
sources. The total area covered is 4.8~deg$^2$ (see Fig~\ref{skyareafig}).

Our data have been combined with both wider and shallower (BSS/HBSS,
Della Ceca et al. \cite{DellaCeca04}, AMSS, Ueda et
al. \cite{Ueda05}), and narrower and deeper (CDF, Bauer et
al. \cite{Bauer04}) surveys to measure the X-ray source counts over as
wide flux range as possible, reaching 4 orders of magnitude in the
soft band ($\sim$2000 sources), $\sim$3 in the hard and XID bands
($\sim$800 and $\sim$1800 sources, respectively), and a bit more than
1 in the (largely unexplored) ultra-hard band ($\sim$140
sources). Data from different observatories and instruments are
mutually compatible, allowing the joint analysis of the source counts.

We have performed maximum likelihood fits to a broken power-law model for the
\ns{} in the four bands (see Table~\ref{MLNSfitTable}). 
We have found that fits to the source counts using fixed spectral slopes
produce similar results to those using individual source spectral slopes (as we
have done here), but give rise to larger error bars.  The source counts in the
soft, hard and XID bands show breaks at fluxes $\sim 10^{-14}$~cgs. Detailed
examination of the ratios between the data and that simple model in the soft
and XID band do not show any significant differences, while in the hard band
there seems to be evidence for a further flux break at $\sim
3\times10^{-15}$~cgs and several changes of slope. Since this is only present
in the hard band, it is difficult to assess whether this behaviour is due to a
too simple characterisation to contributions from different populations at
different redshifts, or it is due to calibration uncertainties. Future large
scale \XMM{} catalogues (such as 2XMM) will be useful to address this
question. The best fit model parameter values are compatible with previous
smaller or similar surveys, but our combination of large number of sources
and wide flux coverage produce in general smaller uncertainties in the best fit
parameters. 

%
%

We have used our best fit \ns{} parameters to calculate both the total
resolved fraction of the XRB (including the contribution from sources
a bright fluxes, see Table~\ref{IntensityTab}), and the relative
contribution of different flux bins to the XRB (see
Fig.~\ref{XRBContribFig}). We have used the estimates of the average
XRB intensity in the soft and hard bands from Moretti et
al. (\cite{Moretti03}), translating them to the XID and ultra-hard
bands using $\Gamma=1.4$ (Marshall et al. \cite{Marshall80}). The
total resolved fraction down to the lowest fluxes of the combined
sample reaches 87\% in the soft band and 85\% in the hard band, where
we have been able to reach deep fluxes using pencil beam surveys
(Bauer et al. \cite{Bauer04}), but it is only $\sim$60\% in the XID
band and about 25\% in the ultra-hard band. The total intensity
produced by extrapolating our best fit soft and hard \ns{} to zero
flux is insufficient to saturate the XRB intensity. Assuming a second
flux break just below our minimum detected fluxes, we have estimated
the minimum slope below this break necessary to produce the whole XRB
with discrete sources, getting 1.85 (1.84) in the soft (hard)
band. Comparing these slopes with those of the fainter galaxy and AGN
source counts from the CDF (Bauer et al. \cite{Bauer04}), reveal that
galaxies could easily provide this re-steepening, while among AGN only
perhaps the absorbed ones could just about do it.

The maximum fractional contribution to the XRB in the soft, hard and
XID bands comes from sources within a decade of $10^{-14}$~cgs
(which is about where the break flux for the broken power-law
lies). This fractional contribution reaches about 50\% of the total in
the soft and hard bands. Medium depth surveys such as AXIS (and indeed
the 1XMM catalogue, and its successor 2XMM) therefore are instrumental
in understanding the evolution of the X-ray emission in the Universe,
at least up to 10~keV.

Hickox \& Markevitch (\cite{HM06}) have recently re-estimated the soft and hard
XRB intensities using CDF data, finding lower resolved fractions of the XRB
than our (and previous) estimates. We have found that the difference with our
results in the soft band (where it is highest) is only at the $\sim$1-$\sigma$
level, and it is mainly due to
the different ways in which the total XRB intensity is calculated. 
HM06 add different contributions at different fluxes, which produces an
``effective'' spectral shape of the XRB ($\Gamma\sim 1.8$) which is much
steeper than the ``canonical'' XRB spectral slope of $\Gamma=1.4$ (which
we have assumed). We have shown that the difference between both estimates of
the soft XRB intensity cannot be removed just by relaxing our assumption.  The
HM06 intensity from resolved sources is very similar to ours. Converting our
results to 1-2~keV, the difference with HM06 all but vanishes.


After excluding two fields in which there is evidence for the presence
of stellar clusters, we have used the AXIS sources to study the
presence of cosmological structure in the X-ray sky, through the
cosmic variance in the number of sources per field, and the
distribution of the angular separations of the sources. The first one
is in principle more sensitive to the presence of significant
over-densities in a few fields, while the second looks for an overall
angular clustering of sources. No cosmic variance is detected at all
in the hard and ultra-hard bands, while some signal at about the 90\%
level is present in the soft and XID bands, probably because they are
the ones with the larger number of sources. Angular clustering is
studied in two ways. The first one is the angular correlation function
$W(\theta)$. Using this method, we detect signal at about the 99\%
level in the soft and XID bands, but not in the hard and ultra-hard
bands (Table~\ref{WThetaResults} and Fig.~\ref{WThetaFig}). The
strength of the clustering signal we have found is intermediate
between those of previous results (e.g. Gandhi et al. \cite{Gandhi06},
Basilakos et al. \cite{Basilakos04}, 2005, Vikhlinin \& Forman
\cite{Vikhlinin95}).

A formally more appropriate method using Poisson statistics detects
clustering at the 99-99.9\% level in the soft and XID samples,
but not in the hard or ultra-hard
bands (Fig.~\ref{PMuThetaFig}). Repeating the test for a randomly
1-in-3 selected sample of soft sources completely destroys the
clustering signal, which we take to imply that there might be a
``real'' angular clustering among hard sources which we have failed to
detect due to the small number of sources. Dividing the soft sample in
several sub-samples reveals that the signal is widespread over the sky,
and not limited to a few deep fields. This means that, if it has a
cosmic origin, it must come from $z\leq 1.5$, the peak of the redshift
distribution of medium flux X-ray surveys (Barcons et
al. \cite{Barcons06b}). We cannot confirm the detection of signal
among hard-spectrum hard sources reported by Gandhi et
al. (\cite{Gandhi06}).

\begin{acknowledgements}
This research has made use of the NASA/IPAC Extragalactic Database (NED; which
is operated by the Jet Propulsion Laboratory, California Institute of
Technology, under contract with the National Aeronautics and Space
Administration) and of the SIMBAD database (operated by CDS, Strasbourg,
France).  Financial support for this work was provided by the Spanish
Ministerio de Educaci\'on y Ciencia under project ESP2003-00812. This research
has made use of the USNO-A2.0 maintained at the U.S. Naval Observatory.
RDC and TM received partial financial support from the Italian Space
Agency under contract ASI-INAF n. I/023/05/0.
\end{acknowledgements}

\appendix

\section{Empirical sensitivity maps}
\label{AppSensMap}

The present version of {\tt esensmap} (the SAS task to calculate
sensitivity maps for the EPIC cameras on board \XMM) assumes pure
Poisson statistics to evaluate the detection sensitivity at different
positions in the field of view. Since the detection of the sources and the
determination of their parameters is a complicated process, which
includes how well does the profile of the count rates match the PSF,
the result is bound to deviate from the na\"ive assumption of pure
Poisson statistics.

We have followed an empirical approach, looking for a simple relationship
between the observed EPIC pn count rates of the detected sources in each band
$cr$ (in cts/s, columns {\tt RATE} in the SAS source lists), and the pure
Poisson count rate $crpoisim$ (in cts/s) as calculated from the total
number of background counts $bgdim$ within a circle of radius $cutrad$ (related
to the column labelled {\tt CUTRAD} in the SAS source lists,  and given in
units of 4$''$ pixels) around the source position, the detection likelihood of
the source $L$ (columns labelled {\tt DET\_ML} in the SAS source lists), and the
average value of the exposure map  (in seconds) within that radius
$expim$, so that

\begin{equation}
-log(P_{bgdim}(>(bgdim+crpoisim\times expim))=L
\end{equation}

\noindent with $P_\lambda(>N)$ as defined in Eq.~\ref{PgtN}.

The values of {\tt CUTRAD} are in principle different for each source
in each field, since they are chosen by {\tt emldetect} to maximise
the signal-to-noise ratio for source detection. We have found that
there is no significant correlation between {\tt CUTRAD} and either
the off-axis angle of the source, {\tt RATE} or {\tt BG\_MAP} (the
average background value in counts per pixel within the extraction
circle), and therefore we have adopted for $cutrad$ in each band the
average of {\tt CUTRAD} for the sources having $L$ between 8 and 20 in
that band. We have checked that there is no significant difference in
the goodness of the fits below between using the actual {\tt CUTRAD}
value for each source or the average $cutrad$.

$expim$ above is the average of the exposure map within a circle of
radius $cutrad$ around the X-ray source positions.  The exposure maps
for the single bands are PPS products. We have created exposure maps
for the composite bands using the SAS task {\tt eexpmap}, which
generates exposure maps for the energy in the middle of the PI
interval, which could be inaccurate for wide bands. To assess the
effect of this approximation we have also calculated ``average''
exposure maps in the composite bands, weighting the single band
exposure maps with the counts in the images pixel by pixel. Again,
there are no significant differences between the results from those
two types of exposure maps.

Unfortunately, the background files are not PPS products. We have
generated them from the individual and composite bands using the SAS
task {\tt esplinemap}, which excludes areas around the sources in an
input source list, and fits the remaining background using a spline
(with 16 nodes in our case). We have then calculated $bgdim$ by
summing the values of the background image over a circle of radius
$cutrad$.

\begin{figure*}
\hbox{
\includegraphics[width=6cm,angle=270.0]{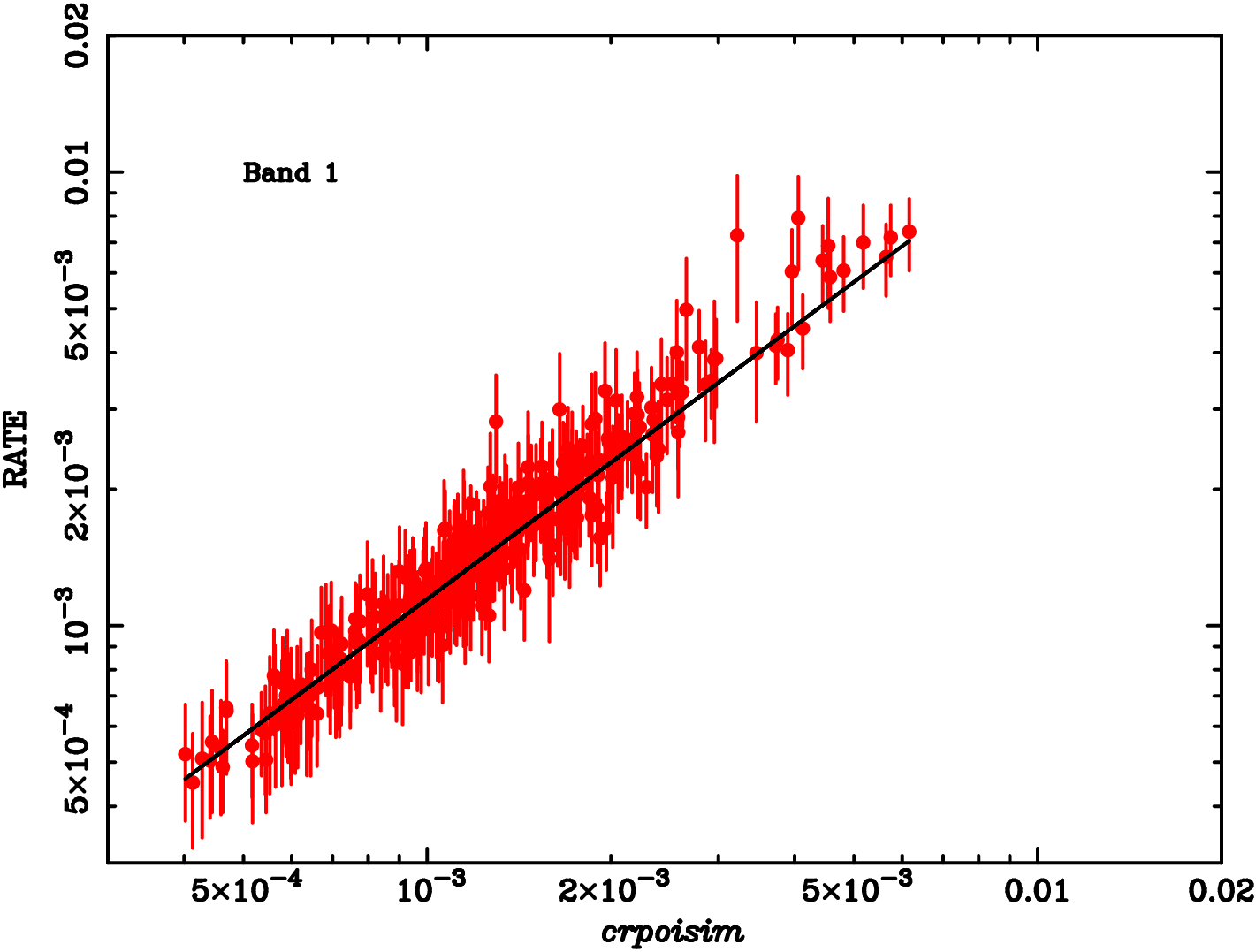}
\includegraphics[width=6cm,angle=270.0]{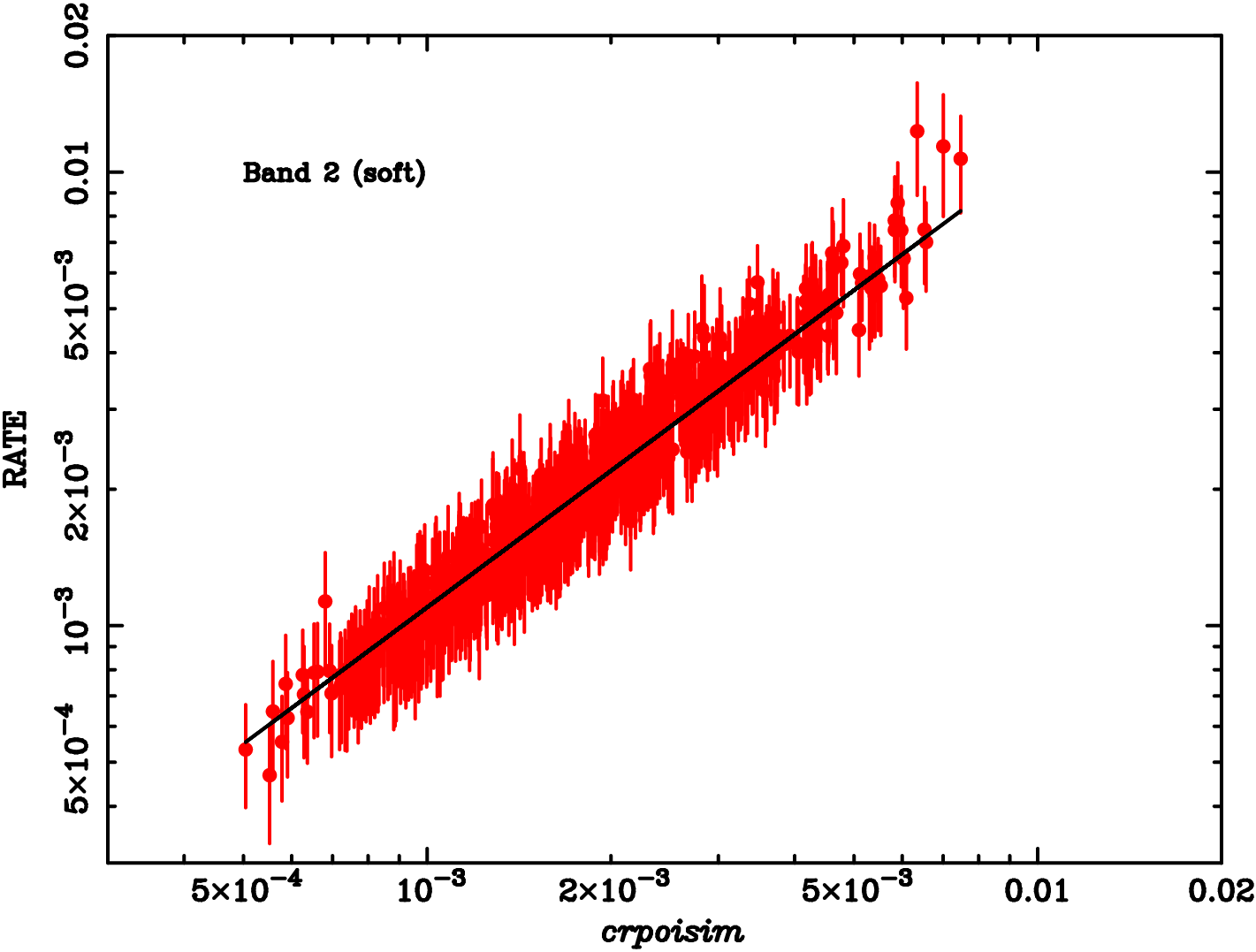}
}
\hbox{
\includegraphics[width=6cm,angle=270.0]{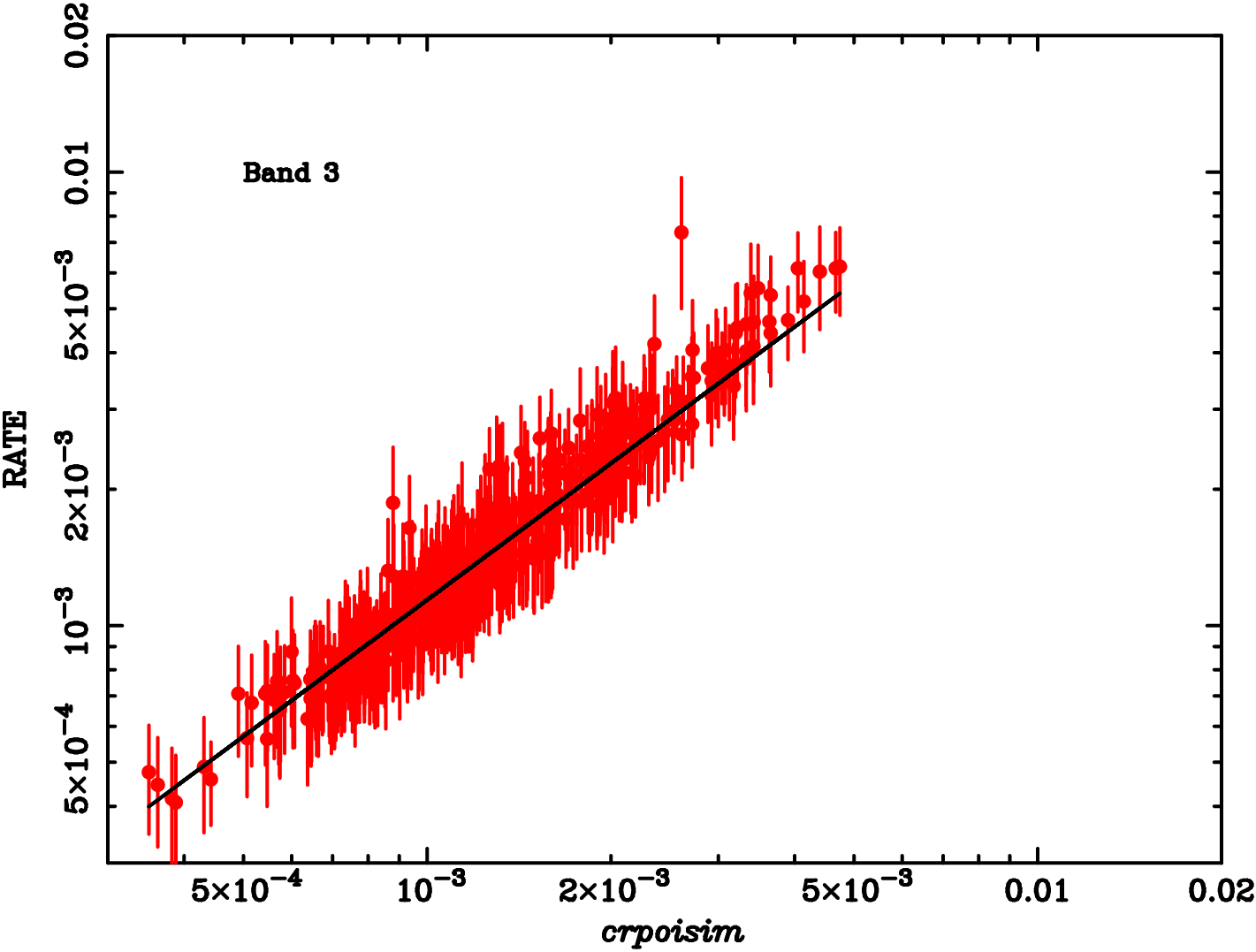}
\includegraphics[width=6cm,angle=270.0]{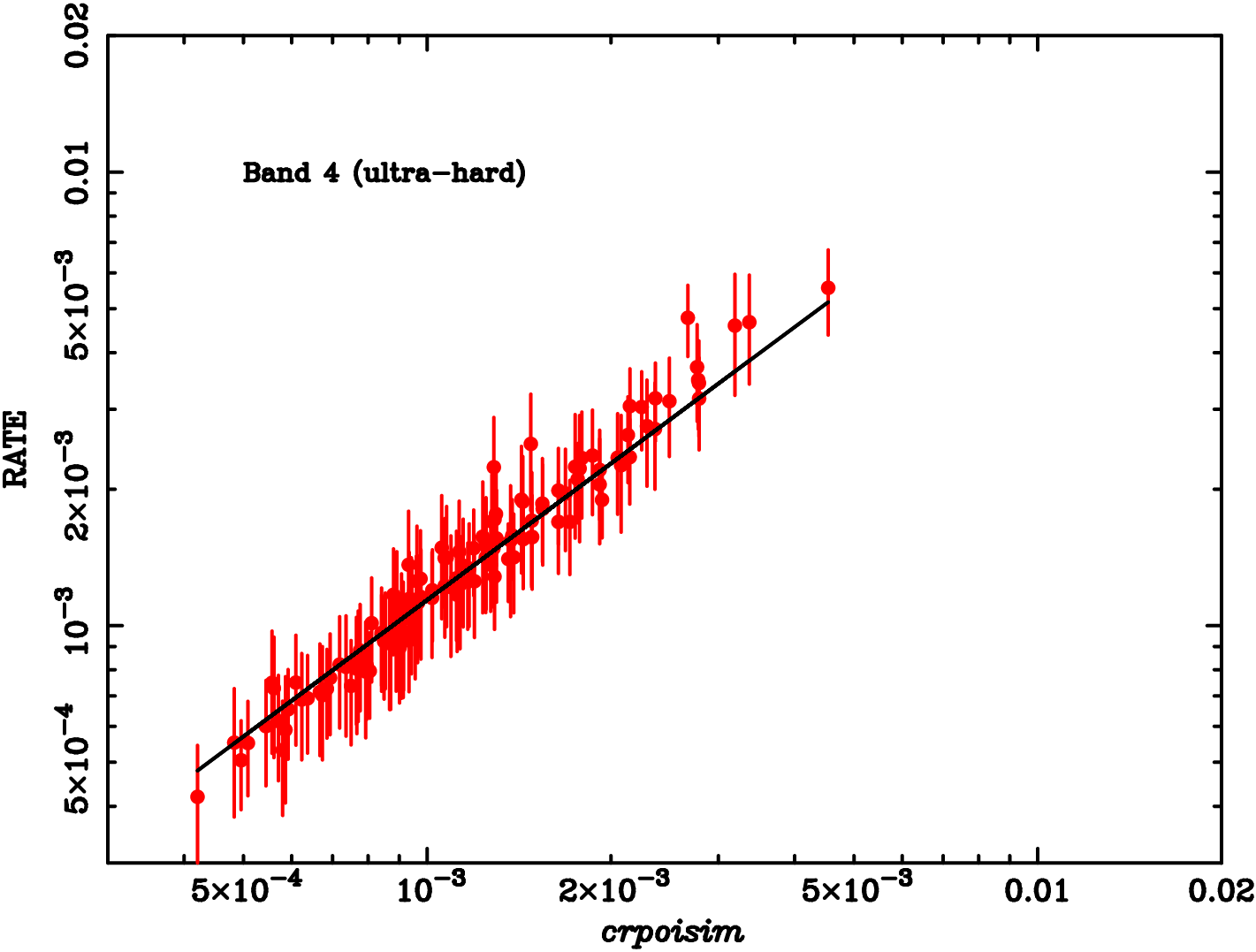}
}
\hbox{
\includegraphics[width=6cm,angle=270.0]{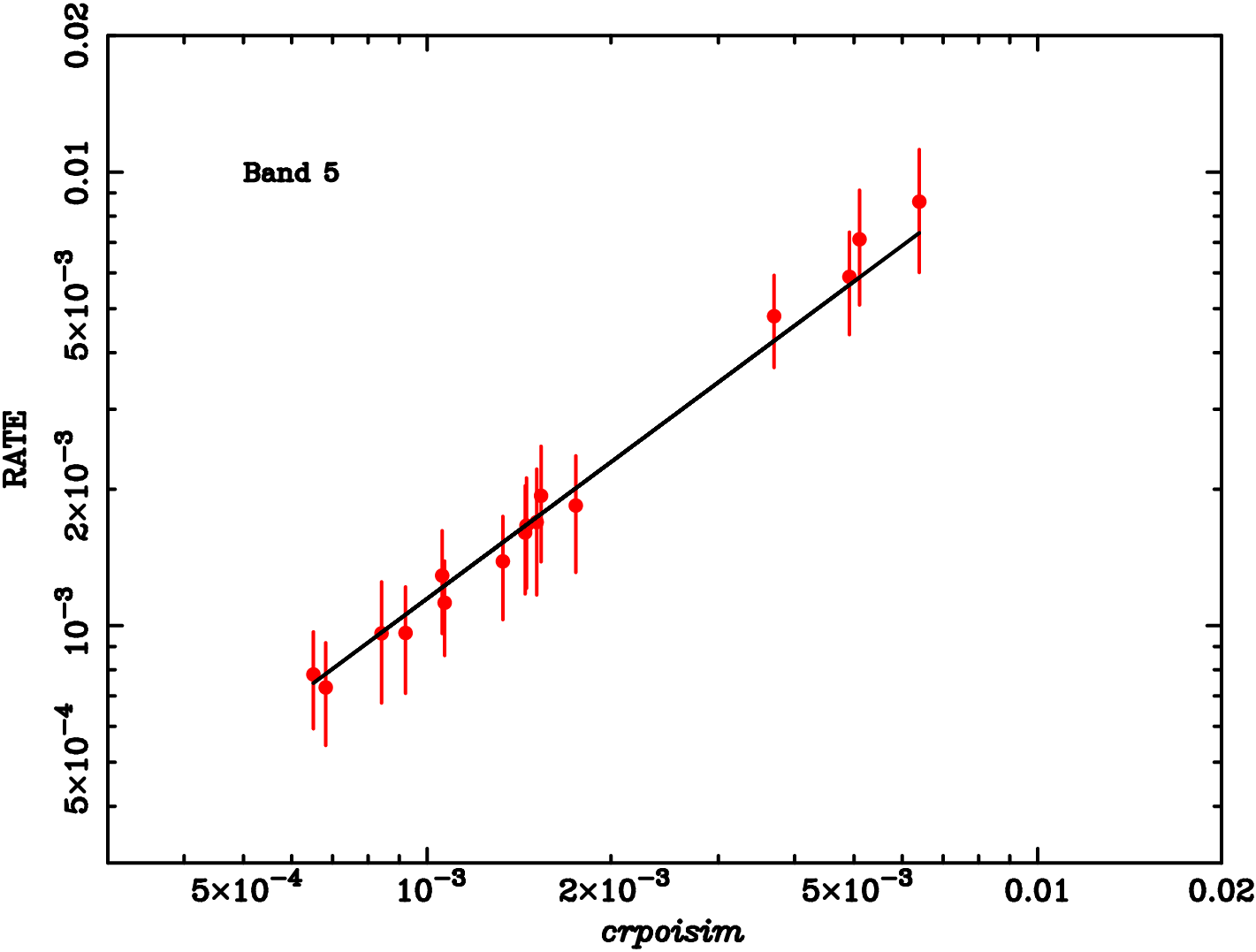}
\includegraphics[width=6cm,angle=270.0]{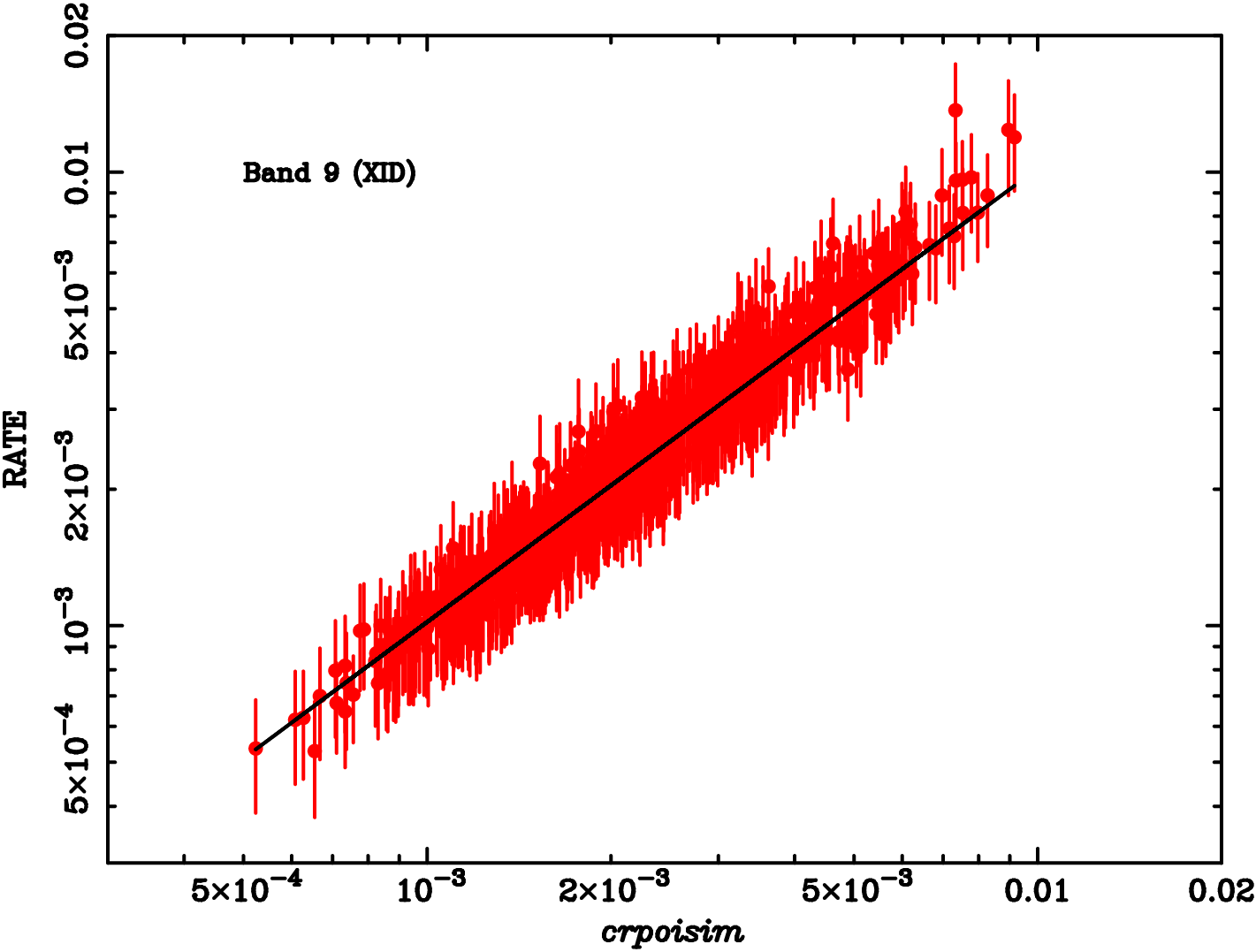}
}
\centering
\includegraphics[width=6cm,angle=270.0]{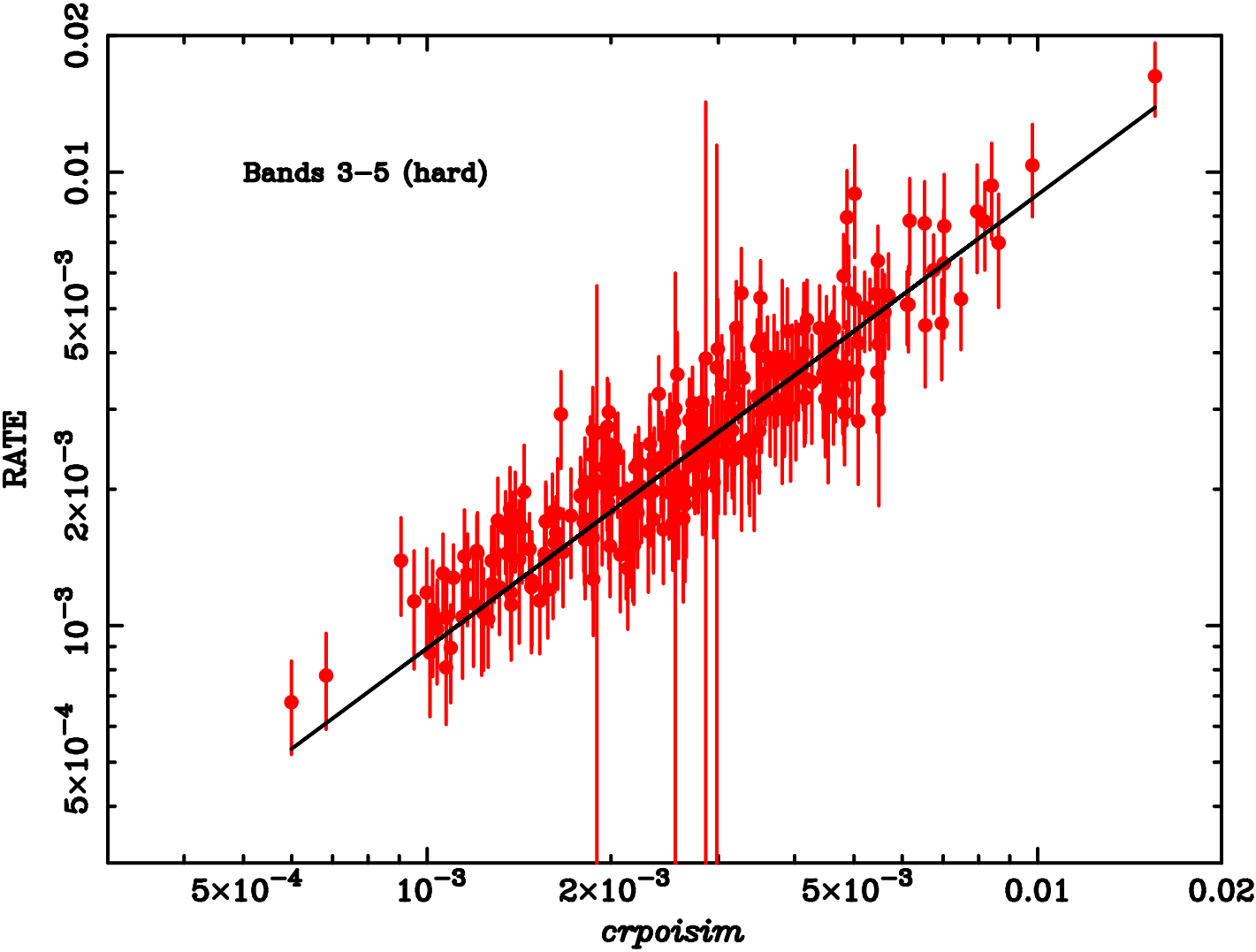}

\caption{{\tt RATE} vs. $crpoisim$ for all bands. Also shown are the best
linear fits.}
\label{crcrpoisFig}
\end{figure*}

We show in Fig.~\ref{crcrpoisFig} plots of {\tt RATE} versus $crpoisim$ for the
single and composite bands, along with the best linear fit model ({\tt
RATE}$=LI\times crpoisim$), for sources having $8\leq L \leq 20$ in each
band. The best fit $LI$ along with the values of $cutrad$ for each band are
given in Table~\ref{crcrpoisTable}. From the number of sources in each fit and
the values of the $\chi^2$ it is clear that the fits are quite good, and more
sophisticated models are not needed. The values of $LI$ are all within
$\sim$10\% of 1, implying that the correction is small (as expected), but from
the improvement of the $\chi^2$ values from using $LI\equiv 1$ (which is
equivalent to {\tt RATE}$\equiv crpoisim$), we conclude that it is also highly
significant.

The recipe for creating a sensitivity map is then:

\begin{enumerate}
\item Create background and exposure maps
\item Chose a likelihood value $L$ for the significance of the
detections. This recipe has only been tested for $8\leq L\leq20$, but it could be in principle valid for likelihood values outside this interval.
\item Choose a source extraction radius $cutrad$ appropriate for the
band you are interested in (Table~\ref{crcrpoisTable})
\item For each pixel $(X,Y)$ in your input image:
   \begin{enumerate}
   \item Calculate the sum of the values of the pixels in the
   background map whose centres are within $cutrad$ of $(X,Y)$: $bgdim$
   \item Calculate the average of the values of the pixels in the exposure
   map whose centres are within $cutrad$ of $(X,Y)$: $expim$
   \item Find $ctspoisim$ such that $-\log(P_{bgdim}(>(ctspoisim+bgdim)))=L$
   \item Calculate $sens(X,Y)=ctspoisim/expim \times$LI.
   \end{enumerate}
\item $sens(X,Y)$ is our empirical estimate of the typical count rate
of a source at $(X,Y)$ detected with likelihood $L$
\end{enumerate}

\begin{table}
\centering
\caption{Summary of the results of the linear fits of $crpoisim$ to
{\tt RATE}: Band is the band used for the fit, $cutrad$ is the source
extraction radius (in units of 4$''$ pixels), $LI$ is the best fit
multiplying constant, $N$ is the number of sources in the fit,
$\chi^2_{LI}$ is the $\chi^2$ value of the best linear fit, while
$\chi^2_0$ is the $\chi^2$ of the fit to $LI\equiv 1$, and $P(F)$ the
F-test probability of letting $LI\neq 1$ not being a
significant improvement in the fit}
\label{crcrpoisTable}
\begin{tabular}{lccrrrc}
\\
\hline
Band           & $cutrad$ & $LI$ &  $N$  &$\chi^2_{LI}$&$\chi^2_0$ & $P(F)$ \\
               & (pixels) &    \\
\hline
1              &  5.12    & 1.14 & 284   & 104.5       & 185.5     & $<10^{-6}$\\
2 (soft)       &  5.08    & 1.10 & 653   & 140.1       & 226.0     & $<10^{-6}$\\
3              &  5.15    & 1.14 & 414   & 105.7       & 208.4     & $<10^{-6}$\\
4 (ultra-hard) &  5.49    & 1.14 & 127   &  25.7       &  56.2     & $<10^{-6}$\\
5              &  5.86    & 1.15 &  16   &   1.7       &   5.5     & $<10^{-4}$\\
9 (XID)        &  5.04    & 1.02 & 722   & 141.0       & 145.2     & $<10^{-5}$\\
\hline		          	         	       		   
3-5 (hard)     &  5.18    & 0.89 & 243   & 176.2       & 237.8     & $<10^{-6}$\\
\hline
\end{tabular}
\end{table}

\section{Evaluating the count rate spectral fit}
\label{CheckSpec}

To check the validity of our procedure we have performed both internal and
external tests. The simplest internal check is to simulate sources with a
single spectral slope ($\Gamma=2$ in our case), fit them, and compare the
output slope values to the input value. We have kept the total
(0.5-10~keV) count rates of the sources fixed to the observed count rates, and
simulated the count rates in each band using a Poisson deviate from the expected
number of counts for the fixed slope and the exposure times $T_{\rm exp}$ in
Table \ref{fields}. The error on the simulated count rates ($\Delta CR$) were a bit more
delicate to estimate. Using simply $\Delta CR=\sqrt{CR/T_{\rm exp}}$ would not
be adequate, because it does not include the contribution from the background
subtraction, and because the statistics of the source detection by {\tt
emldetect} are very close to, but not exactly, Poissonian. We have used the
selected sources to look for an empirical relationship between the observed
$\Delta CR$ and the naive $\sqrt{CR/T_{\rm exp}}$. We have found that a linear
relationship is a good visual fit, with the parameters given in Table
\ref{ecrobsecrtheo}. This is the recipe we have used to simulate the errors in
the count rates in our simulations. The weighted averages of the fitted slopes
are given in Table \ref{nXSouSim}, showing that any internal deviations are
small ($<0.1$), and that there are no obvious biases for sources selected in
different bands. We have repeated the simulations for $\Gamma\equiv1.8$
and $\Gamma\equiv1.7$, with similar results and conclusions.

\begin{table}
   \caption[]{Empirical relationship between the observed errors on the
   count rates $\Delta CR$ and the naive $\sqrt{CR/T_{\rm exp}}$. We give the
   parameters $a$ and $b$ of a linear relation $\Delta CR =
   a\times\sqrt{CR/T_{\rm exp}}+b$} \label{ecrobsecrtheo}

   \begin{tabular}{ccc}
       \hline
       \noalign{\smallskip}
band & $a$& $b$ \\
       \noalign{\smallskip}
       \hline
      \noalign{\smallskip}
2    & 1.597 & - \\
3    & 1.419 & $3.824\times10^{-5}$ \\
4    & 1.452 & $6.166\times10^{-5}$ \\
5    & 2.350 & $1.184\times10^{-4}$ \\
       \noalign{\smallskip}
       \hline
   \end{tabular}
\end{table}


%
%
%

\begin{figure}
   \centering
   \includegraphics[width=6.5cm,angle=270.0]{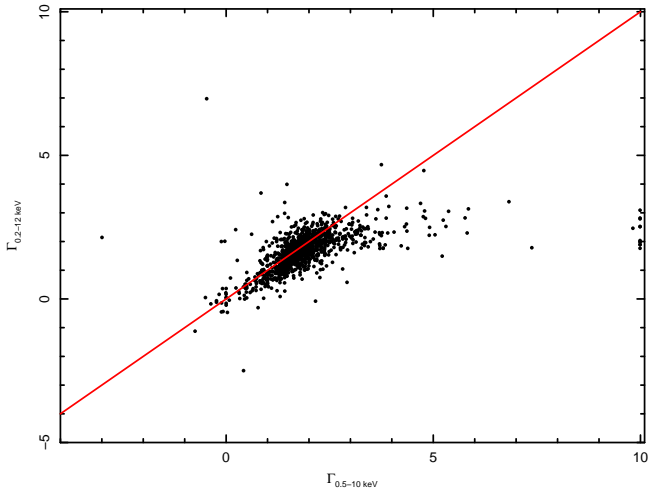}
      \caption{Full band power-law slope from Mateos et al. (2005) vs. our 0.5-10~keV slope for
               the common accepted sources. The solid line shows the 1:1 relation.
              }
         \label{GmateosGus}
   \end{figure}

\begin{figure}
   \centering
   \includegraphics[width=6.5cm,angle=270.0]{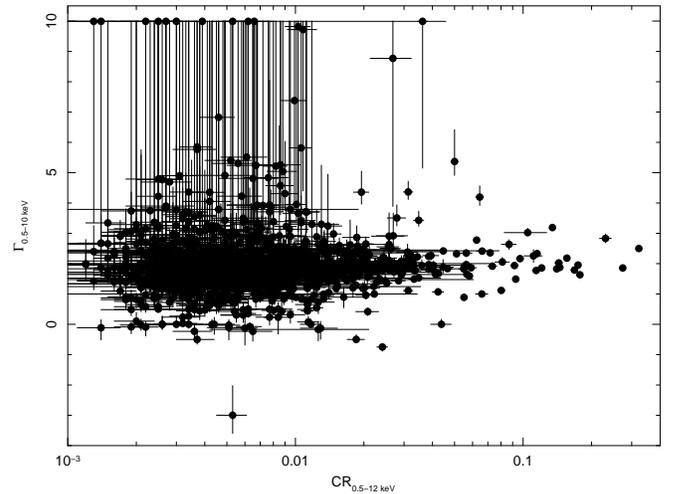}
      \caption{Our 0.5-10~keV slope versus the 0.5 to 12 count rate for our 1433
sources.}
         \label{CRtotGtot}
   \end{figure}

We have also performed an external check against the single power law fits to
the ``properly'' extracted spectra by Mateos et al. (2005). In that work the
fits were done using {\tt xspec} with response matrices and effective areas
generated for each source with the corresponding SAS tasks. Since the Mateos et
al. (2005) spectra were extracted between 0.2 and $\sim$12~keV, we have fitted
our sources using bands 2, 3, 4, and 5. In Fig. \ref{GmateosGus} we show their
best fit single power law slopes vs. our best fit slopes for the common
accepted sources (a total of 1143 sources, which include some sources fitted by
Mateos et al. but not included in their paper, Mateos private
communication). The best $\chi^2$ fit proportionality constant using errors in
both the X and Y axis is 0.980$\pm$0.003. This result makes us confident in our
fitting method, showing that we haven't missed significant effects that were
not already corrected in the {\tt emldetect} count rates.  A number of sources
deviate appreciably from the 1:1 relation in Fig.~\ref{GmateosGus}, the effect
being more noticeable to the eye for values of $\Gamma_{\rm 0.2-12\,keV}$
between about 1.7 and 3. Within that interval, the number of sources with
$\Gamma_{\rm 0.5-10\,keV}>3$ is only 14\% of the total number of sources in
that interval, this simple test already indicates that the number of sources
with pathological spectral fits is a small minority. We will discuss the
influence of these sources in more detail below. We show in
Fig.~\ref{CRtotGtot} our best fit slopes versus our 0.5-12~keV
count rates. There are some sources with very steep fitted slopes (because they
have only been significantly detected in one band), but the bulk of our sources
have reasonably ``standard'' spectral slopes, with no significant trends
towards too steep or too faint slopes at lower count rates.

Individually the fits are therefore good, but we have also assessed
whether there are significant biases in the average slope, and whether
the error bars on the slopes are adequate. We have first studied the
distribution of the differences between the slopes divided by the
quadratic sum of the errors (which would be expected to follow a
$N(0,1)$ distribution in the Gaussian case). There is a small bias in
these ``normalised'' differences of $\sim$20\% of the combined error
bar (such us we tend to find steeper slopes than Mateos et al.), with a
significant dispersion above the expected value. Fitting a Gaussian to
this distribution (which is not a Gaussian) finds a similar central
value, but a much smaller dispersion, indicating that much of the
excess variance is due to a limited number of significant outliers.

Assuming Gaussian statistics, the error on the mean can be obtained
dividing the standard deviation by the square root of the number of
sources ($\sim$1000 in our case). If we perform this calculation for
our normalised differences, the value is much smaller than the
$\sim$20\% relative bias above, which means that the bias is significant (many
times the ``error on the mean''), despite its low relative value.

Studying the distribution of the absolute differences between the
slopes (without dividing by the combined error bar), again there is a
bias towards softer slopes in our sample of $\Delta\Gamma\sim$0.12,
with a very significant dispersion, which again is much reduced in the
Gaussian fit. Our conclusion is similar to above.

If we compare our weighted average slopes with Mateos et al, there is also a
difference, but in the sense of our slopes being flatter than theirs. Since the
weighted average essentially discards the sources with large error bars,
this means that some of the outliers discussed above are soft sources with
large error bars which bias the arithmetic average towards softer slopes.

\begin{table}
   \caption[]{ Number of sources selected in different bands $N$,
   weighted average slopes $\langle \Gamma\rangle$ and errors (taking
   into account both the error bars in the individual $\Gamma$ and the
   dispersion around the mean), and number of sources used in the
   average $N_{\rm ave}$,  for the
   simulated sources (see Section \ref{SecXSp}). The soft, hard, XID and
   ultra-hard bands are as defined in the text. ``Soft and hard''
   refers to sources selected simultaneously in the soft and hard
   bands, ``Only soft'' refers to sources selected in the soft band
   but not in the hard band, and ``Only hard'' refers to sources
   selected in the hard band and not in the soft band} \label{nXSouSim}

   \begin{tabular}{lrccccc}
       \hline
       \noalign{\smallskip}
    &    & \multicolumn{2}{c}{0.5-2~keV} & & \multicolumn{2}{c}{2-10~keV} \\
\cline{3-4}\cline{6-7}
Selection & $N$ & $N_{\rm ave}$ & $\langle \Gamma\rangle$ &&
                  $N_{\rm ave}$ & $\langle \Gamma\rangle$ \\
       \noalign{\smallskip}
       \hline
      \noalign{\smallskip}
Soft          & 1267 & 1267 & 2.008$\pm$0.004 && 1259 & 2.029$\pm$0.007 \\
Hard          &  397 &  397 & 2.008$\pm$0.005 &&  396 & 2.039$\pm$0.010 \\
XID           & 1359 & 1359 & 2.007$\pm$0.004 && 1352 & 2.030$\pm$0.007 \\
Ultra-hard    &   91 &   91 & 2.014$\pm$0.008 &&   91 & 2.019$\pm$0.013 \\
Soft and hard &  345 &  345 & 2.009$\pm$0.005 &&  345 & 2.038$\pm$0.010 \\
Only soft     &  922 &  922 & 2.007$\pm$0.006 &&  914 & 2.005$\pm$0.013 \\
Only hard     &   52 &   52 & 1.99 $\pm$0.03  &&   51 & 2.10 $\pm$0.05  \\      
       \noalign{\smallskip}
       \hline
   \end{tabular}
\end{table}


\end{document}